\documentclass[aps,10pt,prd,preprintnumbers,amsmath,amssymb,nofootinbib,superscriptaddress,a4paper,showpacs]{revtex4-1}
\pdfoutput=1
\usepackage[T1]{fontenc}

\usepackage{tabu}
\usepackage{comment}
\usepackage{wasysym}
\usepackage{color}
\usepackage[caption=false]{subfig}
\usepackage[usenames,dvipsnames,table]{xcolor}
\usepackage{graphicx,amsmath,amssymb,amsthm,multirow,array,bm,bbm,esint,amsmath}
\usepackage[mathscr]{eucal}
\usepackage[bbgreekl]{mathbbol}
\usepackage{epsf,amsfonts}
\usepackage{float}
\usepackage{slashed}
\newcommand{\cN}{{\cal N}}
\newcommand{\psib}{{\overline{\psi}}} 
\newcommand{\Tr}{{\rm Tr\;}}
\newcommand{\hf}{\frac{1}{2}}

\usepackage{soul}

\def\nn{\nonumber}
\def\bec{\begin{center}}
\def\eec{\end{center}}
\def\beq{\begin{equation}}
\def\eeq{\end{equation}}
\def\bea{\begin{eqnarray}}
\def\eea{\end{eqnarray}}

\begin{document}

\title{Complex Langevin Dynamics in Large $N$ Unitary Matrix Models}

\author{Pallab Basu}
\email{pallab.basu@icts.res.in}

\author{Kasi Jaswin}
\email{jaswin@icts.res.in}

\author{Anosh Joseph}
\email{anosh.joseph@icts.res.in}

\affiliation{International Centre for Theoretical Sciences (ICTS-TIFR), \\ Tata Institute of Fundamental Research, \\ Bangalore, 560089 INDIA}

\date{\today}

%%%%%%%%%%%%%%%%
\begin{abstract}
%%%%%%%%%%%%%%%% 

Using complex Langevin dynamics we examine the phase structure of  complex unitary matrix models and compare the numerical results with analytic results found at large $N$. The actions we consider are manifestly complex, and thus the dominant contribution to the path integral comes from the space of complexified gauge field configuration. For this reason, the eigenvalues of unitary matrix lie off the unit circle and venture out in the complex plane. One example of a complex unitary matrix model, with Polyakov line as the unitary matrix, is an effective description of a QCD at finite density and temperature with $N$ number of colors and $N_f$ number of quark flavors defined on the manifold $S^1 \times S^3$. A distinct feature of this model, the occurrence of a series of Gross-Witten-Wadia transitions, as a function of the quark chemical potential, is reproduced using complex Langevin simulations. We simulate several other observables including Polyakov lines and quark number density, for large $N$ and $N_f$ and found excellent match with the analytic results.

%%%%%%%%%%%%%%
\end{abstract}
%%%%%%%%%%%%%%


\maketitle

%%%%%%%%%%%%%%%%%%%%%%%%%%%%%%%%%%%%%%%%%%%%%%%%%%%%%%%%%%%
\section{Introduction}
\label{sec:intro}
%%%%%%%%%%%%%%%%%%%%%%%%%%%%%%%%%%%%%%%%%%%%%%%%%%%%%%%%%%%

A nonperturbative study of the phase structure of QCD at finite temperature and nonzero baryon chemical potential still remains an outstanding problem \cite{Muroya:2003qs, deForcrand:2010ys}. This is due to the fact that the fermion determinant becomes complex and the theory has a sign problem. The standard methods to study the theory, lattice QCD algorithms based on importance sampling, fail to produce reliable simulations. There have been recent developments in tackling this problem. One method is the use of complex Langevin dynamics with stochastic quantization \cite{Klauder:1983sp, Parisi:1984cs}. This method is not based on importance sampling but instead on a stochastic exploration of an enlarged (complexified) field configuration space. Another recently proposed method is the Lefschetz thimble method \cite{Cristoforetti:2012su, Fujii:2013sra, DiRenzo:2015foa, Tanizaki:2015rda, Fujii:2015vha, Alexandru:2015xva}, which is also based on complexification of the original real field variables. 

The complex Langevin method was proposed in the early 1980s by Klauder \cite{Klauder:1983nn, Klauder:1983zm, Klauder:1983sp} and Parisi \cite{Parisi:1984cs}. Though it became popular in the beginning certain problems were found immediately after. First one was the problem of runaways, where the simulations would not converge and the second one was the problem of convergence to a wrong limit. In recent years the complex Langevin method has been revived, with sometimes cases of impressive success \cite{Berges:2005yt, Berges:2006xc, Berges:2007nr, Bloch:2017sex, Aarts:2008rr, Pehlevan:2007eq}. It has been shown recently that complex Langevin simulations produce seemingly correct answer, even when the fermion sign problem is severe, for one-, three- and four-dimensional field theories with nonzero chemical potential \cite{Aarts:2008wh, Aarts:2009hn, Aarts:2010gr, Aarts:2011zn}. There have also been studies of supersymmetric matrix models based on complex Langevin dynamics. See Refs. \cite{Ito:2016efb,Ito:2016hlj,Anagnostopoulos:2017gos}.      

In this paper, we consider a large $N$ unitary matrix model at low temperature with a finite quark chemical potential and quark mass. This model is obtained from the one-loop formulation of QCD on $S^1 \times S^3$ at finite temperature with finite quark chemical potential $\mu$, quark mass $m$, and with $N$ number of colors and $N_f$ number of quark flavors. After integrating out the quark and gauge degrees of freedom we obtain the model of our interest -- a conventional unitary matrix model with a complex action. The unitary matrix $U$ in this model is the holonomy (Wilson loop) of the gauge field around the thermal time circle in Euclidean space. We can use the expectation value of the trace of Polyakov line in the fundamental representation as order parameter for the phase transitions. It is zero in the confined phase and non-zero in the deconfined phase. The model is interesting as it exhibits a rich thermal phase structure. When the chemical potential passes one of the quark energy levels there is a third order Gross-Witten-Wadia (GWW) transition from a confined to a deconfined phase and back again. This model also exhibits another interesting feature known as the {\it Silver Blaze} behavior. When the quark mass is nonvanishing the bulk observables of the model are nearly zero until the onset transition to the deconfined phase, which occurs when the chemical potential reaches the value of the lightest quark mass. 
 
In the matrix model with complex action, the dominant contributions to the functional integral come from complexified gauge field configurations. Due to this reason, the saddle point eigenvalues of the unitary matrix $U$ lie off the unit circle, on a contour in the complex plane. The eigenvalues of $U$ can be written as $\exp(i\theta_i)$ with $\theta_i$ the angle variables and $i = 1, \cdots, N$. We can make a change of variables such that the functional integral reduces to an integral over $\{ \theta_i \}$. At large $N$, the functional integral is dominated by a single saddle point but since the action is complex this saddle point configuration lies out in the complex plane where the $\theta_i$ are no longer real. As a consequence, the Polyakov line and the inverse Polyakov line are not equal, that is, $\langle P \rangle \neq \langle P^{-1} \rangle$. Through complex Langevin simulations we indeed confirm this behavior. In fact the behavior of inverse Polyakov line precedes that of the Polyakov line as a function of chemical potential. This feature was observed analytically in an earlier work by Hands {\it et al.} in Ref. \cite{Hands:2010zp}.

In this paper, we examine this large $N$ unitary matrix model using complex Langevin simulations. It is possible to generate representative field configurations by integrating a stochastic differential equation, known as the complex Langevin equation. The drift terms arising from the complex action force the field variables to evolve in an extended (complexified) field space, in which the large regions where the observables are plagued by phase fluctuations are avoided \cite{Aarts:2008rr}.

When $N$ is large, we can consider the gauge field, corresponding to the angles of the Polyakov line, as a distribution on a contour. From the equation of motion, the saddle point distribution of the Polyakov line eigenvalues can be calculated analytically and plotted by mapping the angles from an arc on the unit circle to a contour over the same range of angles in the complex plane \cite{Hands:2010zp}. The theory is said to be in a confined phase when the contour on which the Polyakov line eigenvalues are distributed is closed. The contour opens up in between quark energy level transitions giving rise to a deconfined phase in the theory. The third derivative of the grand potential is discontinuous at each energy level crossing. These are characteristic features of a third order, GWW transition \cite{Gross:1980he, Wadia:1980www, Wadia:1980cp}.

This paper is organized as follows. In Sec. \ref{sec:cLd} we give a brief outline of the complex Langevin dynamics and stochastic quantization. In Sec. \ref{sec:ab-model} we discuss a simple yet nontrivial matrix model called the ab-Model, which is a complexified version of the Gross-Witten-Wadia (GWW) model. This model has two phases, confined and deconfined, and it exhibits a third-order phase transition. In Sec. \ref{sec:gt-to-u-m-model} we discuss another interesting large $N$ unitary matrix model, which arises in the one-loop formulation of QCD on compact spaces. This model possess a tower of quark energy levels due to compactification and is defined for positive and negative chemical potential values. We then focus on to a truncated cousin of this model - a single quark energy level matrix model with positive chemical potential. This model also has a complex action and captures the physics we are interested in without loss of generality. We can define a transition parameter (which is function of the temperature and chemical potential) in this model and as we change this parameter, the model exhibits confinement/deconfinement phase transitions. We show the eigenvalue distributions corresponding to the confined (closed) and deconfined (gapped) phases of the theory using complex Langevin simulations. We also simulate the behaviors of Polyakov lines and fermion number density as a function of the transition parameter. We simulate the model for a range of temperatures and chemical potentials to study its phase structure. We also show the phase diagram of the model, at low temperature, on the $(\mu, \beta)$ plane, in the vicinity where quark energy level equals the chemical potential. We then simulate the model at large quark mass and show that the bulk observables exhibit the Silver Blaze behavior -- the observables are roughly zero until the onset transition to the deconfined phase, which occurs when the chemical potential equals quark mass. We then move on to discuss the single-level model with a simple nontrivial gauge interaction turned on. We study the behavior of observables as a function of the interaction parameter. We see that the model prefers to stay in the confined phase as the interaction strength is increased. In Sec. \ref{sec:conclusions} we provide conclusions and discussions. In Appendix. \ref{sec:qcd-finite-cp} we use complex Langevin dynamics to simulate QCD on $S^1 \times S^3$ at finite chemical potential and low temperature. We are able to reproduce the series of GWW transitions, as a function of the chemical potential, as described in Ref. \cite{Hands:2010zp}. Our simulations also reproduce the level structure feature of the bulk observables - fermion number density, pressure and energy - of the model. In Appendix \ref{sec:appendix_c} we investigate the reliability of complex Langevin method by studying the probability distribution for the magnitude of the drift term and the Langevin runtime history of the unitarity norm. We note that the probability distribution for the magnitude of the drift term falls of (possibly) with a power law even though the simulations show excellent agreement with analytical results. We think that these diagnostics need further investigations and we save it for future work.  

%%%%%%%%%%%%%%%%%%%%%%%%%%%%%%%%%%%%%%%%%%%%%%%%%%%%%%%%%%%%%%%%%%
\section{Complex Langevin Dynamics}
\label{sec:cLd}
%%%%%%%%%%%%%%%%%%%%%%%%%%%%%%%%%%%%%%%%%%%%%%%%%%%%%%%%%%%%%%%%%%

The central idea of stochastic quantization is that expectation values of observables are obtained as equilibrium values of a stochastic process \cite{Parisi:1980ys, Damgaard:1987rr}. In order to achieve this we evolve the system in a fictitious time $\tau$, subject to a stochastic noise. That is, the system evolves according to Langevin dynamics. When the action is complex it is still possible to consider Langevin dynamics. The force (gradient of the action) becomes complex in this case making the fields also complex during the evolution.   

In this work we make use of complex Langevin dynamics with stochastic quantization to study large $N$ unitary matrix models with complex actions. They exhibit sign problem due to the fact that the action is complex. Standard Monte Carlo methods fail to produce the correct equilibrium distributions of these models. We can use discretized complex Langevin equation with Euler method (which is a first order algorithm) to find the equilibrium field distributions of these models.

We note that in unitary models with real action the domain of the angular variables $\theta_i$, with $i = 1, \cdots, N$, is $[0, 2\pi)$. After complexification the domain becomes a strip with the the domain $[0, 2\pi)$ along the real directions and $(-\infty, \infty)$ along the imaginary directions. The range of $e^{i \theta_i}$, that is, the complexified eigenvalues of $U$ has the whole complex plane as the range. Let us take $\theta_i(\tau)$ as the complexified angle variables of the gauge link $U(\tau)$ at a Langevin time $\tau$. (From now on we take $\theta_i$ to be complex, in this paper, unless otherwise specified.) We have the discrete Langevin evolution equation
\bea
\label{eq:lang-1}
\theta_i(\tau + \Delta \tau) &=& \theta_i(\tau) - \left[\frac{\partial S}{\partial \theta_i(\tau)}\right] \Delta \tau + \sqrt{\Delta \tau}~\eta_i(\tau),
\eea 
where $\Delta \tau$ is the Langevin time step, and $\eta_i(\tau)$ is a Gaussian random variable satisfying the conditions
\bea
\label{eq:randoms}
\langle \eta_i(\tau) \rangle = 0,~~~ \langle \eta_i(\tau) \eta_j(\tau') \rangle = 2 \delta_{ij} \delta_{\tau \tau'}.
\eea

If the action $S$ is of the order $N^2$, then strictly at infinite $N$ the fluctuation term in Eq. (\ref{eq:lang-1}) could be safely dropped. Moreover, to reduce excursions in the imaginary directions of the field configurations, which would spoil the validity of the method, we should use real Gaussian random variables \cite{Aarts:2009uq, Aarts:2011ax, Aarts:2013uza}.

We also need to impose the $SU(N)$ constraint on the complexified angular variables after each Langevin time step. That is, we need
\beq
\label{eq:SU-N-constraint}
\sum_{i=1}^N \theta_i(\tau) = 0.
\eeq

This can be easily implemented by subtracting the average value $\theta_{\rm av}(\tau)$ from each $\theta_i(\tau)$ variable, i.e.
\beq
\label{eq:SU-N-step}
\theta_i \rightarrow \theta_i-\frac{1}{N}\sum_{i=1}^N \theta_i(\tau).
\eeq

Note that this condition is implemented in a holomorphic way. That is, both of the real and imaginary parts of $\theta_{\rm av}(\tau)$ are subtracted. Ideally, one should eliminate one variable (say $\theta_1$) using the constraint Eq. \eqref{eq:SU-N-constraint} and stochastically quantize the remaining variables. To proceed we need to justify that our method of imposing the constraint after each time step leads to the same result.
	
A set of stochastic flow equations involving the gradient of the action like the one given in  Eq. \eqref{eq:lang-1} is invariant under the orthogonal transformation of variables
\beq
\tilde \theta_i = \sum_j O_{ij} \theta_j,
\eeq
where $O$ is an orthogonal matrix. In terms of the transformed variables, the  set of equations is
\begin{align}
d\tilde \theta_i &= -\sum_j O_{ij} \left[\frac{\partial S}{\partial \theta_j(\tau)}\right] d \tau + \sqrt{d \tau}~ \sum_j O_{ij} \eta_j(\tau) \\
&=-\sum_k \sum_j O_{ij} O_{kj} \left[\frac{\partial S}{\partial \tilde \theta_k(\tau)}\right] d \tau+ \sqrt{d \tau}~ \sum_j O_{ij} \eta_j(\tau) \\
&= -\left[\frac{\partial S}{\partial \tilde \theta_i(\tau)}\right] d \tau+ \sqrt{d \tau}~ \tilde \eta_i(\tau),
\end{align}
where we have used the orthogonality of matrix $O$. Orthogonality also guarantees that new random variables $\tilde \eta$s satisfy the condition Eq. \eqref{eq:randoms}.

Now, we can always choose an $O$ such that $\tilde \theta_1 =\frac{1}{\sqrt{n}} \sum_i {\theta_i}$. In terms of the transformed variables it is easy to understand why our method works. The constraint Eq. \eqref{eq:SU-N-constraint} is now rewritten simply as $\tilde \theta_1 = 0$. If we start with a set of variables which already satisfies this constraint then a valid Langevin time evolution step may be performed by simply discarding any evolution in $\tilde \theta_1$. This is precisely our method of imposing constraint after each time step, rewritten in terms of the new variables. To emphasize, one can straight forwardly argue that in terms of old variables, this step is same as Eq. \eqref{eq:SU-N-step}. Our argument works for any arbitrary linear constraint.

We note that there also exists another complementary method in which one could implement complex Langevin dynamics directly on the matrix variables $U(\tau)$. In this case the evolution equation takes the form
\beq
U(\tau + \Delta \tau) = R(\tau) U(\tau)
\eeq
where the matrix $R$ is a stochastic unitary matrix. We note that this method can be used for studying similar models in higher spacetime dimensions.

In this paper, we use the first method described above where the link field $U$ is diagonalized and the $SU(N)$ constraint has been imposed.

We note that the complexification of the dynamical variables in the theory can change the Langevin evolution drastically. There can be unstable directions on the complexified field configuration space and the Langevin evolution can converge to wrong limits. One should be aware that the numerical integration must be performed carefully when the Langevin trajectory makes a large excursion into imaginary directions. One could, in principle, use a small step size but it still has two problems: $(i)$ it does not solve instabilities in all directions and $(ii)$ it will result in a slow evolution, which can be computationally very inefficient. In order to take care of both of these problems we follow the algorithm given by Aarts {\it et al.} in Ref. \cite{Aarts:2009dg}. We consider an adaptive step size in the discretized complex Langevin equations. We compute the absolute value of the maximum drift, $K_{\max}$, at a given Langevin time $\tau$
\beq
K_{\max}(\tau) \equiv \operatorname*{max}_i \sqrt{\left(\left[\frac{\partial S}{\partial \theta_i(\tau)}\right]^R\right)^2 + \left(\left[\frac{\partial S}{\partial \theta_i(\tau)}\right]^I\right)^2},
\eeq
and the stepsize for the next evolution step is taken to be 
\beq
\Delta \tau = \frac{\gamma}{K_{\max}(\tau)},
\eeq
where $\gamma$ is a number chosen according to the model we want to simulate. In our simulations we typically take $\gamma$ to be ${\cal O}(1)$.

%%%%%%%%%%%%%%%%%%%%%%%%%%%%%%%%%%%%%%%%%%%%%%%%%%%%%%%%%%%%
\section{ab-Model}
\label{sec:ab-model}
%%%%%%%%%%%%%%%%%%%%%%%%%%%%%%%%%%%%%%%%%%%%%%%%%%%%%%%%%%%%

To demonstrate the effectiveness of Complex Langevin Dynamics, we begin by studying a simple, yet nontrivial model -- a complexified version of Gross-Witten-Wadia (GWW) Model \cite{Wadia:1980cp, Gross:1980he, Wadia:1980www,Buividovich:2015oju}. We refer to our model as \textit{ab-Model}. It has two phases, confined and deconfined, exhibiting a third-order phase transition. The action is given by
\begin{equation}
S = N \left(a \Tr U + b \Tr U^{\dagger}\right),
\label{abmodel}
\end{equation}
where $a,b \in \mathbb{C} $, $U$ is an element of $SU(N)$, and when $a=b$ it becomes the Gross-Witten-Wadia model.

Before proceeding further let us make a few generic comments. A linear term in $\Tr U$ breaks the center symmetry. Furthermore, the above action (or other polynomial generalization of it) is complex. If $a \neq b$, then the $\mathbb{Z}_2$ symmetry $U \rightarrow U^{\dagger}$ is broken. This implies $\langle \Tr U \rangle \neq \langle \Tr U^{\dagger} \rangle$. One may ask, that what it means in terms of manifestly gauge invariant operators. This means that the contribution from baryon and anti-baryon is different. Another related observation is one may naively expand Eq. (\ref{abmodel}) in a series

\begin{align}
Z = \int DU e^{-S}
= \int DU \left( 1 + N ab \Tr U \Tr U^{\dagger} + N^2 (ab)^2 (\Tr U \Tr U^{\dagger})^2 \cdots \right) + \\ \nn 
\left( N^N a^N \Tr U^N + N^N b^N \Tr U^{\dagger N} \right) + \cdots.
\end{align}

Here we have separated the ``mesonic'' and ``baryonic'' contributions. Due to the center symmetry only a center symmetry invariant combination of $\Tr U$ and $\Tr U^{\dagger}$ contributes. By mesonic contribution we mean product of traces for which sum of powers all the occurrence of unitary matrix and its inverse sum to zero. For a baryonic operator, the sum is only zero up to modulo $N$, i.e., proportional to a non-zero integral power of $N$. If baryonic contributions are neglected then Eq. (\ref{abmodel}) is equivalent to a model with parameters, $a = b = \sqrt{(ab)}$. We will later see that for center symmetry invariant operators, this equivalence is actually held in the ungapped phase. 

Expressing the action in diagonal gauge, the effective action becomes
\bea
S_{eff} &=& S_{\rm Vdm} + i N \mathcal{M} \sum_{i=1}^N \theta_i + N \left(a \sum_{i=1}^N e^{i\theta_i} + b \sum_{i=1}^N e^{-i\theta_i} \right),
\eea 
where the first term is the Vandermonde piece
\beq
S_{\rm Vdm} = \sum_{i,j = 1, i \neq j}^N -\dfrac{1}{2} \ln \left( \sin^2 \left( \frac{\theta_i - \theta_j}{2}\right)\right),
\eeq
and $\mathcal{M}$ is the Lagrange multiplier which ensures that $\det(U) = 1$. 
 
At large $N$, the theory is dominated by the saddle-point equation
\beq
\dfrac{\partial S_{eff}}{\partial \theta_i} = 0,
\eeq
which gives the equation of motion 
\beq
i \mathcal{M} + i \left( ae^{i\theta_i} - be^{-i\theta_i} \right) = \dfrac{1}{N} \sum_{j \neq i} \cot \left(\frac{\theta_i - \theta_j}{2}\right).
\eeq

On substituting $z_i = e^{i\theta_i}$ the equation of motion becomes
\beq
i \mathcal{M} + i az_i - i \left(\frac{b}{z_i} \right) = \frac{i}{N} \sum_{j \neq i} \left( \dfrac{z_i + z_j}{z_i - z_j} \right),
\eeq
and $\mathcal{M}$ is given by 
\beq
\mathcal{M} = \dfrac{1}{N}\sum_{i=1}^N \left(\dfrac{b}{z_i} - az_i \right).
\eeq
In the saddle point, $\mathcal{M}$ may have a nonzero value and could be thought as effective baryon number.

At $N \rightarrow \infty$ limit, we can replace the summation by an integral over a nondecreasing function 
\beq
\dfrac{1}{N} \sum_{i=1}^N \rightarrow \int_{-\pi}^\pi \dfrac{ds}{2\pi}, 
\eeq
and performing a change of variables from $s$ to complex variables $z(s)$
\beq
\label{eq:change-var1}
\dfrac{ids}{dz} = \rho(z),
\eeq
the equation of motion becomes 
\beq
\label{eq:eqofmotion}
\mathcal{M} + a z - \left( \dfrac{b}{z} \right) = P \oint_c \dfrac{d\omega }{2\pi i} \rho(\omega) \left( \dfrac{z+\omega}{z-\omega} \right),
\eeq
and $P$ implies we are taking the principal value of the integral.

%%%%%%%%%%%%%%%%%%%%%%%%%%%% 
\subsection{Ungapped Phase}
%%%%%%%%%%%%%%%%%%%%%%%%%%%%

In the GWW model, it is known that for small potential, i.e., $a < 0.5$, the theory is in an ungapped phase. Assuming a similar picture also holds for the \textit{ab-model}, we solve it by taking an ansatz for $\rho(z)$ in ungapped phase as, 
\beq
\rho(\omega) = A_0 + \frac{A_1}{\omega} + \frac{A_2}{\omega^2} + \cdots  
\eeq
then 
\beq
P \oint_C \dfrac{d\omega}{2\pi i} \rho(w)  \left( \dfrac{z + \omega}{z - \omega} \right) = -A_0 z + \dfrac{A_2}{z} + \cdots
\eeq

Comparing with the left hand side of Eq. (\ref{eq:eqofmotion}) we have
\beq
A_0 = -a~{\rm and}~A_2 = -b.
\eeq

Therefore $\rho$ becomes, 
\beq
\rho(z) = - a + \dfrac{A_1}{z} - \dfrac{b}{z^2} + \cdots. 
\eeq

We also find 
\beq
\label{eq:lagrange_multiplier}
\mathcal{M} = \oint_C \dfrac{dz}{2\pi i} \rho(z)\left( \dfrac{b}{z} - az \right) = 0,
\eeq
which indicates that the theory is in an ungapped phase. Demanding normalization of $\rho(z)$ 
\beq
\oint \dfrac{dz}{2\pi i} \rho(z) = 1, 
\eeq
we fix $A_1 = 1$. 

Therefore, 
\beq
\label{eq:ungapped_density}
\rho(z) = \dfrac{1}{z} - a - \dfrac{b}{z^2} + \cdots. 
\eeq

We can solve for the contour, where $\rho(z)$ is positive definite, by integrating Eq. \eqref{eq:change-var1} 
\beq
\label{eq:change-var2}
is = \ln(z) - az + \dfrac{b}{z} + c.
\eeq

Since $s$ is purely real, and assuming that 
\beq
z = r(\theta) e^{i\theta} ,~~a = |a|e^{i \phi_1}~{\rm and}~b = |b|e^{i\phi_2},
\eeq
the above equation is satisfied only if the real part of the right hand side is zero. That is,  
\beq
\ln(r(\theta)) - |a| r(\theta) \cos(\theta + \phi_1)  + \dfrac{|b|}{r(\theta)} \cos(\theta - \phi_2) + Re(c) = 0.
\eeq
 
To fix $c$, we invoke the condition that $\det(U) =1$, i.e., $\sum_{i=1}^N \theta_i =0$, which translates to 
\beq
\int_C \dfrac{dz }{2\pi i}\ln(z) \rho(z) = 0,
\eeq
where the branch-cuts are taken from $z = 0$ to the point $z(\pm \pi) $. Replacing $\ln(z)$ using Eq. \eqref{eq:change-var2}, the above equation becomes
\bea
&& \int \dfrac{dz}{2\pi i} \left(is + az - \dfrac{b}{z} - c \right) \rho(z) =  0 \nn \\
&&\quad \quad \Rightarrow - c + \oint \dfrac{dz}{2\pi} \rho(z) s  = 0 \nn \\ 
&&\quad \quad \Rightarrow - c + i \int_{-\pi}^{\pi} \dfrac{ds}{2\pi} s = 0 \nn \\ 
&&\quad \quad \Rightarrow c  =  0.
\eea

Hence the contour is got by solving the transcendental equation
\beq
\label{eq:ungapped_contour}
\ln(r(\theta)) - |a| r(\theta) \cos(\theta + \phi_1)  + \dfrac{b}{r(\theta)} \cos(\theta - \phi_2) = 0.
\eeq

\begin{figure}[H]
	\centering
	\includegraphics[scale=0.4]{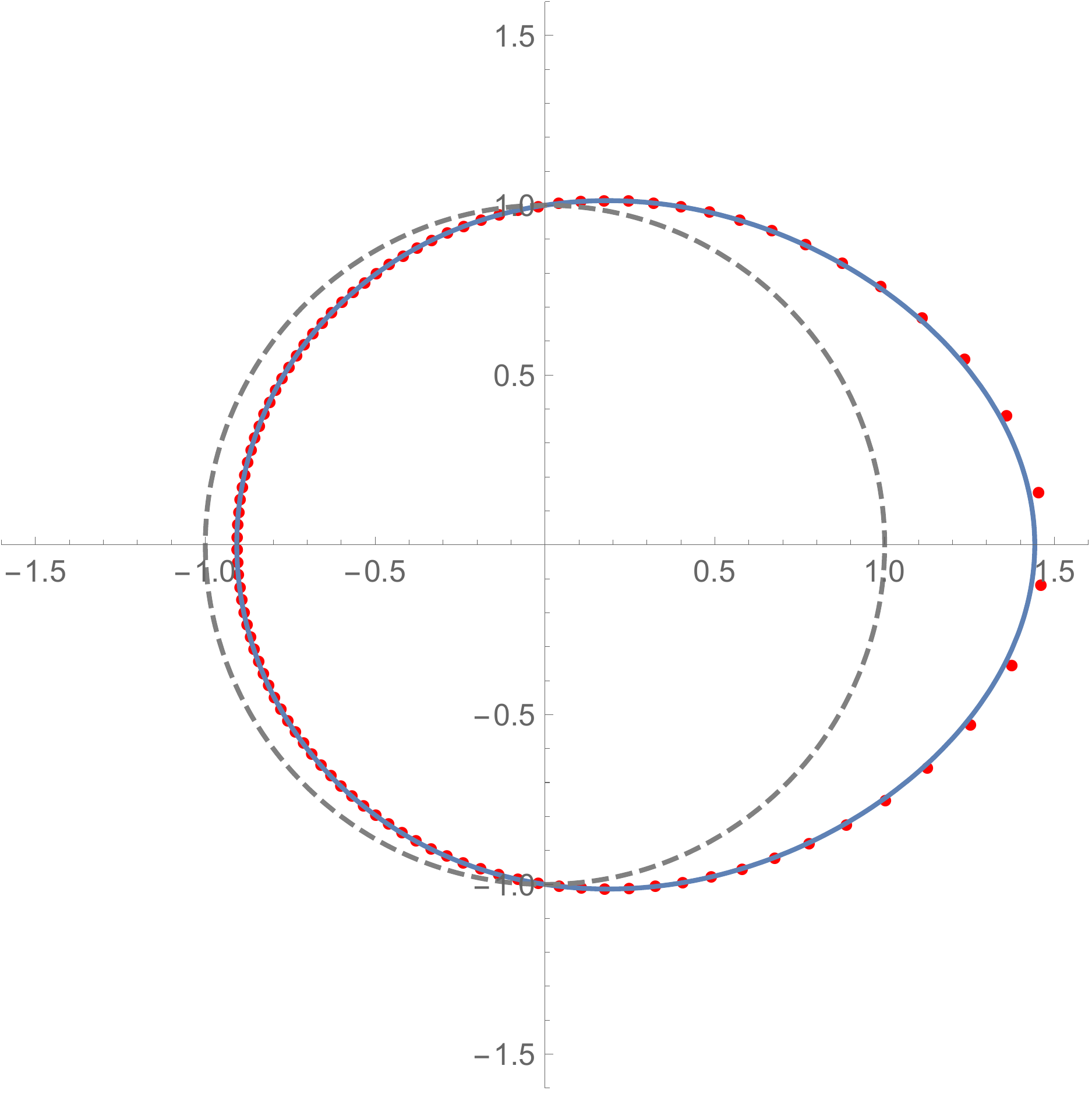}
	\caption{The distribution of eigenvalues of $ab$-model with parameters $a = 0.35$, $b = 0.2$ and $N = 100$. The solid curve is the analytical result. The data are obtained through complex Langevin simulations without noise. We used a fixed Langevin step size $\Delta \tau = 0.00001$ and evolved the system for $45000$ steps. The dashed unit circle is guide to the eye.
}
	\label{fig:ab-model_a0p35_b0p2_no_noise}
\end{figure}

\begin{figure}[H]
	\centering
	\includegraphics[scale=0.5]{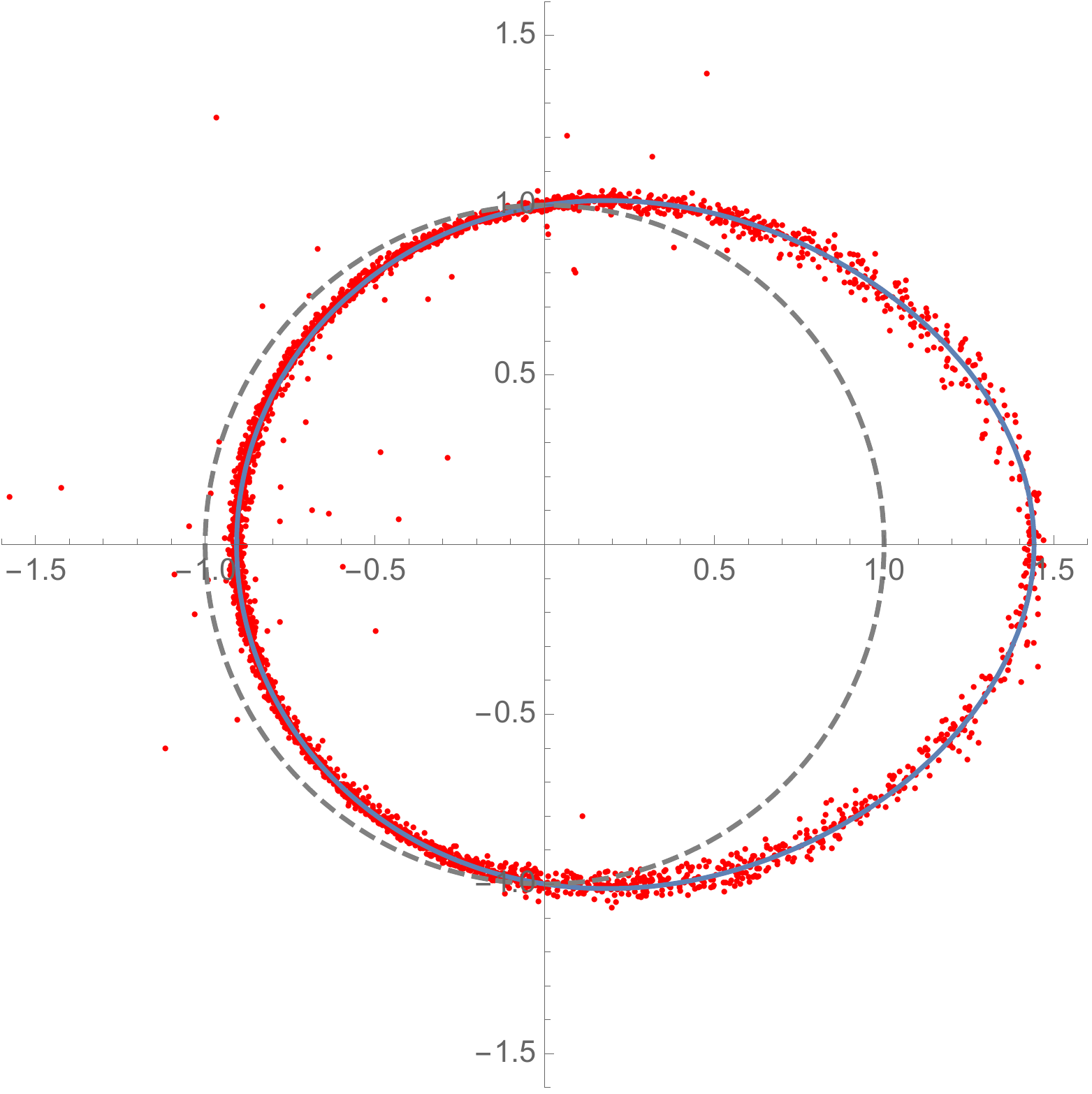}
	\caption{The distribution of eigenvalues of $ab$-model with parameters $a = 0.35$, $b = 0.2$ and $N = 100$. The solid curve is the analytical result. The data are obtained through complex Langevin simulations with fixed Langevin step size $\Delta \tau = 0.00001$, thermalization steps $N_{\rm therm} = 45000$, generation steps $N_{\rm gen} = 5000$ and with measurements performed with an interval of $250$ steps. The dashed unit circle is guide to the eye.
}
	\label{fig:ab-model_a0p35_b0p2_with_noise}
\end{figure}

Now we can compare the distribution of eigenvalues from complex Langevin dynamics with the analytic result for any $(a, b)$ combination. In Fig. \ref{fig:ab-model_a0p35_b0p2_no_noise} we show the analytical result and the data obtained through complex Langevin simulations without noise for parameters $a = 0.35$, $b = 0.2$ and $N = 100$. In Fig. \ref{fig:ab-model_a0p35_b0p2_with_noise} we show the result with Gaussian noise turned on. We see an excellent agreement between the analytical and numerical results. 

%%%%%%%%%%%% FIG %%%%%%%%%%%%%%%%%%%
%%%%%%%%%%%% FIG %%%%%%%%%%%%%%%%%%%

%%%%%%%%%%%% FIG %%%%%%%%%%%%%%%%%%%
%%%%%%%%%%%% FIG %%%%%%%%%%%%%%%%%%%

\begin{figure}[H]
	\centering
	\includegraphics[scale=0.45]{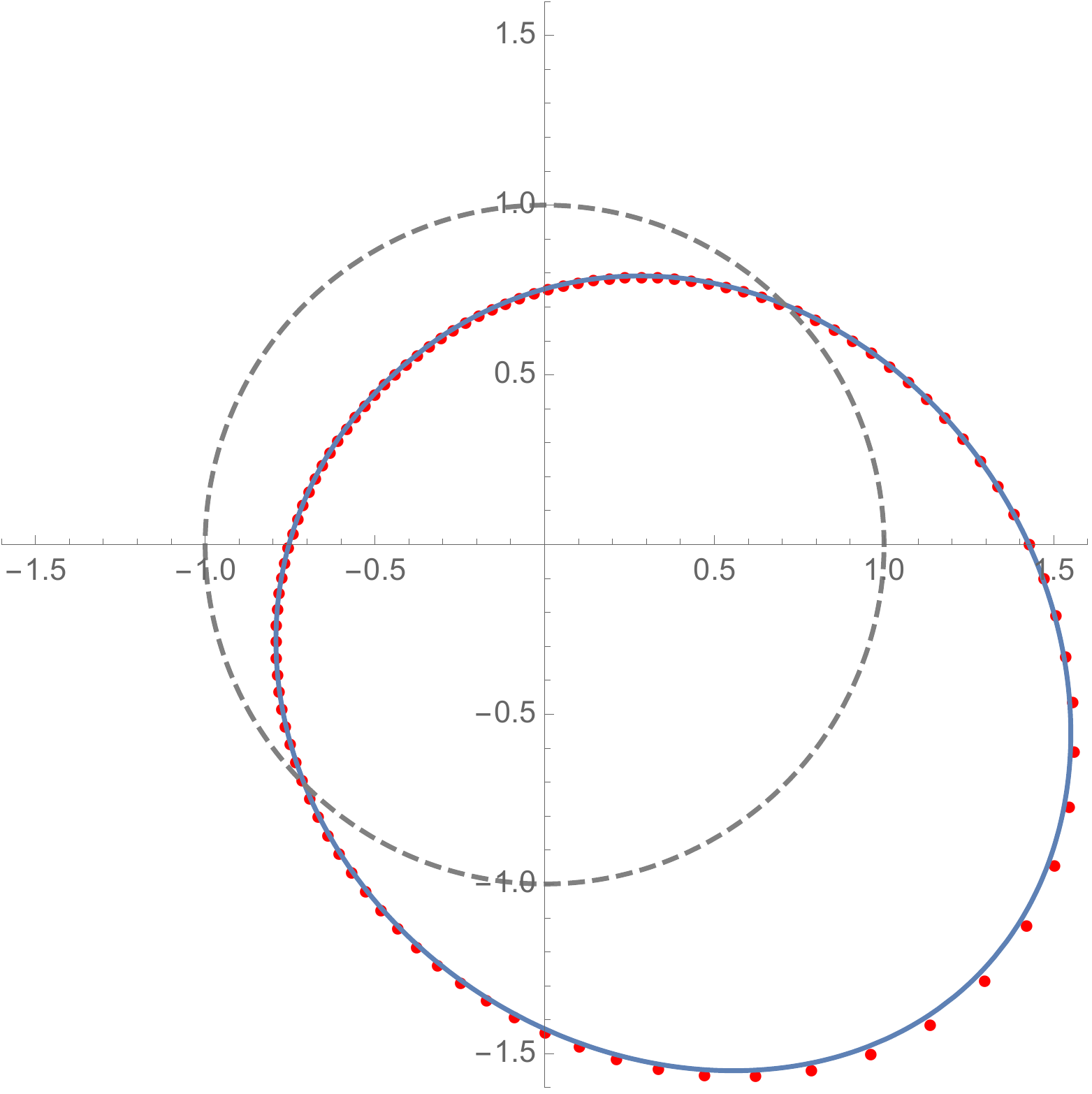}
	\caption{The distribution of eigenvalues of $ab$-model with parameters $a = 0.2 + i 0.2$, $b = -0.1 + i 0.1$ and $N = 100$. The solid curve is the analytical result. The data are obtained through complex Langevin simulations without noise. We used a fixed Langevin step size $\Delta \tau = 0.00001$ and evolved the system for $45000$ steps. The dashed unit circle is guide to the eye.}
	\label{fig:ab-model_a0p2i0p2_bm0p1i0p1_no_noise}
\end{figure}

\begin{figure}[H]
	\centering
	\includegraphics[scale=0.5]{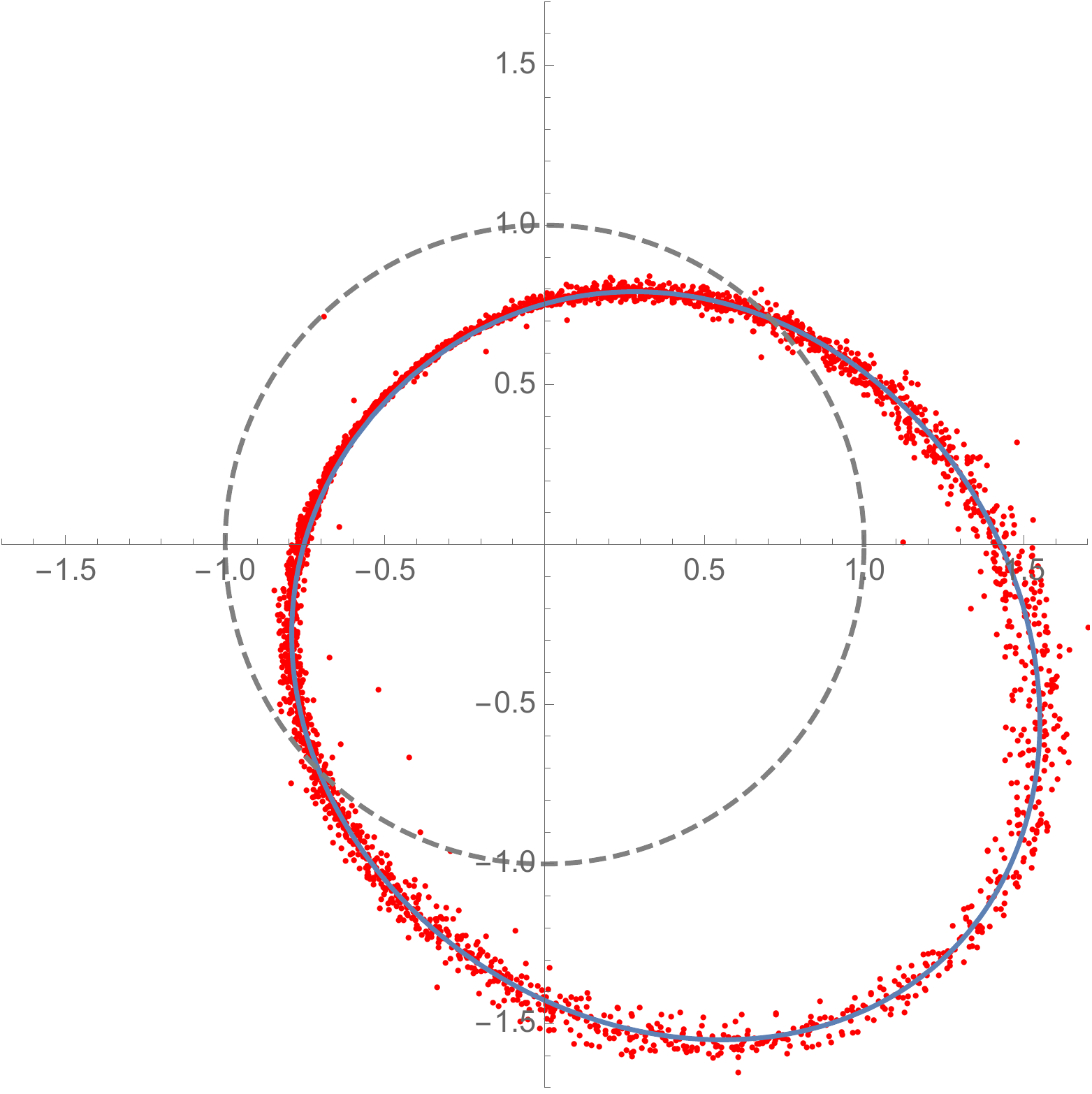}
	\caption{The distribution of eigenvalues of $ab$-model with parameters $a = 0.2 + i 0.2$, $b = -0.1 + i 0.1$ and $N = 100$. The solid curve is the analytical result. The data are obtained through complex Langevin simulations with fixed Langevin step size $\Delta \tau = 0.00001$, thermalization steps $N_{\rm therm} = 45000$, generation steps $N_{\rm gen} = 5000$ and with measurements performed with an interval of $250$ steps. The dashed unit circle is guide to the eye.
}
	\label{fig:ab-model_a0p2i0p2_bm0p1i0p1_with_noise}
\end{figure}

We also note that the complex Langevin simulations show excellent agreement with analytical results when the parameters are also complex. In Fig. \ref{fig:ab-model_a0p2i0p2_bm0p1i0p1_no_noise} we show the analytical result and the data obtained through complex Langevin simulations without noise for parameters $a = 0.2 + i 0.2$, $b = -0.1 + i 0.1$ and $N = 100$. In Fig. \ref{fig:ab-model_a0p2i0p2_bm0p1i0p1_with_noise} we show the result with Gaussian noise turned on. 

%%%%%%%%%%%% FIG %%%%%%%%%%%%%%%%%%%
%%%%%%%%%%%% FIG %%%%%%%%%%%%%%%%%%%

%%%%%%%%%%%% FIG %%%%%%%%%%%%%%%%%%%
%%%%%%%%%%%% FIG %%%%%%%%%%%%%%%%%%%

%%%%%%%%%%%%%%%%%%%%%%%%%
\subsection{Gapped Phase}
%%%%%%%%%%%%%%%%%%%%%%%%%

In the gapped phase, similar to GWW model, the eigenvalues lie on an open contour $C$.

To study this phase, we employ resolvent/spectral-curve method used in Ref. \cite{Hands:2010zp}, and reviewed in Ref. \cite{Marino:2004eq}. The resolvent is defined as 
\beq
\omega(z) = - \dfrac{1}{N} \sum_j \left( \dfrac{z + z_j}{z - z_j} \right).
\eeq

At large $N$ limit, $\omega(z)$ is analytic everywhere in the complex plane, except along a square-root branch cut running along $C$, and expressed as 
\beq
\label{eq:resolventdefn}
\omega(z) = - \int_C \dfrac{d z'}{2\pi i} \rho(z') \dfrac{z+ z'}{z-z'}.
\eeq 

For a given potential $V(z)$, the equation of motion (similar to Eq. \eqref{eq:eqofmotion})
\beq
z V'(z) = P \oint_C \dfrac{dz'}{2\pi i} \rho(z') \dfrac{z + z'}{z - z'} 
\eeq
can be expressed in terms of $\omega(z)$ using the Plemelj formulae
\beq
z V'(z) = \dfrac{1}{2} \left[ \omega(z + \epsilon) + \omega(z - \epsilon) \right],~~z \in C,
\eeq
where $z \pm \epsilon$ lies on either side of the branch cut and $\epsilon \rightarrow 0$ limit is taken.

We can also express $\rho(z)$ as the discontinuity of $\omega(z)$ across the cut $C$ as 
\beq
\label{eq:density_gapped}
z \rho(z) = \dfrac{1}{2} \left[ \omega(z + \epsilon) - \omega(z - \epsilon)\right].
\eeq

The expectation value of any function $G(z)$ can be found as 
\beq
\label{eq:average_value}
\int_C \dfrac{dz}{2\pi i} \rho(z) G(z) = \oint_{\tilde{C}} \dfrac{dz}{4\pi i z} \omega(z) G(z).
\eeq

For $ab$-model
\beq
\label{eq:eqofmotiongapped}
\omega(z) = - \mathcal{M} -az + \dfrac{b}{z} + f(z) \sqrt{ (z-\tilde{z})(z-\tilde{z}^* )}~,
\eeq
where $\tilde{z}, \tilde{z}^*$ are the end points of branch cut $C$ and $f(z)$ is an unknown function, which remains to be fixed. Since $\omega(z)$ has to be regular over the entire plane except along $C$ and the origin we can fix the form of $f(z)$ as 
\beq
f(z) = c + \dfrac{d}{z}.
\eeq 

Therefore $\omega(z)$ becomes (substituting $\tilde{z} = R e^{i \phi}$)  
\beq
\omega(z) = - \mathcal{M} - az + \dfrac{b}{z}  +  \left( c + \dfrac{d}{z} \right)\sqrt{z^2 + R^2 - 2Rz \cos (\phi)}.
\eeq

Normalization of $\rho(z)$, from Eq. \eqref{eq:resolventdefn}, translates to
\beq
\lim_{|z| \rightarrow 0} \omega (z) = 1
\eeq
and
\beq
\lim_{|z| \rightarrow \infty} \omega (z) = -1.
\eeq

This fixes $f(z)$ as
\beq
f(z) = a - \dfrac{b}{R z}.
\eeq

We also get two more relations between $R$, $\mathcal{M}$ and $\cos(\phi)$
\beq
\label{eq:cond1}
aR + \dfrac{b \cos(\phi)}{R} = 1+ \mathcal{M},
\eeq
and
\beq
\label{eq:cond2}
a \cos(\phi) R + \dfrac{b}{R} = 1 - \mathcal{M}.
\eeq 

\begin{figure}[H]
\bec
\begin{minipage}[h]{.35\textwidth}
  \includegraphics[width= 1.75in, height = 1.75in]{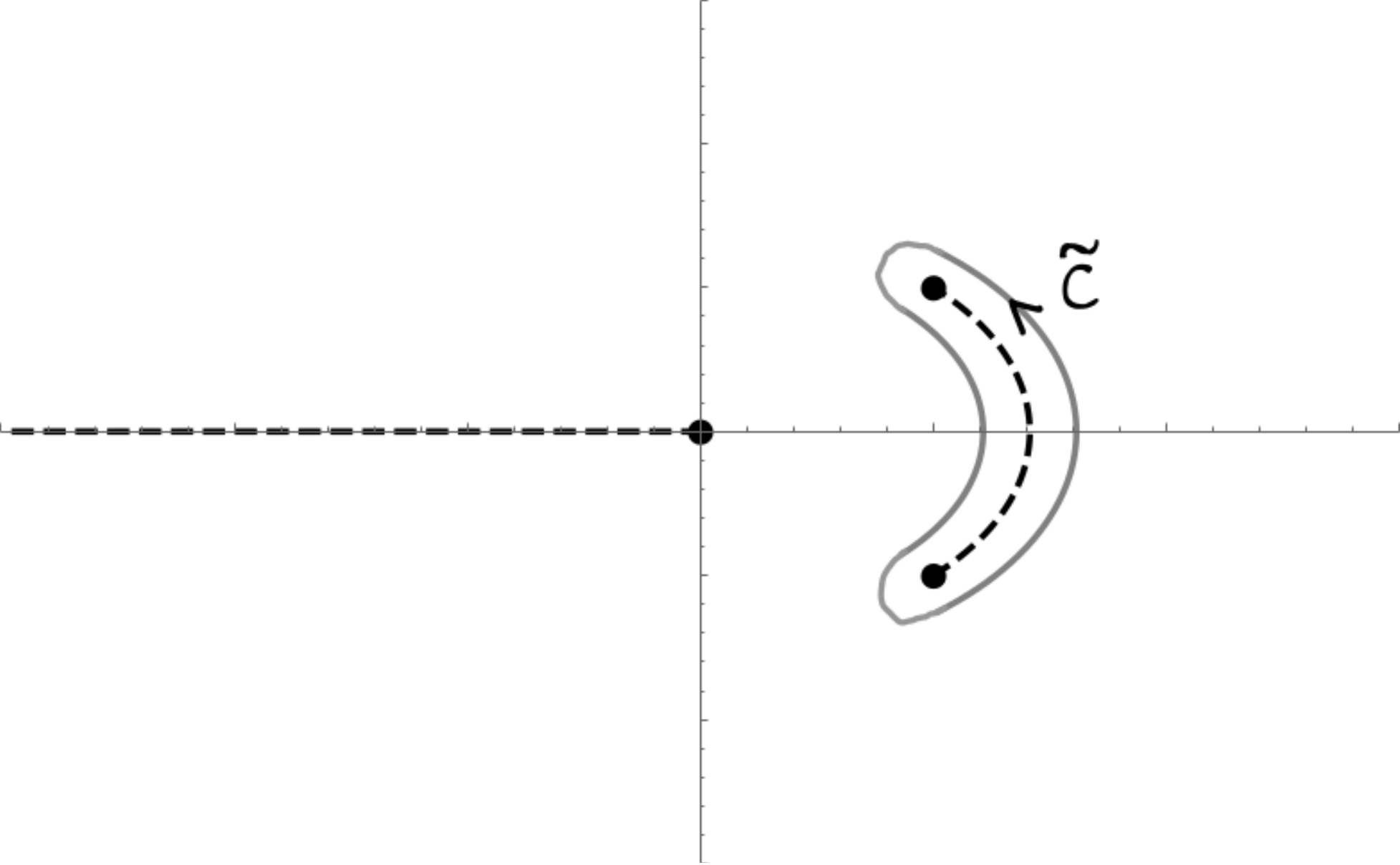}
  \caption{Actual contour over which Eq. \eqref{eq:gapped_log_integral} needs to be performed.}
  \label{fig:original_contour}
\end{minipage}
\hspace{0.05\textwidth}
\begin{minipage}[h]{.35\textwidth}
  \includegraphics[width= 1.75in, height = 1.75in]{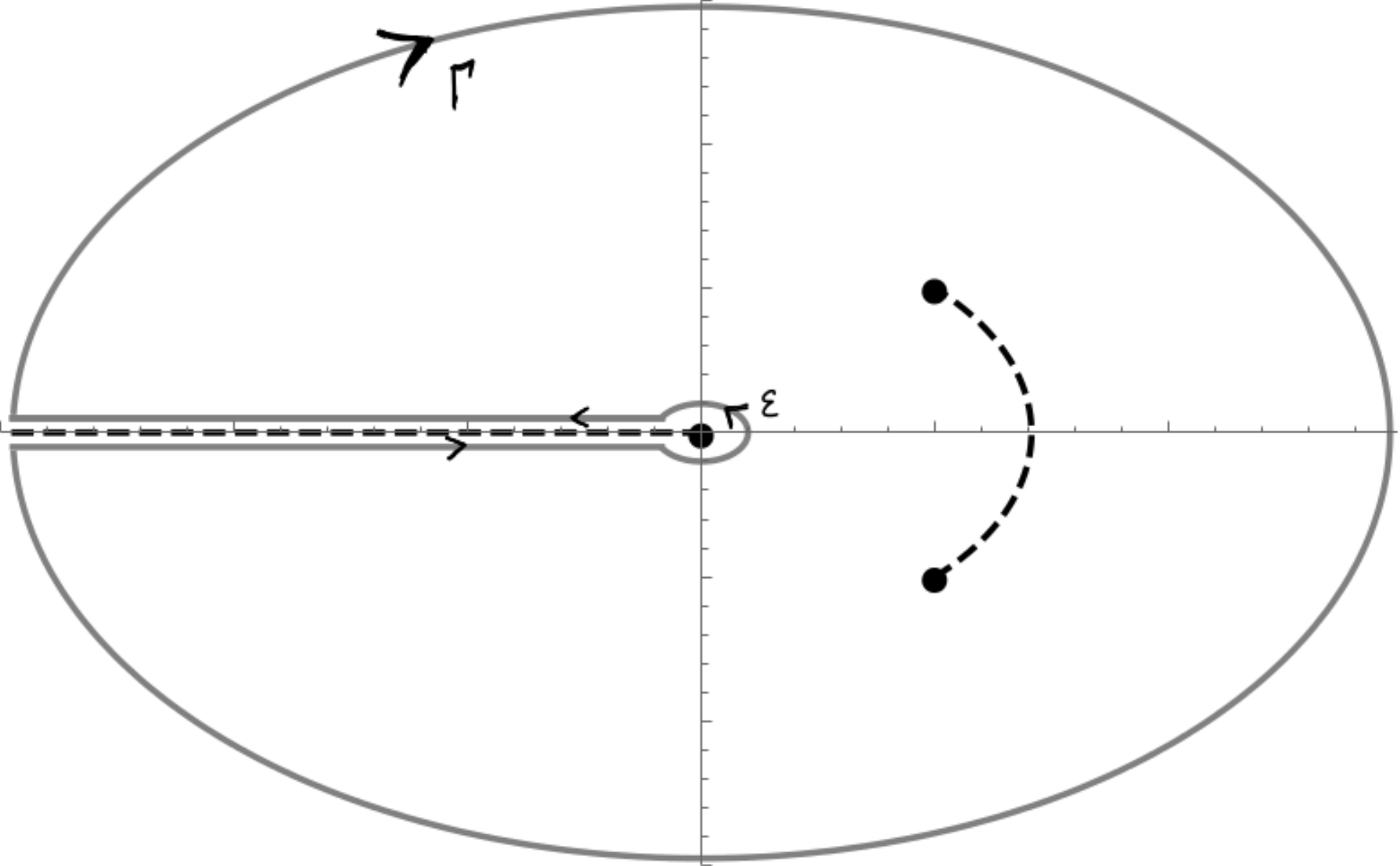}
  \caption{Deformed contour over which the integral was performed.}
  \label{fig:deformed_contour}
\end{minipage}
\eec
\end{figure}

To fix the three unknowns completely, we need a third equation, which comes from invoking the $\det(U) = 1$ condition, from Eq. \eqref{eq:average_value}
\beq
\label{eq:gapped_log_integral}
\int_{\tilde{C}} \dfrac{dz}{4\pi i z} \omega(z) \ln (z) = 0,
\eeq
where $\tilde{C}$ is a contour encircling the branch cut $C$, and the branch cut of $\ln(z)$ ranges from $(-\infty, 0)$. Deforming the contour Fig. \ref{fig:original_contour} to the one in Fig. \ref{fig:deformed_contour} and evaluating in $\epsilon \rightarrow 0$ and $\Gamma \rightarrow \infty$ limits, we find that the divergences arising from the cutoffs $\Gamma$ and $\epsilon$ cancel separately and we arrive at the following condition  
\bea
\label{eq:cond3}
&&\left(a R - \dfrac{b}{R}\right) \left[ \left( \dfrac{1-\cos(\phi)}{2} \right) \ln \left( \dfrac{1 - \cos(\phi)}{2} \right) + \left(\dfrac{1 + \cos(\phi)}{2}\right) \right] \quad \quad \quad \nn \\
&&\quad \quad \quad \quad \quad = \left(a R + \dfrac{b}{R} \right) \left( \dfrac{1 + \cos (\phi)}{2} \right) \ln (R).  
\eea
 
%%%%%%%%%%%% FIG %%%%%%%%%%%%%%%%%%%
%%%%%%%%%%%% FIG %%%%%%%%%%%%%%%%%%%

Now for a given $a, b$ we can numerically solve the Eqs. \eqref{eq:cond1}, \eqref{eq:cond2}, and \eqref{eq:cond3} for $R$, $\mathcal{M}$ and $\cos(\phi)$, and hence fix $\omega(z)$ completely. Also from Eq. \eqref{eq:density_gapped} we can fix $\rho(z)$ 
\beq
\rho(z) = \left(\dfrac{a}{z} - \dfrac{b}{R z^2} \right) \sqrt{z^2 + R^2 - 2Rz \cos (\phi)}.
\eeq
 
\begin{figure}[H]
\centering
\includegraphics[scale=0.25]{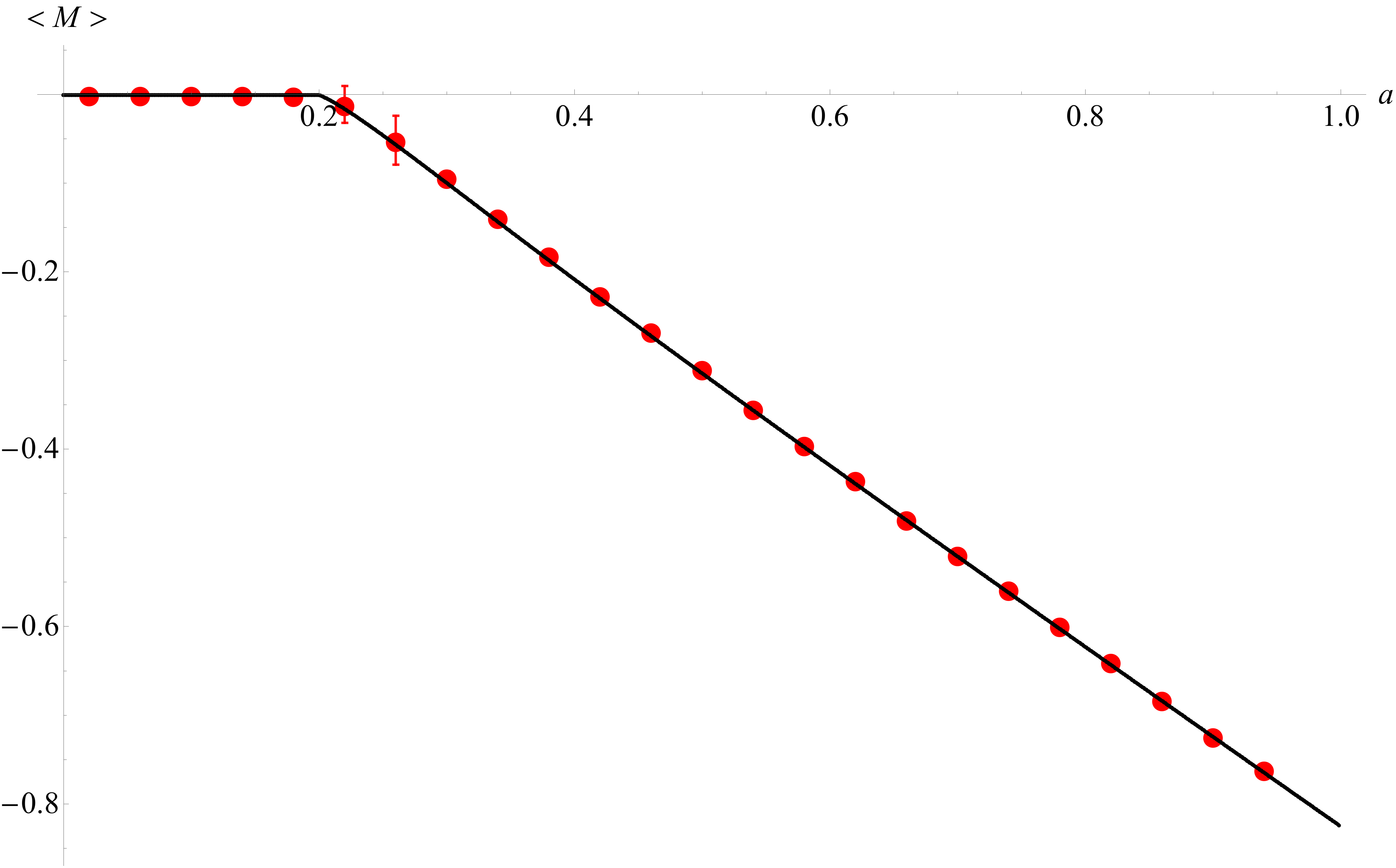}
\caption{The value of $\mathcal{M}$ at $(a, 2a)$ for the $ab$-model with $N = 100$. The solid curve is the analytical result. The data are obtained through complex Langevin simulations with adaptive step size $\Delta \tau \leq 0.00005$, thermalization steps $N_{\rm therm} = 35000$, generation steps $N_{\rm gen} = 250000$ and with measurements performed with an interval of $500$ steps.
}
\label{fig:value-of-M-at-a-2a}
\end{figure}

From Eq. \eqref{eq:lagrange_multiplier}, we can numerically compute $\mathcal{M}$, both in ungapped and gapped phases, and compare it against analytical results. Choosing $b = 2.0 a$ and varying $a$ from $0$ to $1.2$, we find that it matches very well both in ungapped and gapped regimes -- see Fig. \ref{fig:value-of-M-at-a-2a}. (Gap opening point can be found from Fig. \ref{fig:phase_space}.)

%%%%%%%%%%%% FIG %%%%%%%%%%%%%%%%%%%
%%%%%%%%%%%% FIG %%%%%%%%%%%%%%%%%%%

Similarly we compare other observables, $\big \langle \Tr(U) \big \rangle $ and $\big \langle \Tr(U^{-1}) \big \rangle $. Analytically $ \big \langle \Tr(U) \big \rangle $ is given by, 

\beq
 \big \langle \Tr(U) \big \rangle = 
  \begin{cases} 
   \oint \dfrac{dz}{2\pi i} \left( \dfrac{1}{z} - a - \dfrac{b}{z^2} \right) z   =  -b & \text{(Ungapped)} \\
   \oint_{\tilde{C}} \dfrac{d z}{4 \pi i} \dfrac{w(z)}{z} z   =  \left( \dfrac{\cos \phi + 1}{4} \right) \left (a (\cos \phi  - 1)R^2 - 2b \right)       & \text{(Gapped)}
  \end{cases}
\eeq

%%%%%%%%%%%% FIG %%%%%%%%%%%%%%%%%%%
\begin{figure}[H]
\centering
\includegraphics[scale=0.25]{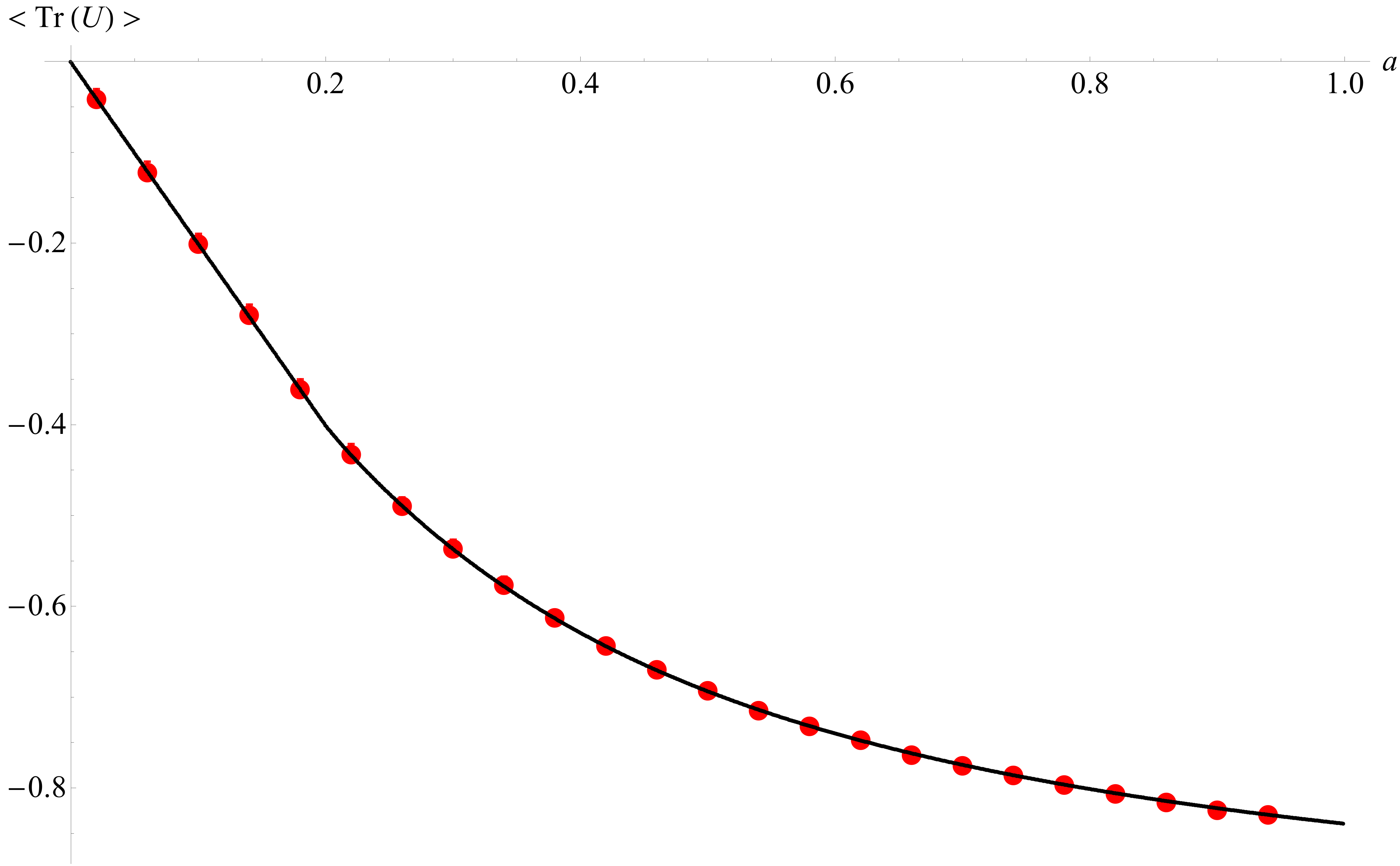}
\caption{The value of $\Tr(U)$ at $(a, 2a)$ for the $ab$-model with $N = 100$. The solid curve is the analytical result. The data are obtained through complex Langevin simulations with adaptive step size $\Delta \tau \leq 0.00005$, thermalization steps $N_{\rm therm} = 35000$, generation steps $N_{\rm gen} = 250000$ and with measurements performed with an interval of $500$ steps.
}
\label{fig:tr-U}
\end{figure}
%%%%%%%%%%%% FIG %%%%%%%%%%%%%%%%%%%

and $\big \langle \Tr(U^{-1}) \big \rangle $ is given by  

\beq
 \big \langle \Tr (U^{-1}) \big \rangle = 
  \begin{cases} 
   \oint \dfrac{dz}{2\pi i} \left( \dfrac{1}{z} - a - \dfrac{b}{z^2}\right) \dfrac{1}{z} = -a & \text{(Ungapped)} \\
   \oint_{\tilde{C}} \dfrac{dz}{4 \pi i} \dfrac{\omega(z)}{z} \dfrac{1}{z} = \left(\dfrac{\cos \phi + 1}{4} \right) \left( \dfrac{b(\cos \phi - 1)}{R^2} - 2 a \right)     & \text{(Gapped)}
  \end{cases}
\eeq

%%%%%%%%%%%% FIG %%%%%%%%%%%%%%%%%%%
\begin{figure}[H]
\centering
\includegraphics[scale=0.25]{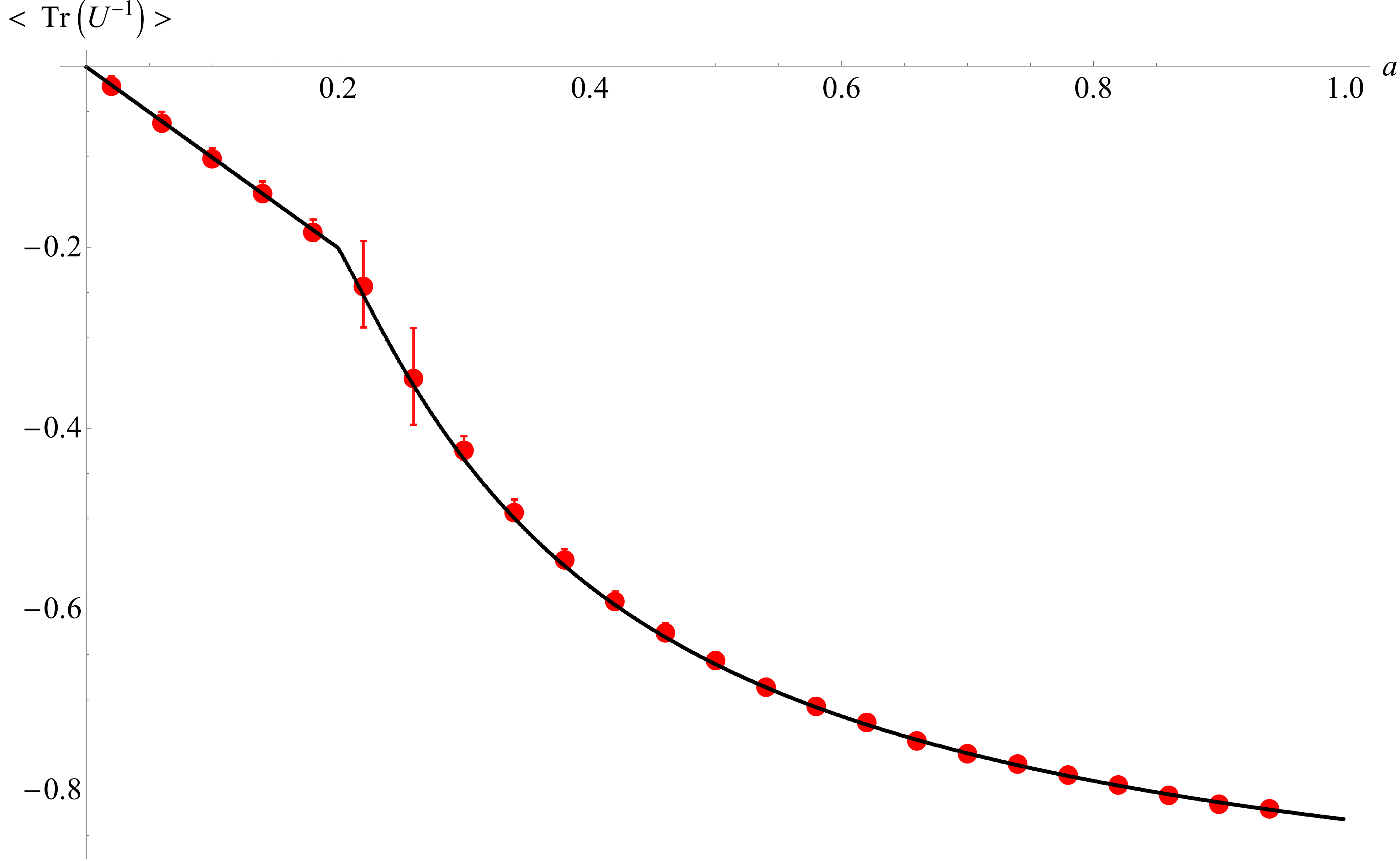}
\caption{The value of $\Tr(U^{-1})$ at $(a, 2a)$ for the $ab$-model with $N = 100$. The solid curve is the analytical result. The data are obtained through complex Langevin simulations with adaptive step size $\Delta \tau \leq 0.00005$, thermalization steps $N_{\rm therm} = 35000$, generation steps $N_{\rm gen} = 250000$ and with measurements performed with an interval of $500$ steps.
}
\label{fig:tr-U-inv}
\end{figure}
%%%%%%%%%%%% FIG %%%%%%%%%%%%%%%%%%%

In Fig. \ref{fig:tr-U} and Fig. \ref{fig:tr-U-inv} we show the observables $\big \langle \Tr(U) \big \rangle $ and $\big \langle \Tr(U^{-1}) \big \rangle$, respectively. We see that the analytical and numerical results show excellent agreement.

%%%%%%%%%%%%%%%%%%%%%%%%%%%%%%%%%%%%%%%%%%%%
\subsection{Phase transition of $ab$-Model}
%%%%%%%%%%%%%%%%%%%%%%%%%%%%%%%%%%%%%%%%%%%%

The eigenvalue density Eq. \eqref{eq:ungapped_density} on contour Eq. \eqref{eq:ungapped_contour}, is proportional to $ds$, which in terms of $r(\theta)$ is given by
\bea
ds &=& \dfrac{d}{d\theta} \left[ \theta - |a| \sin(\theta + \phi_1) r(\theta) - \dfrac{|b|}{r (\theta) } \sin (\theta - \phi_2) \right] d \theta  \nn \\ 
&=& \Big[1 - |a| \cos(\theta + \phi_1) r(\theta) - |a| \sin(\theta + \phi_1) r'(\theta) \nn \\
&&~~~~~~~~ - \dfrac{|b|}{r(\theta)} \cos(\theta - \phi_2)  + \dfrac{|b| r'(\theta)}{r(\theta)^2}\sin(\theta - \phi_2) \Big]  d\theta 
\eea
which is not positive definite for all $(a, b)$ combinations. It fails to do so, when the function inside the brackets, $\left[ \dots \right]$, becomes negative. Restricting to $a, b \in \mathbb{R}$, the condition simplifies as the gap opens about $\theta = 0$
\bea
\label{eq:gap_opening_condition}
& 1 - ar(0) - \dfrac{b}{r(0)} & \leq 0 \nn \\ 
\Rightarrow & \exp{\left(ar(0) + \dfrac{b}{r(0)}\right)} & \geq e. \\ \nn
\eea

From Eq. \eqref{eq:ungapped_contour} $r(0)$ is given by
\beq
\label{eq:gap_opening_contour}
r(0) = \exp{\left(a r(0) - \frac{b}{r(0)} \right)}.
\eeq

\begin{figure}[H]
	\centering
	\includegraphics[scale=0.3]{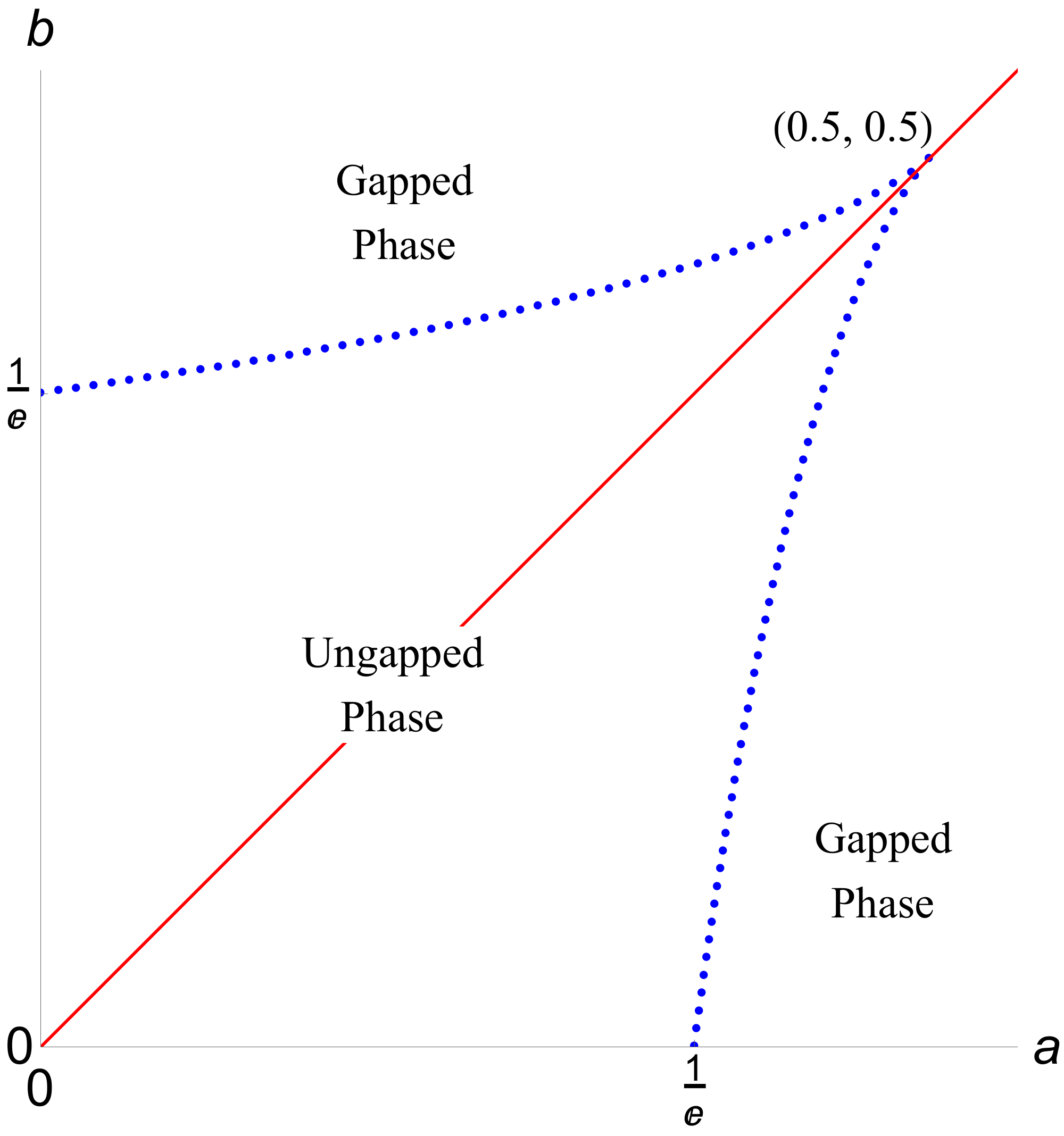}
	\caption{Phase diagram of the $ab$-model in the positive $ab$-plane. The solid red line indicates that the phase diagram is symmetric under the exchange of $a$ and $b$.}
	\label{fig:phase_space}
\end{figure}

The phase diagram of the model is shown in Fig. \ref{fig:phase_space}.
 
%%%%%%%%%%%% FIG %%%%%%%%%%%%%%%%%%%
%%%%%%%%%%%% FIG %%%%%%%%%%%%%%%%%%%

It would be interesting to know how quantities change across the gap opening transition and also the order of the phase transition. To study that we first restrict ourselves to a special case, $b = 0$ in our model. Then from Eqs. \eqref{eq:gap_opening_condition} and \eqref{eq:gap_opening_contour} the gap opens about $a = \frac{1}{e}$, $R = e$, and since the ungapped phase has no branch cuts in the eigenvalue distributions, $\phi$ should start from zero, about the gap-opening point. And the conditions Eqs. \eqref{eq:cond1} and \eqref{eq:cond2} simplifies to 
\beq
\label{eq:acond1}
aR \left(\dfrac{\cos(\phi)+1}{2}\right) = 1
\eeq
and Eq. \eqref{eq:cond3} to 
\beq
\label{eq:acond2}
\left(\dfrac{1 - \cos (\phi)}{2} \right) \ln \left( \dfrac{1 - \cos (\phi)}{2} \right) = \left( \dfrac{1 + \cos(\phi)}{2}\right) \ln \left( \dfrac{R}{e} \right).
\eeq

The observable $\big \langle \Tr(U) \big \rangle $ becomes 

\beq
 \big \langle \Tr(U) \big \rangle = 
  \begin{cases} 
   0 & \text{(Ungapped)} \\
   \left(\dfrac{a  (\cos(\phi)^2 -1 )R^2}{4} \right)       & \text{(Gapped)}
  \end{cases}
\eeq

Since the first derivative of free-energy $F[a]$ is the expectation value of $\Tr(U)$ 
\bea
\dfrac{\partial F[a]}{\partial a} &=& \dfrac{\partial \ln Z[a]}{\partial a} \nn \\
&=& \dfrac{1}{Z[a]} \int [DU] \Tr (U)\exp(a \Tr (U)) \nn \\
&=& \big \langle \Tr (U) \big \rangle,  
\eea
we find that it is continuous across the gap. 

Upon expanding about 
\beq
a = \dfrac{1}{e} + \delta a,~~\cos (\phi) = 1 - 2\delta p~{\rm and}~R = e + \delta R 
\eeq
the variation of $\delta \big \langle \Tr(U) \big \rangle $ is given by 
\beq
\delta \big \langle \Tr(U) \big \rangle  = \delta \left(\dfrac{a  (\cos(\phi)^2 -1 )R^2}{4} \right) = - e \delta p.
\eeq

From Eqs. \eqref{eq:acond1} and \eqref{eq:acond2} we get 
\beq
\delta p = e \delta a + \dfrac{\delta R}{e},
\eeq
and
\beq
\label{eq:dpdRrel}
\delta p \ln (\delta p) = \dfrac{\delta R}{e}. 
\eeq

Eliminating $\delta R$ from above two equations we get the equation
\beq
\delta p (1 - \ln(\delta p)) = e \delta a. 
\eeq

To invert the above equation let us substitute $\delta p \rightarrow e^{k}$. Then we have  
\beq
(k-1) e^{(k - 1)} = - \delta a.
\eeq

The above equation is of the form, $x e^{x} = y$, which can be inverted to express $x$ as a function of $y$ and it is known as the Lambert-W function \cite{lambertW}. (It is often expressed as $W_{c}(y)$.) This function is in general a multivalued-complex function, where $ c \in \mathbf{Z}$, chooses each branch. Since $\delta a > 0$ and $\delta p \in \mathbf{R}$ we have two real valued branches: $W_{0}(y)$ (the principal branch) and $W_{-1}(y)$. 

Therefore, 
\beq
\delta p = e^{W_0(-\delta a) + 1}~\text{or}~e^{W_{-1}(-\delta a) + 1}.
\eeq

For small values of $\delta a$ we know that

\begin{subequations}
\begin{align}
        \lim_{\delta a \rightarrow 0} W_{0}(-\delta a) &= 0,\\
        \lim_{\delta a \rightarrow 0} W_{-1}(-\delta a) &\approx\ln (\delta a).
\end{align}
\end{subequations}

Therefore, $\delta p$ will vanish as $\delta a \rightarrow 0$ only if we choose the second branch, i.e., $\delta p = e^{W_{-1}(-\delta a) + 1}$. Hence 
\beq
\delta \big \langle \Tr(U) \big \rangle = -e^{W_{-1}(-\delta a) + 2}.
\eeq
 
Now the second derivative of free energy 
\beq
\dfrac{\partial^2 F }{\partial (\delta a)^2} = \dfrac{\partial \left(\delta \big \langle \Tr(U) \big \rangle \right)}{\partial (\delta a)} \dfrac{e^2}{W_{-1}(-\delta a) + 1}
\eeq
goes to zero as $\delta a \rightarrow 0$ and is continuous across the gap. However, the third derivative 
\beq
\dfrac{\partial^3 F}{\partial (\delta a)^3 } = -\dfrac{e^2 W_{-1} (-\delta a)}{\delta a (W_{-1} (-\delta a) + 1)^3} 
\eeq
diverges as $\delta a \rightarrow 0$. Hence it has a third order phase transition. It can also be shown that similar arguments hold in the generic case $b \neq 0$. Thus we conclude that the $ab$-model displays a third order phase transition. 

%%%%%%%%%%%%%%%%%%%%%%%%%%%%%%%%%%%%%%%%%%%%%%%
\section{Gauge Theory to Unitary Matrix Model}
\label{sec:gt-to-u-m-model}
%%%%%%%%%%%%%%%%%%%%%%%%%%%%%%%%%%%%%%%%%%%%%%%

A unitary matrix model arises in a one-loop formulation of QCD [and analogous $SU(N)$ gauge theories] on compact spaces (often $S^1 \times S^3$). This was originally derived in Refs. \cite{Sundborg:1999ue, Aharony:2003sx,AlvarezGaume:2005fv,AlvarezGaume:2006jg} for theories with more general matter content. 

The one-loop effective action of QCD on $S^1 \times S^3$ with inverse temperature $\beta$, chemical potential $\mu$ and quark mass $m$ has the following form \cite{Hands:2010zp}, with thermal Polyakov line as the unitary matrix model
\bea
\label{eq:qcd-s1-s3-action}
S &=& \sum_{n=1}^\infty \frac{1}{n} z_b\left(\frac{n\beta}{R}\right) \Tr U^{ n} ~\Tr U^{\dagger n} \nn \\
&& \quad \quad \quad \quad + \sum_{n=1}^\infty \frac{(-1)^n}{n} N_f z_f\left(\frac{n\beta}{R}, mR\right)  \left[e^{n\beta \mu} \Tr U^{n} + e^{-n\beta \mu } \Tr U^{\dagger n} \right],
\eea
where $R$ is the radius of $S^3$ and $N_f$ is the number of flavors of fundamental fermions.

The quadratic term in Polyakov loop is the contribution from adjoint fields and the linear term is the contribution from the fundamental matter fields. Here, we have taken the adjoint contribution to be bosonic and the the contribution from fundamental fields to be fermionic. 

To be noted is that in the free theory the effective action is determined in terms of single particle (bosonic and fermionic) partition functions
\beq
z_b\left(\frac{\beta}{R}\right) = 2 \sum_{l=1}^\infty l(l+2) e^{-\beta(l+1)/R},
\eeq
and
\beq
z_f\left(\frac{\beta}{R}, mR\right) = 2 \sum_{l=1}^\infty l(l+1) e^{-\frac{\beta}{R}\sqrt{(l + \hf)^2 + m^2 R^2}}.
\eeq

Also note that we will be using dimensionless variables $\beta/R$, $\mu R$ and $mR$ in numerical simulations. 

An analogous action, for the simpler $0+1$ dimensional case would be,
\beq
z_b = 0,
\eeq
and
\beq
z_f = 2 e^{-\beta m},
\eeq
where the parameter $m$ is the mass of the fundamental fermions.

In the low temperature limit, $\beta \to \infty$, we have $z_b(\infty) = 0$ and so the gluonic contribution is negligible. Thus the action is
\beq
S = S_{\rm Vdm} + S_f,
\eeq 
where $S_{\rm Vdm}$ is the Vandermonde piece of the action and $S_f$ is the fundamental fermionic contribution. The fermionic part could be summed in a logarithm
\beq
\label{eq:action-multi-level}
S[U] = - \sum_{l=1}^\infty \sigma_l \left( \log \left[ \det\left(1 + e^{\beta(\mu - \epsilon_l)} U\right) \det\left(1 + e^{\beta(-\mu - \epsilon_l)} U^{-1}\right) \right] \right),
\eeq
where
\bea
\sigma_l &=& 2l (l+1) \frac{N_f}{N}, \\
\epsilon_l &=&  \sqrt{m^2 + \left(l + \hf \right)^2 R^{-2}}.
\eea

%%%%%%%%%%%%%%%%%%%%%%%%%%%%%%%%%%%%%%%%%%%%%%%%%%%%%%%%%%%%%%%%%%
\subsection{Observables}
%%%%%%%%%%%%%%%%%%%%%%%%%%%%%%%%%%%%%%%%%%%%%%%%%%%%%%%%%%%%%%%%%%

We would like to simulate the action given in Eq. \eqref{eq:action-multi-level} using complex Langevin method. We can study several interesting observables in this model. We briefly describe them below

\begin{enumerate}
	
	\item[1.]  Polyakov line $P$ and inverse Polyakov line $P^{-1}$

  These are the most natural set of observables to study the confined/deconfined phases in the theory.
	
	\item[3.] Fermion number $f_N$
	
	It gives the number of fermions minus the number of anti-fermions in a given volume
	\bea
	f_N &=& \frac{1}{\beta} \left( \frac{\partial \log Z}{\partial \mu} \right).
	\eea
	
	In the model we study here we have a single chemical potential $\mu$. In general there can be chemical potential for each fermion flavor.  
	
	The quark number susceptibility $\chi_f$ measures the response of the fermion number density to infinitesimal changes in the chemical potential, 
	\beq
	\chi_f = \frac{1}{\beta} \frac{\partial f_N}{\partial \mu}.
	\eeq
	This observable follows the behavior of the Polyakov line. Thus, it also serves as an indicator of confinement-deconfinement transitions for nonzero chemical potential. 
	
	\item[4.] Pressure $p$
	
	\bea
	p &=& \frac{1}{\beta} \left( \frac{\partial \log Z}{\partial V_3} \right),
	\eea
	with $V_3$ denoting the spatial volume.
	
	\item[5.] Energy $E$
	
	It can be constructed from pressure and fermion number density
	
	\bea
	E &=& - p V_3 + \mu f_N.
	\eea
	
\end{enumerate}

It is also possible to compute the chiral condensate and average phase, though we will not compute them in this work. The chiral condensate $\langle \psib \psi \rangle$ is given by
\bea
\langle \psib \psi \rangle &=& - \frac{1}{\beta V_3} \lim_{m \to 0}\left( \frac{\partial \log Z}{\partial m} \right),
\eea
and the average phase $\langle e^{i\phi}\rangle_{pq}$ has the form
\bea
\langle e^{i\phi}\rangle_{pq} &=& \frac{Z}{Z_{pq}},
\eea
where $pq$ refers to the phase quenched theory.

%%%%%%%%%%%%%%%%%%%%%%%%%%%%%%%%%%%%%%%%%%%%%%%%%%%%%%%%%%%%%%%%%
\subsection{Single Level Model with Positive Chemical Potential}
\label{sec:single-level-model}
%%%%%%%%%%%%%%%%%%%%%%%%%%%%%%%%%%%%%%%%%%%%%%%%%%%%%%%%%%%%%%%%%

We can truncate the action given in Eq. (\ref{eq:action-multi-level}) in a double scaling limit:
\begin{align}
\beta \rightarrow \infty, \\
\nn \mu \rightarrow \epsilon_0, \\
\nn \exp(\beta(\mu-\epsilon_l))=\xi,
\end{align}
where $\epsilon_0$ is a fixed quark energy level and we call $\xi$ the transition parameter.

Only contribution from a single level survives here and the action takes the form
\beq
\label{eq:act-single-level}
S[U] = - \sigma \log \left( 1 + \xi U \right).
\eeq  

The effective action on the complexified angle variables include the Vandermonde piece and a Lagrange multiplier.

In the large $N$ limit, the integral over the angles is dominated by a saddle point obtained by solving the equation of motion that follows from the effective action involving Eq. (\ref{eq:act-single-level})
\beq
\frac{\partial S}{\partial \theta_i} = i N \cN - \frac{i N \sigma \xi e^{i \theta}}{\left(1 + \xi e^{i \theta_i}\right)} - \sum_{j (\neq i)}^N \cot \left(\frac{\theta_i - \theta_j}{2}\right).
\eeq

Here also the action is not hermitian, giving rise to the {\it sign problem} in the presence of a chemical potential. As a result the saddle point configuration will lie out in the complex plane. If we define $z_i = \exp(i \theta_i)$ then in the presence of the non-real potential the $z_i$ will move off the unit circle in the $z$-plane.

We can explore the nature of eigenvalue distribution in the complex plane for various values of transition parameter $\xi$. We find that when $\xi$ is either very small or large, the potential vanishes and so we expect the $\{z_i\}$ to be uniformly distributed around the unit circle. Thus, when $\mu$ varies from $\mu \ll \epsilon$ to $\mu \gg \epsilon$ the quark energy level becomes occupied and the effective fermion umber jumps by factor $\sigma$. In Ref. \cite{Hands:2010zp} the authors provide a detailed description of this transition.

Let us look at the various regimes of $\xi$ and see how it affects the eigenvalue distribution, following the analytical study given in Ref. \cite{Hands:2010zp}.

\begin{enumerate}
	
	\item[1.] {\it The small $\xi$ confined phase}
	
	In the small $\xi$ confining phase the effective fermion number vanishes, $\cN = 0$, and the Polyakov line expectation values are 
	\beq
	P = 0,~~P^{-1} = \sigma \xi.
	\eeq
	
	Thus we have $P \neq P^{-1}$, as a result of the complex action.
	
	As $\xi$ is increased the contour of eigenvalue distribution opens into an arc, just as the matrix model solved by Gross and Witten \cite{Gross:1980he} and Wadia \cite{Wadia:1980www, Wadia:1980cp}.
	
	The line of phase transitions in the $(\mu, T)$ plane corresponds to the straight line
	\beq
	\mu = \epsilon - T \Big[ (1 + \sigma) \log(1 + \sigma) - \sigma \log \sigma \Big].
	\eeq
	Note that is approximation is valid only in the low temperature ($\beta \to \infty$) limit.
	
	\item[2.] {\it The large $\xi$ confined phase}
	
	In this phase the effective fermion number is 
	\beq
	\cN = \sigma,
	\eeq
	indicating that the level is now occupied.
	
	The Polyakov line expectation values are
	\beq
	P = \frac{\sigma}{\xi},~~P^{-1} = 0.
	\eeq
	
	Comparing with the previous case the behavior of $P$ and $P^{-1}$ swaps over along the replacement $\xi \to \xi^{-1}$.
	
	The large $\xi$ confined phase persists until the value
	\beq
	\xi = \xi_2 = \frac{(1 + \sigma)^{1 + \sigma}}{\sigma^\sigma}.
	\eeq
	
	For smaller values of $\xi$ the contour of eigenvalue distribution is not closed and the phase does not exist. The points of transition $\xi = \xi_1$ and $\xi = \xi_2$ satisfy $\xi_1 \xi_2 = 1$. 
	
	In the $(\mu, T)$ plane the boundary lies along the straight line
	\beq
	\mu = \epsilon + T \Big[(1 + \sigma) \log(1 + \sigma) - \sigma \log \sigma \Big],
	\eeq
	again valid in the low temperature limit.
	
	\item[3.] {\it The deconfined phase}
	
	In the region $\xi_1 \leq \xi \leq \xi_2$, experience with GWW matrix model suggests that the eigenvalue distribution exhibits the shape of an open contour. 
	
	In this regime we get a condition
	\beq
	\xi = \frac{(\sigma - \cN)^{\sigma - \cN} (1 + \cN)^{1 + \cN}}{\cN^\cN (1 + \sigma - \cN)^{1 + \sigma - \cN}}.
	\eeq
	
	This equation determines $\cN$ as a function of $\xi$. 
	
	From the above equation it follows that across the transitions at $\xi = \xi_1$ and $\xi = \xi_2$, fermion number density $\cN$ and its first derivative $\partial \cN / \partial \mu$ are continuous, however higher derivatives are discontinuous. Since $\cN$ is the effective fermion number, the first derivative of the grand potential, it follows that the transitions are third order, just as in the original GWW model. 
	
	For a single winding, the Polyakov lines are
	\beq
	P = \frac{\cN}{\sigma + 1 - \cN} \frac{1}{\xi},~~P^{-1} = \frac{\sigma - \cN}{1 + \cN} \xi.
	\eeq
	
\end{enumerate}

\begin{figure}[h!]
	\centering
	\includegraphics[width=5.5in]{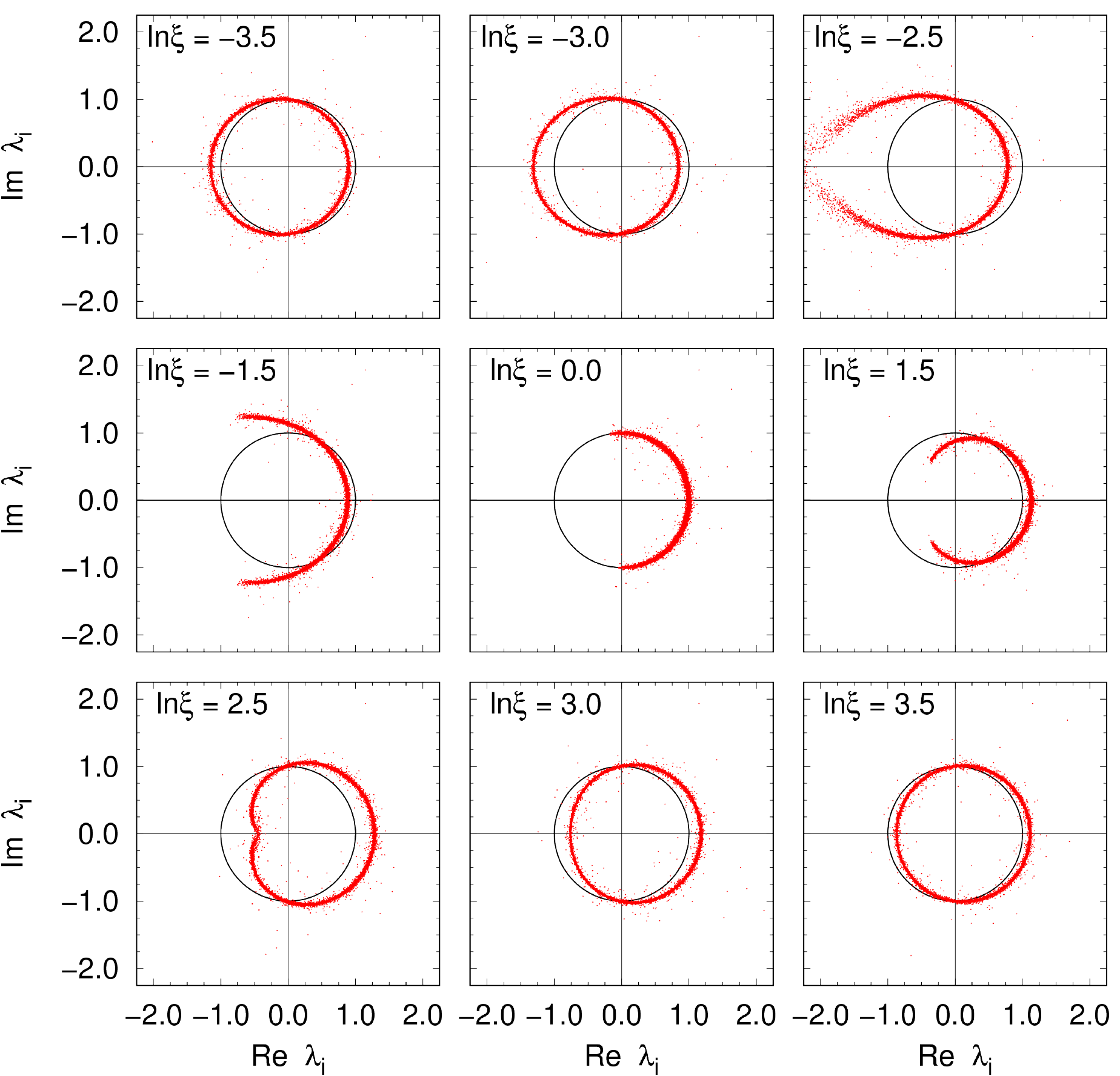}
	\caption{The eigenvalue distributions in the confined and deconfined phases as a function of $\log \xi$ for the single level matrix model with positive chemical potential. (See Eq. \eqref{eq:act-single-level} for the form of the action.) Here $N = N_f = 500$ and quark mass $m = 0$. The data are obtained through complex Langevin simulations with adaptive Langevin step sizes $\Delta \tau \leq 0.00005$, thermalization steps $N_{\rm therm} = 18000$, generation steps $N_{\rm gen} = 2000$ and with measurements performed with an interval of $100$ steps. The solid unit circles are guide to the eye.}
	\label{fig:eigs-nc500-nf500-m0-b30_noise}
\end{figure}

Using complex Langevin dynamics we have simulated the single level matrix model given by the action in Eq. \eqref{eq:act-single-level}. In Fig. \ref{fig:eigs-nc500-nf500-m0-b30_noise} we show the eigenvalue distributions of the Polyakov line in the confined and deconfined phases as a function of the logarithm of the transition parameter, $\log \xi$, for $SU(N)$ case with $N = N_f = 500$ and quark mass $m = 0$. We see that the eigenvalue distributions start with a closed contour (confined phase), passes through an open contour (deconfined phase) and again goes into a closed contour. (This figure can be compared with Fig. 12 in Sec. 4.1 of Ref. \cite{Hands:2010zp}, where it was obtained through analytical methods.)

%%%%%%%% FIGS %%%%%%%%%%%%%%%%
%%%%%%%% FIGS %%%%%%%%%%%%%%%% 

\begin{figure}[h!]
	\centering
	\includegraphics[width=4.5in]{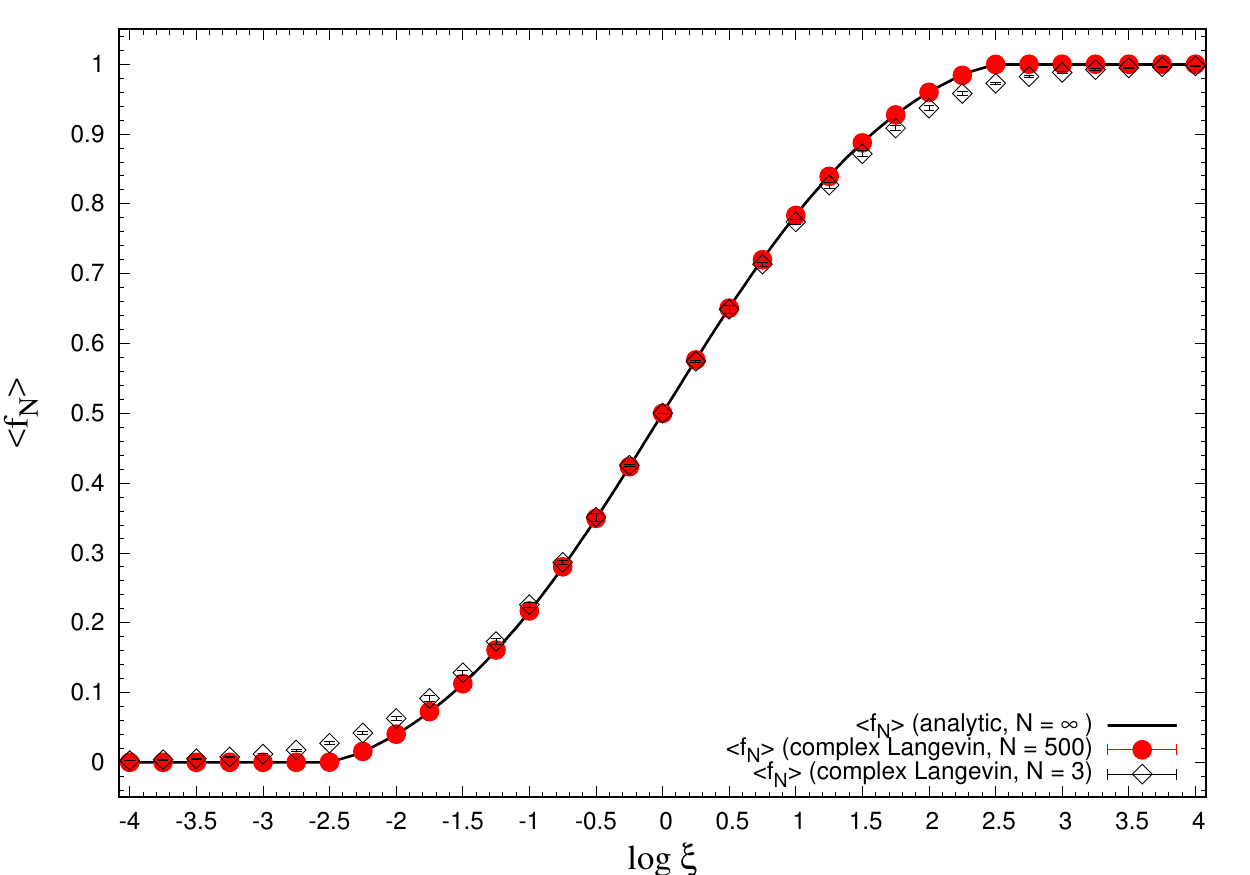}
	\caption{The (normalized) effective fermion number $\langle f_N \rangle$ across the pair of GWW transitions from the small $\xi$ confined phase through the deconfined phase to the large $\xi$ confined phase for the single level model with positive chemical potential. (See Eq. \eqref{eq:act-single-level} for the form of the action.) The solid curve is the analytical result ($N = \infty$). The data points are obtained through complex Langevin simulations. We used adaptive Langevin step sizes $\Delta \tau \leq 0.00005$, thermalization steps $N_{\rm therm} = 10000$, generation steps $N_{\rm gen} = 10000$ and measurements are performed with an interval of $100$ steps. We show simulation data for quark mass $m = 0$ and for $N = N_f = 500$ and $N = N_f = 3$.}
	\label{fig:fN_N500}
\end{figure}

\begin{figure}[h!]
	\centering
	\includegraphics[width=4.5in]{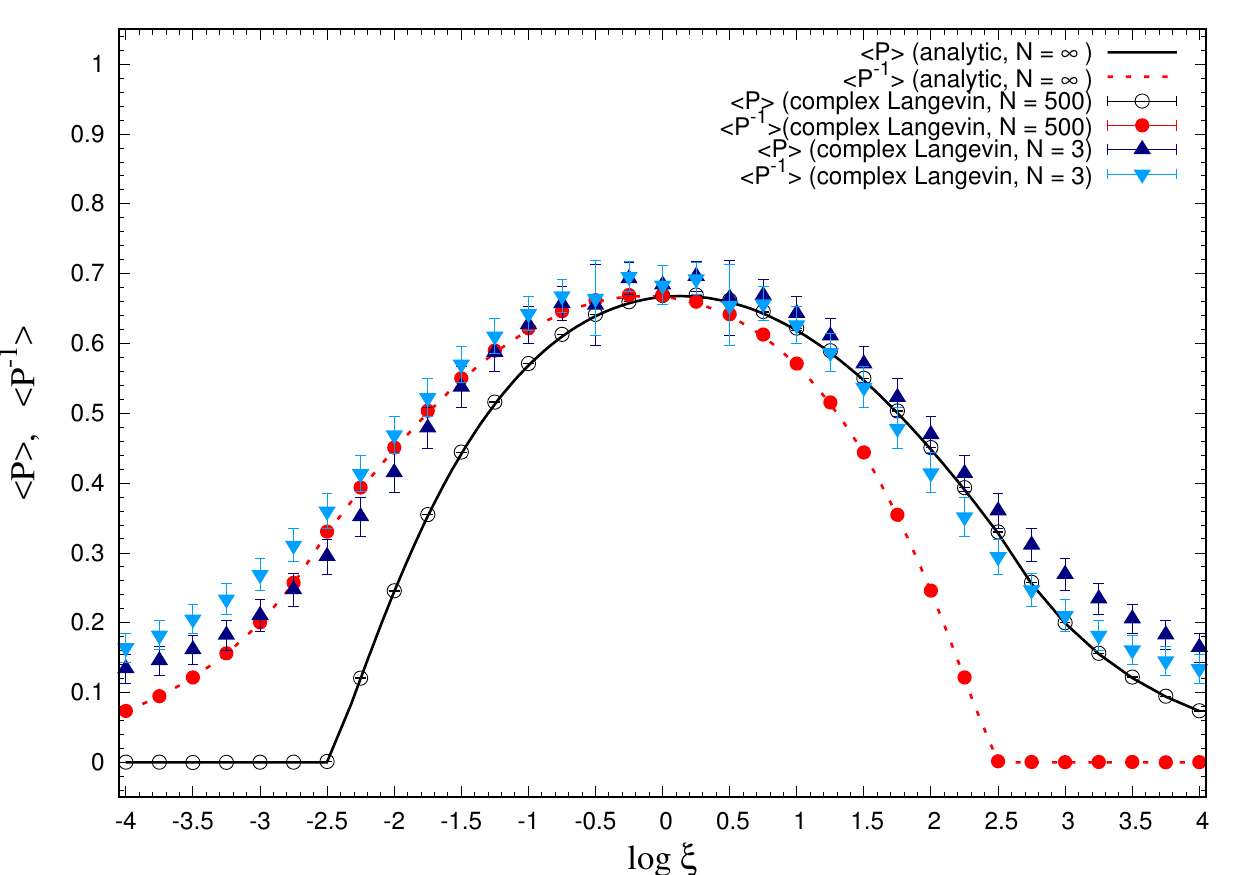}
	\caption{The Polyakov line $\langle P \rangle$ and inverse Polyakov line $\langle P^{-1} \rangle$ across the pair of GWW transitions from the small $\xi$ confined phase through the deconfined phase to the large $\xi$ confined phase for the single level model with positive chemical potential. (See Eq. \eqref{eq:act-single-level} for the form of the action.) The transitions from confined/deconfined phases occur when either $\langle P \rangle$ or $\langle P^{-1} \rangle$ vanish. The solid and dotted curves are the analytical results ($N = \infty$) for $\langle P \rangle$ and $\langle P^{-1} \rangle$, respectively. The data points are obtained through complex Langevin simulations. We used adaptive Langevin step sizes $\Delta \tau \leq 0.00005$, thermalization steps $N_{\rm therm} = 10000$, generation steps $N_{\rm gen} = 10000$ and measurements are performed with an interval of $100$ steps. We show simulation data for quark mass $m = 0$ and for $N = N_f = 500$ and $N = N_f = 3$.}
	\label{fig:P_inv_P_N500}
\end{figure}

In Fig. \ref{fig:fN_N500} we provide the (normalized) effective fermion number $\langle f_N \rangle$, and in Fig. \ref{fig:P_inv_P_N500} the Polyakov line expectation value $\langle P \rangle$ and the inverse Polyakov line expectation value $\langle P^{-1} \rangle$ across a pair of GWW transitions from the small $\xi$ confined phase through the deconfined phase to the large $\xi$ confined phase. The transitions from confined/deconfined phases occur when either $\langle P \rangle$ or $\langle P^{-1} \rangle$ vanish. The parameters used are: $N = N_f = 3~{\rm and}~500$ and quark mass $m = 0$. The simulations show excellent agreement with the analytical results in the large $N$.

%%%%%%%% FIGS %%%%%%%%%%%%%%%%
%%%%%%%% FIGS %%%%%%%%%%%%%%%% 

%%%%%%%% FIGS %%%%%%%%%%%%%%%%
%%%%%%%% FIGS %%%%%%%%%%%%%%%% 

\begin{figure}
	\centering
	\includegraphics[width=5.0in]{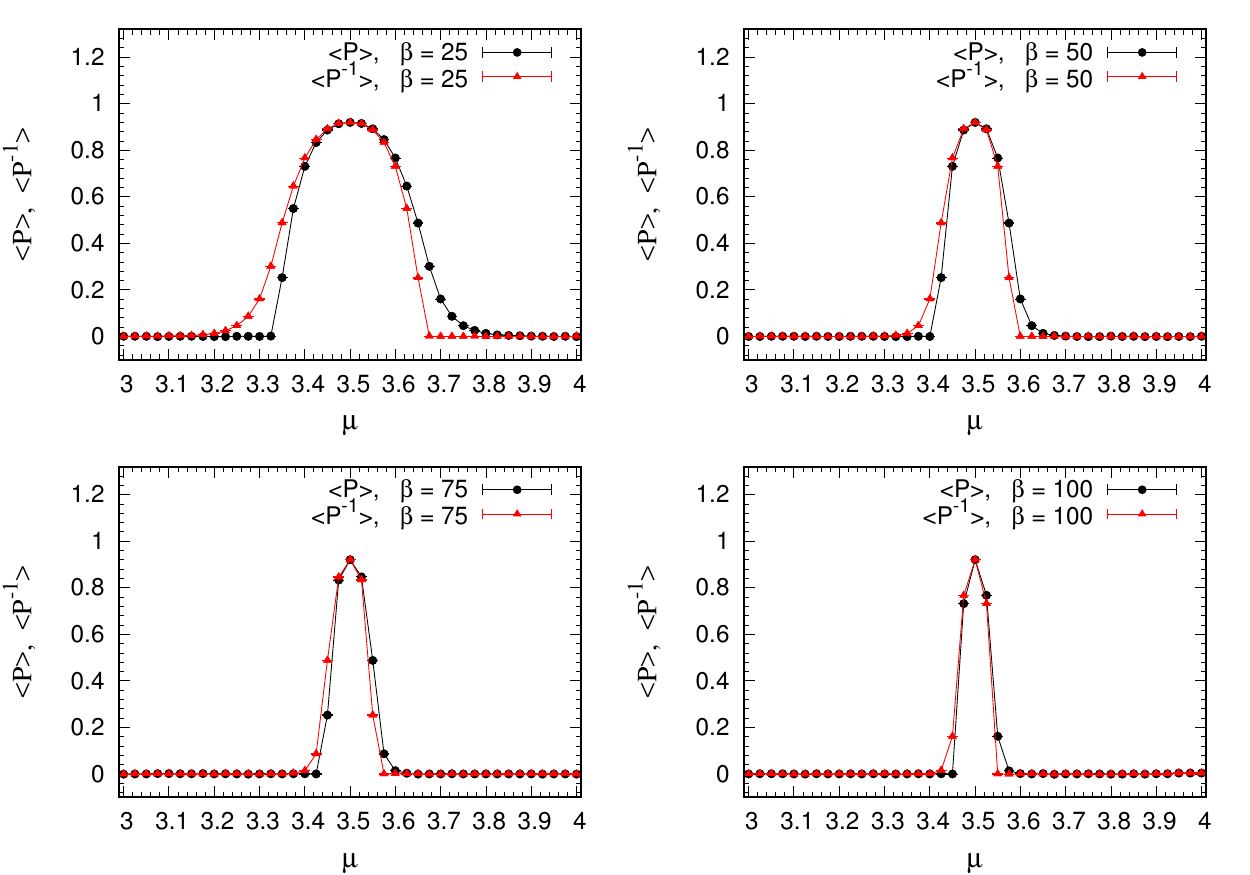}
	\caption{Polyakov line $\langle P \rangle$ and inverse Polyakov line $\langle P^{-1} \rangle$ as a function of chemical potential for single level matrix model with quark energy level $\epsilon \equiv \epsilon_{(l=3)} = 3.5$ and quark mass $m = 0$. (See Eq. \eqref{eq:act-single-level} for the form of the action.) Here $N = N_f = 500$ and $\beta = 25, 50, 75, 100$. The data are obtained through complex Langevin simulations with adaptive Langevin step sizes $\Delta \tau \leq 0.000005$, thermalization steps $N_{\rm therm} = 5000$, generation steps $N_{\rm gen} = 5000$ and with measurements performed with an interval of $50$ steps.}
	\label{fig:single-level-p-invp}
\end{figure}

\begin{figure}[htp]

\subfloat[$\beta = 50$ with $\mu = 3.0, 3.3, 3.5$]{\includegraphics[width=6.0in]{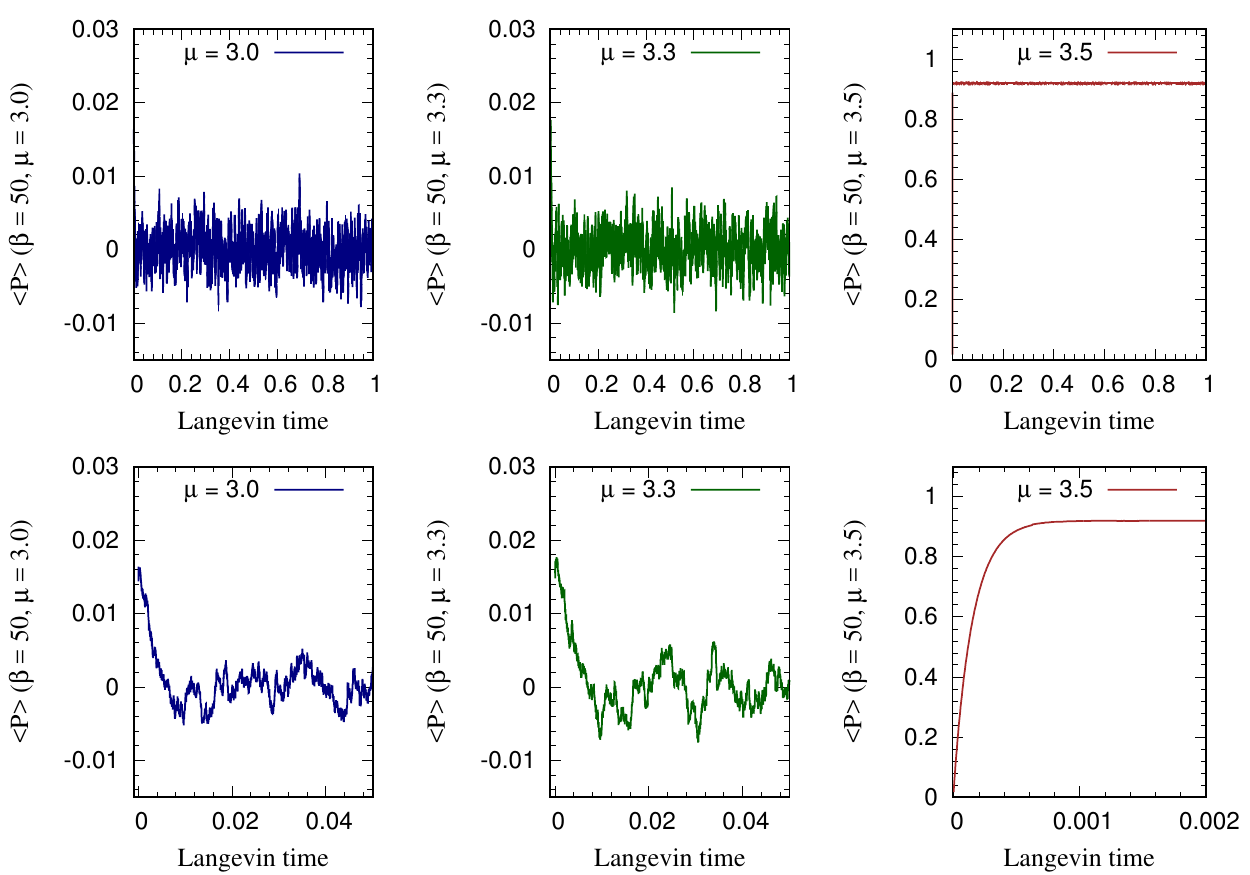}}

\subfloat[$\beta = 75$ with $\mu = 3.0, 3.3, 3.5$]{\includegraphics[width=6.0in]{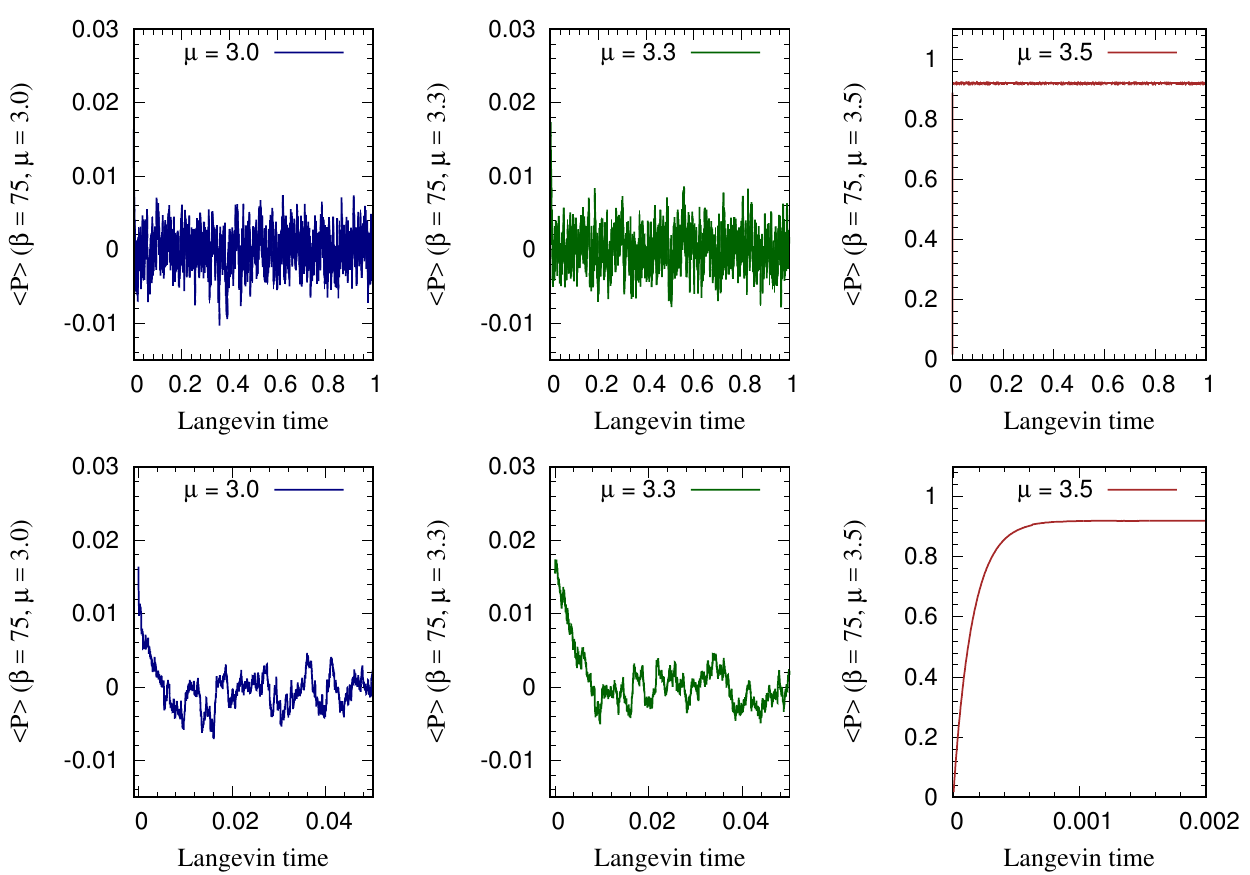}}

	\caption{The Langevin time evolution of the Polyakov loop observable for the single level matrix model with quark energy level $\epsilon \equiv \epsilon_{(l=3)} = 3.5$ and quark mass $m = 0$. (See Eq. \eqref{eq:act-single-level} for the form of the action.) Here $N = N_f = 500$. The simulations were performed using adaptive Langevin step sizes $\Delta \tau \leq 0.000005$ and for $200000$ evolution steps. (a) The plots are for $\beta = 50$ with $\mu = 3.0, 3.3, 3.5$ (left to right). The plots on the bottom row are for the same parameters but shows the initial stages of the evolution, focusing on the saturation of observables. (b) The plots are for $\beta = 75$ with $\mu = 3.0, 3.3, 3.5$ (left to right). The plots on the bottom row are again for the same parameters but shows the initial stages of the evolution, focusing on the saturation of observables.}
	\label{fig:runtime-hist-b50-75}
\end{figure}

\begin{figure}
	\centering
	\includegraphics[width=5.0in]{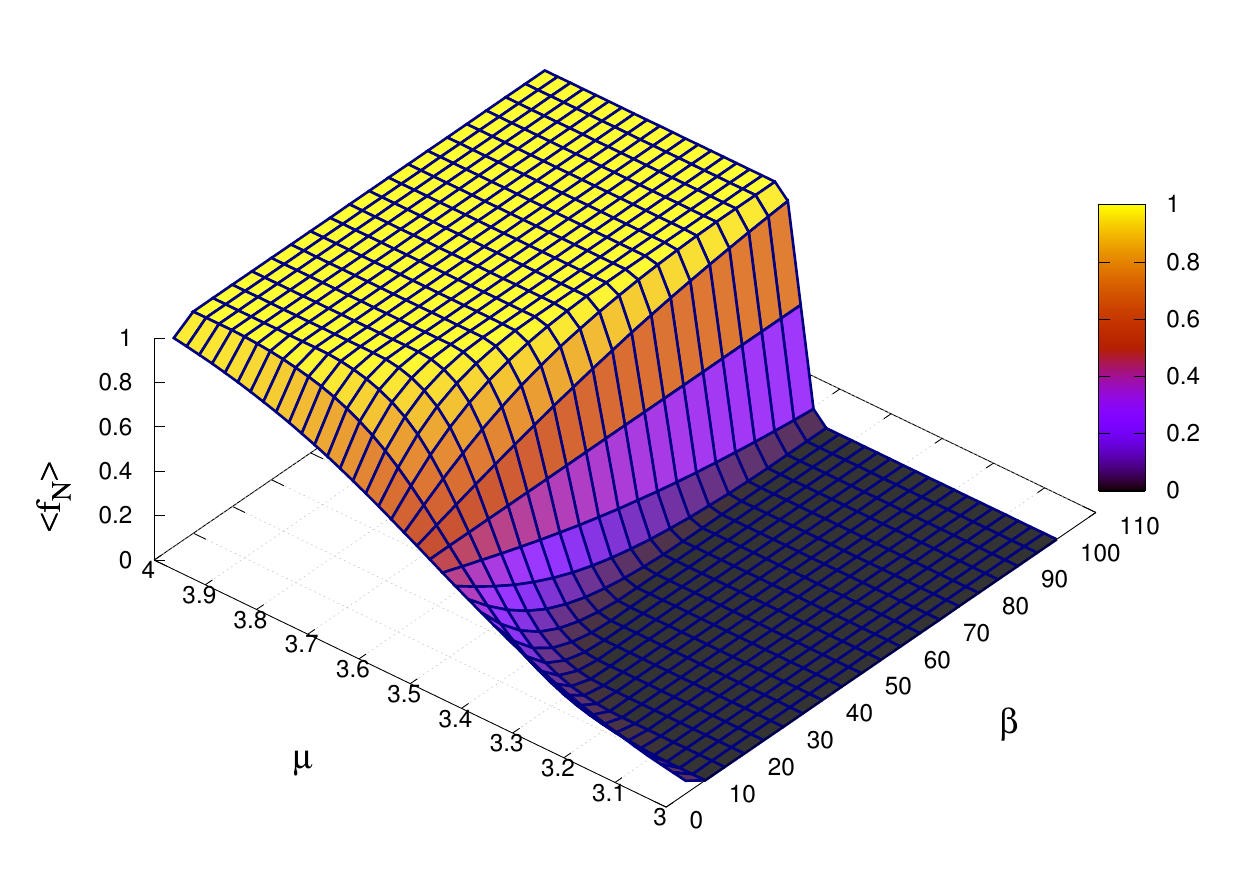}
	\caption{The (normalized) fermion number density $\langle f_N \rangle$ as a function of chemical potential $\mu$ and inverse temperature $\beta$ for single level matrix model with quark energy level $\epsilon \equiv \epsilon_{(l=3)} = 3.5$ and quark mass $m = 0$. (See Eq. \eqref{eq:act-single-level} for the form of the action.) Here $N = N_f = 500$. The model is in a deconfined phase when $0 < \langle f_N \rangle < 1$. The data are obtained through complex Langevin simulations with adaptive Langevin step sizes $\Delta \tau \leq 0.000005$, thermalization steps $N_{\rm therm} = 5000$, generation steps $N_{\rm gen} = 5000$ and with measurements performed with an interval of $50$ steps.}
	\label{fig:single-level-fn}
\end{figure}

In Figs. \ref{fig:single-level-p-invp} and \ref{fig:single-level-fn} we show the Polyakov lines and fermion number density for a range of simulation parameters of the single level matrix model (see Eq. \eqref{eq:act-single-level} for the form of the action): $\beta = \{10, 15, \cdots, 100\}$ and $\mu = \{3.0, 3.025, 3.05, \cdots, 4.0\}$. The quark energy level of the model is fixed to the third level $\epsilon \equiv \epsilon_{(l=3)} = 3.5$. The Polyakov loops peak around $\mu = 3.5$ in this model. In Fig. \ref{fig:single-level-p-invp} we show the behavior of Polyakov and inverse Polyakov loops for $\beta = \{25, 50, 75, 100\}$. It is clear that the widths of the Polyakov loops decrease as the temperature is reduced (large $\beta$) and the behavior of inverse Polyakov line precedes that of the Polyakov line as a function of $\mu$. In Fig. \ref{fig:runtime-hist-b50-75} we show the Langevin evolution history of the Polyakov loop observable in this model for $\beta = 50, 75$ and with $\mu = 3.0, 3.3, 3.5$ for each $\beta$ value. We note that the observables saturate to their equilibrium values rather quickly in this model. In Fig. \ref{fig:single-level-fn} we show the behavior of the (normalized) fermion number density $\langle f_N \rangle$ as a function of chemical potential and inverse temperature. The transition in fermion number becomes sharper as the temperature is decreased (high $\beta$). The model is in a deconfined phase when $ 0 < \langle f_N \rangle < 1$. 

%%%%%%%% FIGS %%%%%%%%%%%%%%%%
%%%%%%%% FIGS %%%%%%%%%%%%%%%%

\begin{comment}
%%%%%%%% FIGS %%%%%%%%%%%%%%%%
\begin{figure}
	\centering
	\includegraphics[width=6.0in]{FIGS/run_hist_B50_mu_P.pdf}
  \includegraphics[width=6.0in]{FIGS/run_hist_B75_mu_P.pdf}
	\caption{The Langevin time evolution of the Polyakov loop observable for the single level matrix model with quark energy level $\epsilon \equiv \epsilon_{(l=3)} = 3.5$ and quark mass $m = 0$. (See Eq. \eqref{eq:act-single-level} for the form of the action.) Here $N = N_f = 500$. The plots on the first raw are for $\beta = 50$ with $\mu = 3.0, 3.3, 3.5$ (left to right). The plots on the second raw are for the same parameters but shows the initial stages of the evolution, focusing on the saturation of observables. Similar figures are given on the third and fourth raws for $\beta = 75$ with $\mu = 3.0, 3.3, 3.5$. The simulations were performed using adaptive Langevin step sizes $\Delta \tau \leq 0.000005$. The Langevin evolution is performed for $200000$ steps.}
	\label{fig:runtime-hist-b50-75}
\end{figure}
%%%%%%%% FIGS %%%%%%%%%%%%%%%%
\end{comment}

%%%%%%%%%%%%%%%%%%%%%%% FIG %%%%%%%%%%%%%%%%%%%%%%%%%%%%%%%%%%%%%%
%%%%%%%%%%%%%%%%%%%%%%% FIG %%%%%%%%%%%%%%%%%%%%%%%%%%%%%%%%%%%%%%

%%%%%%%% FIGS %%%%%%%%%%%%%%%%
%%%%%%%% FIGS %%%%%%%%%%%%%%%%

\begin{figure}
	\centering
	\includegraphics[width=5.0in]{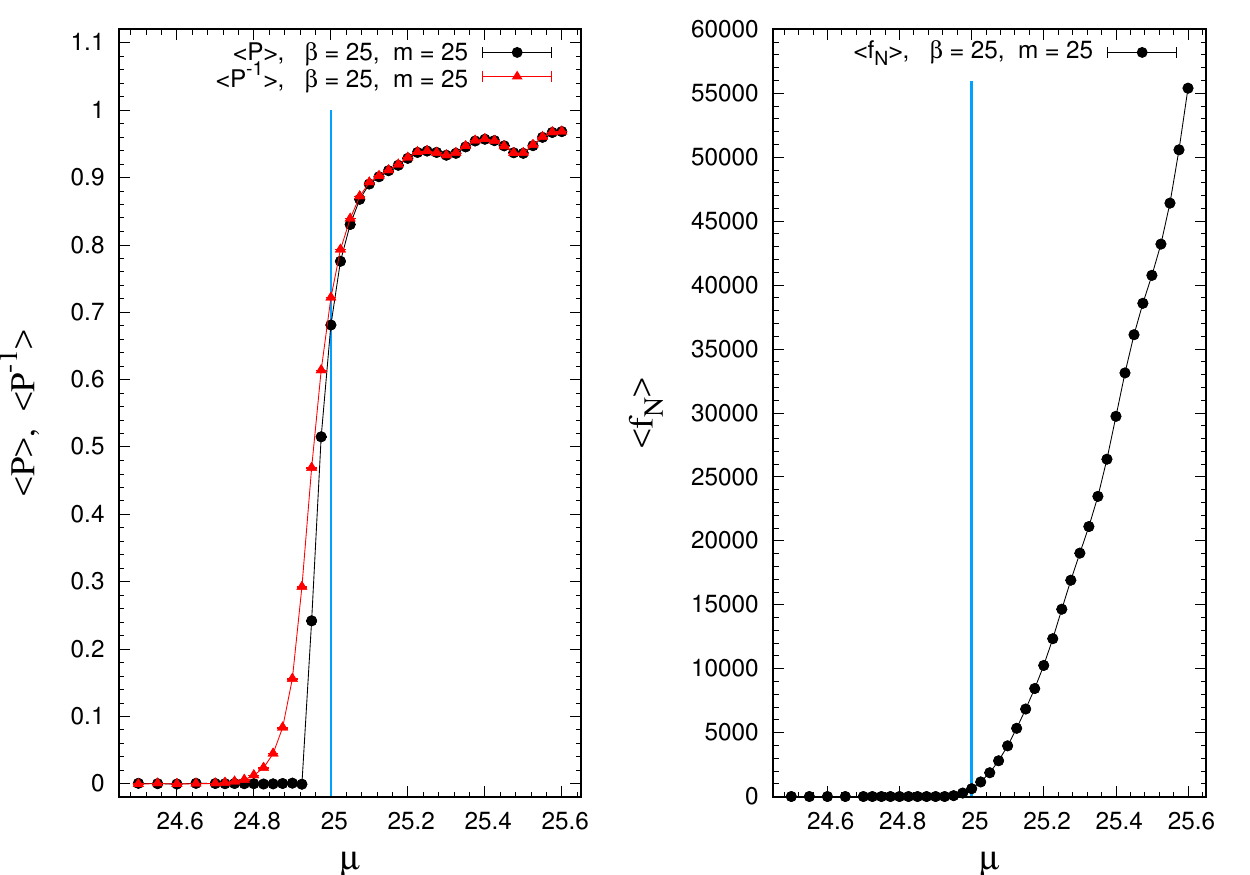}
	\caption{The Silver Blaze behavior of observables $\langle P \rangle$ and $\langle P^{-1} \rangle$, and $\langle f_N \rangle$ at non-zero quark mass $m$ for the model given by the action in Eq. \eqref{eq:action-multi-level}. (Left) Polyakov line $\langle P \rangle$ and inverse Polyakov line $\langle P^{-1} \rangle$ and (Right) fermion number $\langle f_N \rangle$ as a function of chemical potential for large quark mass near onset at $\mu = m = 25$ (marked by the solid vertical lines in the figures). Here $N = N_f = 500$ and $\beta = 25$ (low $T$). The data are obtained through complex Langevin simulations with an adaptive Langevin step sizes $\Delta \tau \leq 0.00005$, thermalization steps $N_{\rm therm} = 5000$, generation steps $N_{\rm gen} = 5000$ and with measurements performed with an interval of $50$ steps.}
	\label{fig:Silver-Blaze}
\end{figure}

When the quark mass is non-vanishing in QCD, the expectation values of bulk observables such as the fermion number density, Polyakov lines and energy, exhibit the `Silver Blaze' behavior. The bulk observables are nearly zero until onset \cite{Cohen:2003kd} to a deconfinement transition, which occurs when the chemical potential increases to the value of the lightest quark mass. We simulate the model given by the action in Eq. \eqref{eq:action-multi-level} to see this phenomenon. In this model the onset occurs at $\mu = m$. The Polyakov line is given in Fig. \ref{fig:Silver-Blaze} (Left) as a function of chemical potential for large quark mass, near the onset $\mu = m = 25$ for $N = N_f = 500$ and $\beta = 25$ (low $T$). In the large $m$ limit, similar to the $m = 0$ case, the behavior of inverse Polyakov line $\langle P^{-1} \rangle$ precedes that of $\langle P \rangle$ as a function of $\mu$. The transition in $\mu$ occurs around onset at $m$. In Fig. \ref{fig:Silver-Blaze} (Right) we show the effective fermion number density as a function of chemical potential. As we can see in the figures the bulk observables are close to zero until the onset transition at $\mu = m$. The observables rise smoothly from the onset and as $\mu$ is increased further from $m$ the observables behave as they would for $m = 0$. This is reflected in the oscillations that appear in the observables at larger $\mu$. The oscillations in the Polyakov and inverse Polyakov loops are clearly visible. In order to see the prominent nature of oscillations in the fermion number density one has to normalize this observable by its Stefan-Boltzmann value. (See Ref. \cite{Hands:2010zp} for a discussion on this.)

%%%%%%%% FIGS %%%%%%%%%%%%%%%%
%%%%%%%% FIGS %%%%%%%%%%%%%%%%

%%%%%%%%%%%%%%%%%%%%%%%%%%%%%%%%%%%%%%%%%%%%%%%%%%%%
\subsection{Single Level Model with $U$ and $U^\dagger$}
\label{sec:single-level-model-pm-mu}
%%%%%%%%%%%%%%%%%%%%%%%%%%%%%%%%%%%%%%%%%%%%%%%%%%%

In this section we consider the phase diagram of the model given by the following action
\beq
\label{eq:act-xi1_xi2}
S[U] = - \sigma \Big[\log \left( 1 + \xi_1 U^{\phantom{\dagger}} \right) + \log \left( 1 + \xi_2 U^\dagger \right)\Big],
\eeq  
where

\begin{subequations}
\begin{align}
        \xi_1 &= e^{\beta(\mu - \epsilon)}, \\
        \xi_2 &= e^{\beta(-\mu - \epsilon)}.
\end{align}
\end{subequations}

Such a model naturally arises from $0+1$-dimensional gauge theory with a fundamental fermion.
 
\begin{figure}
	\centering
	\includegraphics[width=4.5in]{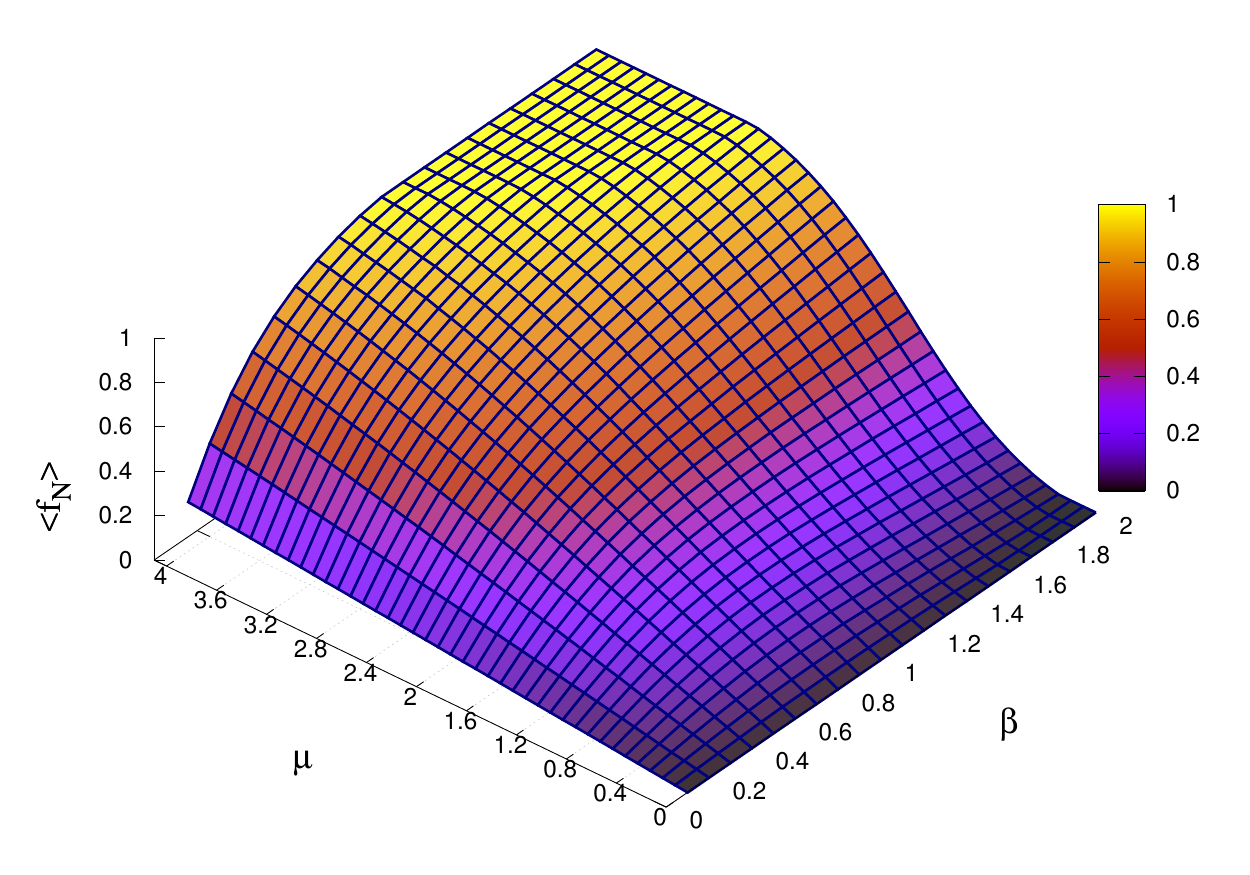}
	\caption{The (normalized) fermion number density $\langle f_N \rangle$ as a function of chemical potential $\mu$ and inverse temperature $\beta$ for the matrix model given by the action in Eq. \eqref{eq:act-xi1_xi2}. The model has fixed quark energy level $\epsilon \equiv \epsilon_{(l=1)} = 1.5$, quark mass $m = 0$ and $N = N_f = 100$. The model is in a deconfined phase when $ 0 < \langle f_N \rangle < 1$. The data are obtained through complex Langevin simulations with adaptive Langevin step sizes $\Delta \tau \leq 0.00005$, thermalization steps $N_{\rm therm} = 10000$, generation steps $N_{\rm gen} = 50000$ and with measurements performed with an interval of $100$ steps. }
	\label{fig:two-fuga-fn_high_T}
\end{figure}

In Fig. \ref{fig:two-fuga-fn_high_T} we provide the phase diagram of this model on the $(\mu, \beta)$ plane for the level $l=1$. (Corresponding to quark energy level $\epsilon = 1.5$ and $\sigma = 4$.) From the behavior of the expectation value of the fermion number density we see that the phase transition from confined to deconfined phase is smooth on the $(\mu, \beta)$ plane even at high temperature ($0.1 \leq \beta \leq 2.0$).

%%%%%%%% FIGS %%%%%%%%%%%%%%%%
%%%%%%%% FIGS %%%%%%%%%%%%%%%%

%%%%%%%%%%%%%%%%%%%%%%%%%%%%%%%%%%%%%%%%%%%%%%%%%%%%%%%%%%%%%%%%%
\subsection{Single Level Model with Interaction}
\label{sec:single-level-model-int}
%%%%%%%%%%%%%%%%%%%%%%%%%%%%%%%%%%%%%%%%%%%%%%%%%%%%%%%%%%%%%%%%%

It would be interesting to consider the single-level matrix model with a nontrivial interaction turned on. We take a Polyakov line interaction term of the form
\beq
S_{\rm int}[U] = g~ (\Tr U) (\Tr U^{-1}).
\eeq
Here $g$ denotes a coupling parameter.

Thus we have
\beq
\label{eq:act-single-level-int}
S[U] = - \sigma \log \left( 1 + e^{\beta(\mu - \epsilon)} U \right) + S_{\rm int}[U].
\eeq  

\begin{figure}
	\centering
	\includegraphics[width=4.5in]{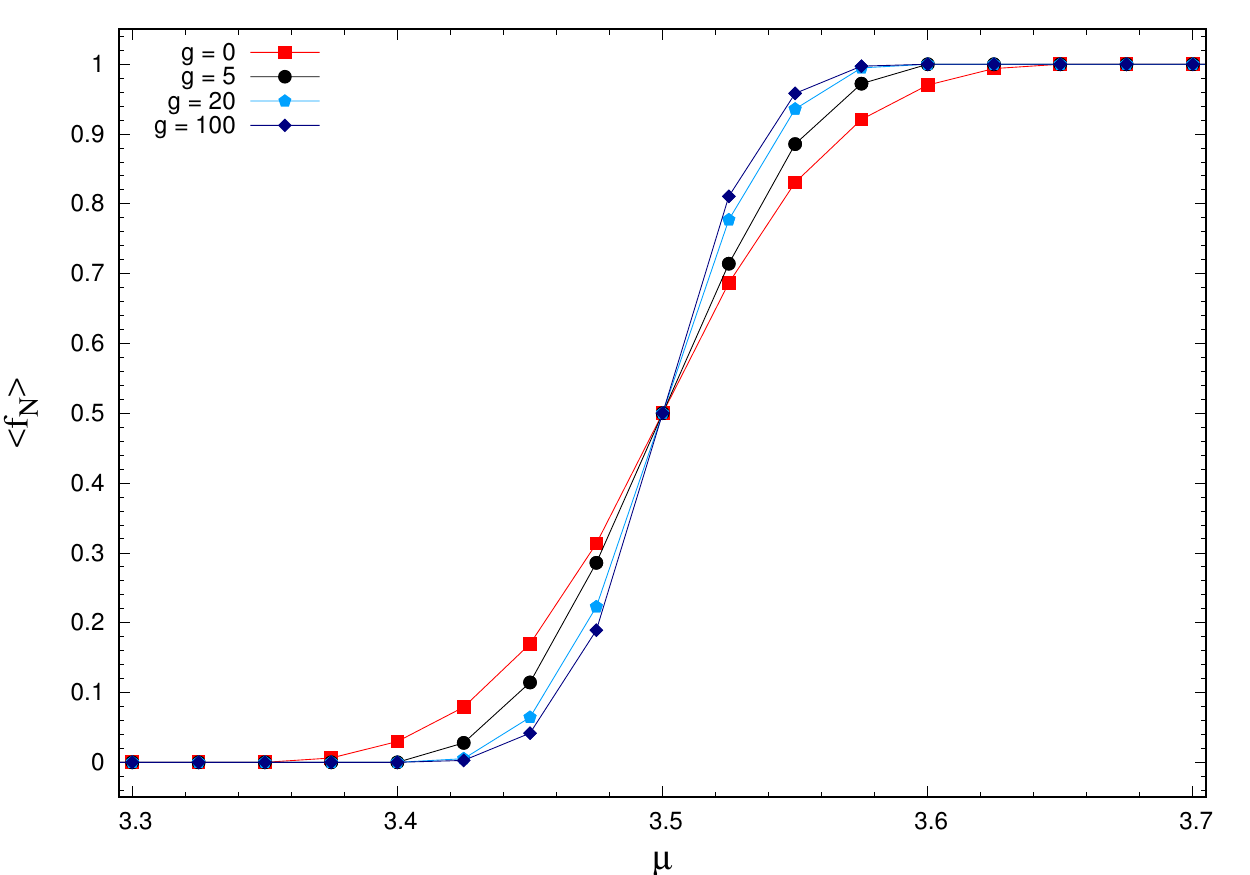}
	\caption{The (normalized) fermion number density $\langle f_N \rangle$ as a function of chemical potential $\mu$ for interacting single-level matrix model, given by the action in Eq. \eqref{eq:act-single-level-int}, with couplings $g = 0, 5, 20$ and $100$. The quark energy level is taken as $\epsilon \equiv \epsilon_{(l=3)} = 3.5$ and quark mass is $m = 0$. Here $N = N_f = 500$. The data are obtained through complex Langevin simulations with adaptive Langevin step sizes $\Delta \tau \leq 0.000005$, thermalization steps $N_{\rm therm} = 5000$, generation steps $N_{\rm gen} = 5000$ and with measurements performed with an interval of $50$ steps. The model is in a deconfined phase when $0 < \langle f_N \rangle < 1$. The data show that the phase transition becomes sharper as the interaction strength $g$ is increased.}
	\label{fig:single-level-fn-int}
\end{figure}

\begin{figure}
	\centering
	\includegraphics[width=4.5in]{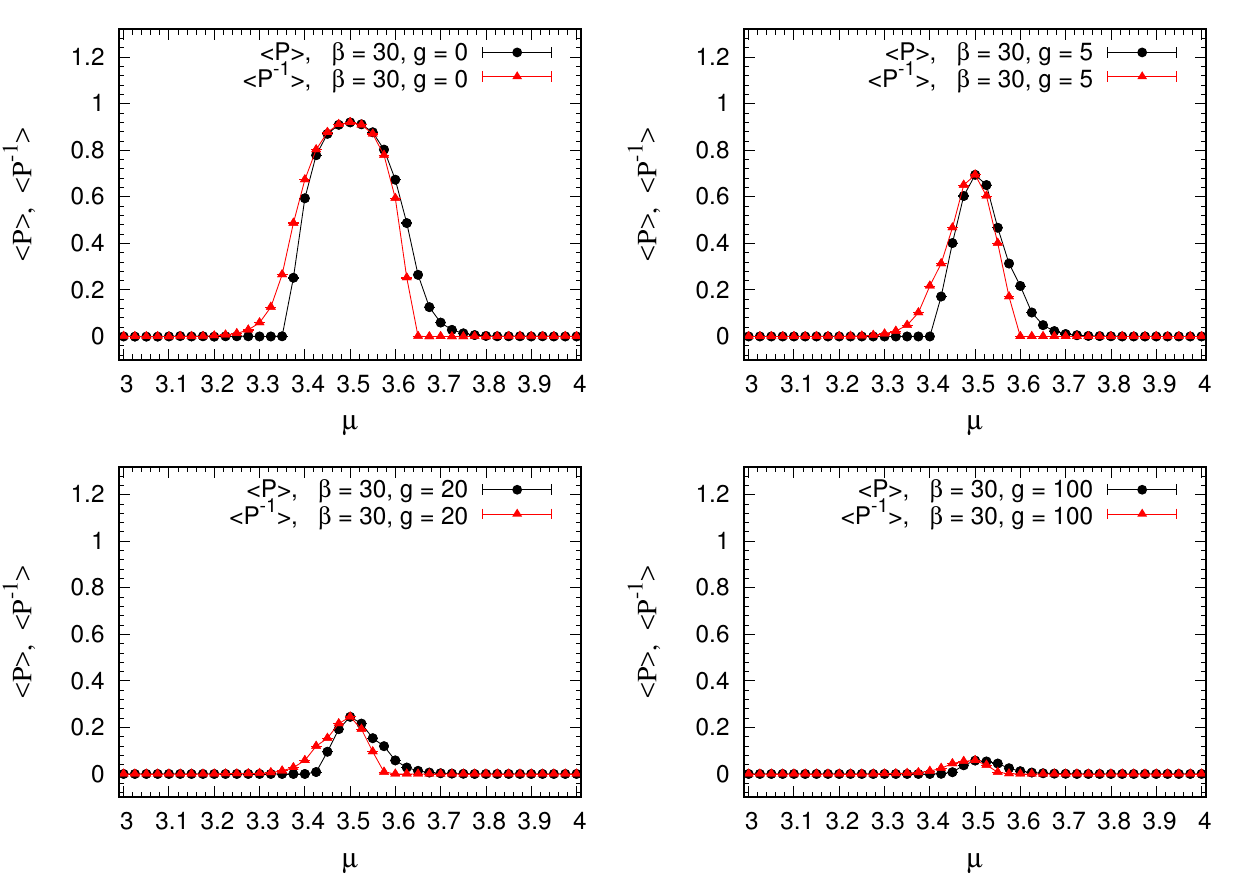}
	\caption{The Polyakov line and inverse Polyakov line across a pair of GWW transitions for the interacting single-level matrix model, given by the action in Eq. \eqref{eq:act-single-level-int}, with a fixed quark energy level $\epsilon \equiv \epsilon_{(l=3)} = 3.5$, quark mass $m = 0$ and $N = N_f = 500$. The data are obtained through complex Langevin simulations with adaptive Langevin step sizes $\Delta \tau \leq 0.000005$, thermalization steps $N_{\rm therm} = 5000$, generation steps $N_{\rm gen} = 5000$ and with measurements performed after every $50$ steps. The solid lines are guid to the eye. The plots indicate that the model prefers to stay in a confined phase as the interaction strength $g$ is increased.}
	\label{fig:single-level-poly-int}
\end{figure}

Here also we take the quark energy level to be fixed at $\epsilon \equiv \epsilon_{(l=3)} = 3.5$. The action is again not hermitian, giving rise to the sign problem in the presence of a chemical potential. In Figs. \ref{fig:single-level-fn-int} and \ref{fig:single-level-poly-int} we plot the fermion number density and the Polyakov lines of the interacting model for various values of the coupling $g = 0, 5, 20, 100$. It is evident that the confinement/deconfinement transition becomes sharper as the interaction strength is increased. The behavior of the Polyakov lines show that the model is in a confined phase for most of the values of the chemical potential.   

%%%%%%%% FIGS %%%%%%%%%%%%%%%%
%%%%%%%% FIGS %%%%%%%%%%%%%%%%

%%%%%%%% FIGS %%%%%%%%%%%%%%%%
%%%%%%%% FIGS %%%%%%%%%%%%%%%%

%%%%%%%%%%%%%%%%%%%%%%%%%%%%%%%%%%%%%%%%%%%%%%%%%%%%%%%%%%%%%%%%%%
\section{Conclusions and Discussions}
\label{sec:conclusions}
%%%%%%%%%%%%%%%%%%%%%%%%%%%%%%%%%%%%%%%%%%%%%%%%%%%%%%%%%%%%%%%%%%

In this work we have successfully used complex Langevin dynamics with stochastic quantization to simulate the thermodynamics of large $N$ unitary matrix models with complex actions. We started with a simple matrix model called the $ab$-model and investigated its phase structure analytically and numerically. The numerical simulations show excellent match with analytical results. We also studied a model obtained from the effective theory of QCD on $S^1 \times S^3$ at low temperature and finite quark chemical potential. At zero quark mass and low temperature our simulations showed a series of GWW confinement-deconfinement phase transitions as a function of the chemical potential. The phases are characterized by the distribution of eigenvalues of the Polyakov line in the complex plane. In the large quark mass regime we were also able to observe the Silver Blaze behavior in that the bulk observables are roughly zero until the onset transition to the deconfined phase, which occurs at $\mu = m$. We also simulated the model with a simple nontrivial Polyakov loop interaction turned on. The model prefers to live in the confined phase as the interaction strength is increased.

We also note that each confinement-deconfinement transition in the Polyakov loop is associated with a quark energy level transition. It is interesting to note that the non-monotonic behavior of Polyakov loops have been observed in lattice simulations of QCD with gauge group $SU(2)$ near its saturation density in Ref. \cite{Hands:2010gd}. 

We successfully applied complex Langevin dynamics to QCD on $S^1 \times S^3$ with finite chemical potential and computed several bulk observables. We provided our simulation results on this in Appendix \ref{sec:qcd-finite-cp}. 

There are several interesting future directions. One could consider complex Langevin simulations of the model with several quark flavors with masses $m_f$ and different chemical potentials $\mu_f$. One could also add other types of nontrivial interaction terms into the model and look for cross-over transitions on the $(\mu, \beta)$ plane \cite{Basu:2008uc}. It would also be interesting to see if there exists an AdS/CFT type gravitational dual of the models we studied here. One could ask the question whether the infinite sequence of GWW transitions that we observe in the matrix model can be seen in the dual gravitational description. 

%%%%%%%%%%%%%%%%%%%%%%%%%%
\acknowledgments 
%%%%%%%%%%%%%%%%%%%%%%%%%%

We gratefully acknowledge support from the International Centre for Theoretical Sciences (ICTS-TIFR), the Infosys Foundation and the Indo-French Centre for the Promotion of Advanced Research (IFCPAR/CEFIPRA). We thank Spenta Wadia, Takehiro Azuma, Jun Nishimura and Andrei Alexandru for a careful reading of the manuscript and providing valuable suggestions. We also thank Gautam Mandal, Antonio Gonzalez-Arroyo and Shiraz Minwalla for valuable comments and discussions. We also thank the organizers of ICTS program ``Nonperturbative and Numerical Approaches to Quantum Gravity, String Theory and Holography'', 2018, where this work was presented. PB thanks TIFR theory group for inviting him to present this work as a part of the Quantum Spacetime Seminars.

%%%%%%%%%%%%%%%%%%%%%%%%%%
\appendix
%%%%%%%%%%%%%%%%%%%%%%%%%%

%%%%%%%%%%%%%%%%%%%%%%%%%%%%%%%%%%%%%%%%%%%%%%%%%%%%%%%%%%%%%%%%%%
\section{QCD on $S^1 \times S^3$ at Finite Chemical Potential}
\label{sec:qcd-finite-cp}
%%%%%%%%%%%%%%%%%%%%%%%%%%%%%%%%%%%%%%%%%%%%%%%%%%%%%%%%%%%%%%%%%%

In this section we discuss the results obtained through complex Langevin simulations of QCD on $S^1 \times S^3$ with finite chemical potential, zero quark mass and at low temperature, given by the action in Eq. (\ref{eq:action-multi-level}).

%%%%%%%%%%%%%%%%%%%%%%%%%%%%%%%%%%%%%%%%%%%%%%%%%%%%%%%%%%%%%%%%%%
\subsection{Fermion number $\langle f_N \rangle$}
%%%%%%%%%%%%%%%%%%%%%%%%%%%%%%%%%%%%%%%%%%%%%%%%%%%%%%%%%%%%%%%%%%

In Fig. \ref{fig:fN-nc3_30-nf3_30-m0-b30} we show $\langle f_N \rangle$ as a function of $\mu$ for low temperatures for $m = 0$. The presence of an occupation level structure is evident. The transitions occur when $\epsilon_l - \mu$ changes sign, that is, when $\mu$ passes a quark energy level.

It is interesting to compare with the results obtained in Ref. \cite{Hands:2010zp}. We also note that in Ref. \cite{Banerjee:2010kc} Banerjee and Chandrasekharan observed the same level structure in the particle number in the nonlinear $O(2)$ sigma model. 

The fermion number can be used as an order parameter of the confinement-deconfinement transitions in the large $N$ theory. The first and second derivatives of the grand potential, $\langle f_N \rangle$ and $\langle \partial f_N/\partial \mu \rangle$ are continuous as a function of the chemical potential but the third derivative $\langle \partial^2 f_N/\partial \mu^2 \rangle$ is discontinuous. This indicates that the transitions are third order, of the GWW type.

%%%%%%%% FIGS %%%%%%%%%%%%%%%%
\begin{figure}[h!]
  \hfill
  \begin{minipage}[t]{.49\textwidth}
    \begin{center}
\includegraphics[width=0.99\textwidth]{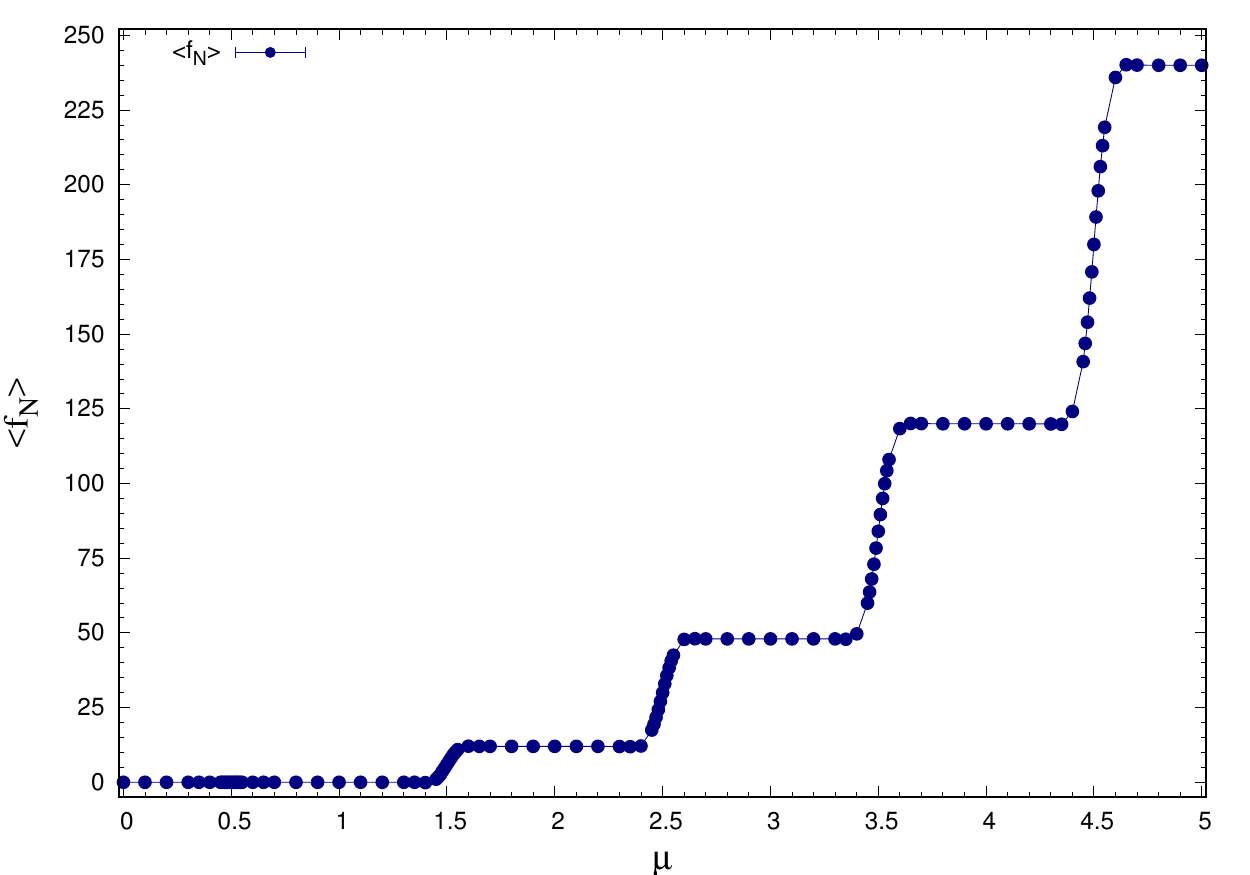}
    \end{center}
  \end{minipage}
  \hfill
  \begin{minipage}[t]{.49\textwidth}
    \begin{center}
\includegraphics[width=0.99\textwidth]{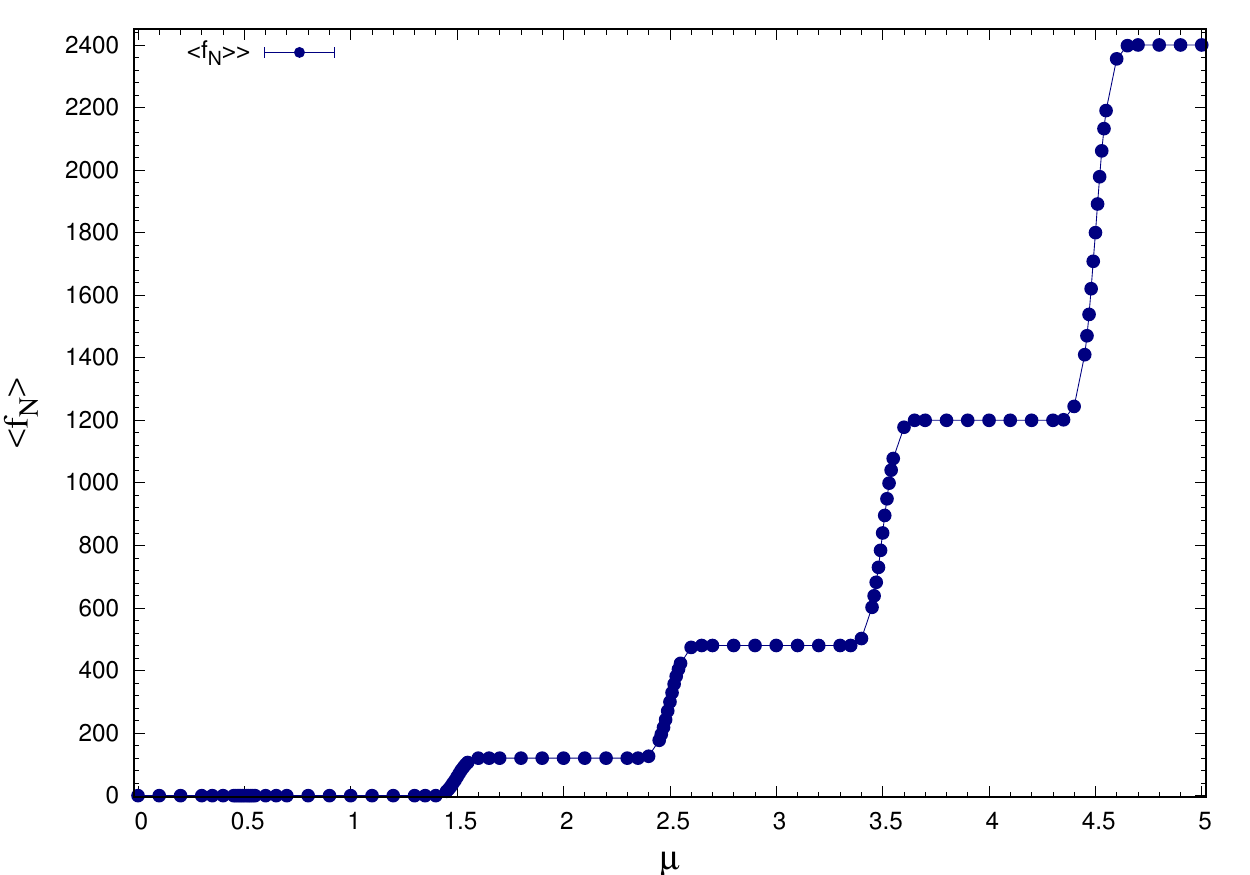}
    \end{center}
  \end{minipage}
  \hfill
\caption{Expectation values of the effective fermion number $\langle f_N \rangle$ as a function of the quark chemical potential for QCD on $S^1 \times S^3$. (See Eq. (\ref{eq:action-multi-level}) for the form of the action.) Here $m = 0$, inverse temperature $\beta = 30$, $N = N_f = 3$ (Left) and $N = N_f = 30$ (Right). The data are obtained through complex Langevin simulations with adaptive Langevin step sizes $\Delta \tau \leq 0.00005$, thermalization steps $N_{\rm therm} = 10000$, generation steps $N_{\rm gen} = 50000$ and with measurements performed with an interval of $100$ steps. The solid lines are to guide the eye.} 
\label{fig:fN-nc3_30-nf3_30-m0-b30}
\end{figure}
%%%%%%%% FIGS %%%%%%%%%%%%%%%%

%%%%%%%%%%%%%%%%%%%%%%%%%%%%%%%%%%%%%%%%%%%%%%%%%%%%%%%%%%%%%%%%%%%%%%%%%%%%%
\subsection{Polyakov Lines $\langle P \rangle$ and $\langle P^{-1} \rangle$}
%%%%%%%%%%%%%%%%%%%%%%%%%%%%%%%%%%%%%%%%%%%%%%%%%%%%%%%%%%%%%%%%%%%%%%%%%%%%%

When the chemical potential is zero the Polyakov line $\langle P \rangle$ and the conjugate Polyakov line $\langle P^{-1} \rangle$ coincide and it is no longer the case for non-zero chemical potential. In Fig. \ref{fig:fN-nc3_30-nf3_30-m0-b30} we show $\langle P \rangle$ and $\langle P^{-1} \rangle$ as a function of $\mu$. Each spike in $\langle P \rangle$ and $\langle P^{-1} \rangle$ corresponds to a level transition in $\langle f_N \rangle$. They exhibit similar behavior as a function of $\mu$ however, the the behavior of $\langle P^{-1} \rangle$ always precedes that of $\langle P \rangle$ at the start and finish of each level transition. We note that the lines peak at $\mu = 1.5, 2.5, \cdots$. We also note that the widths of deconfined regions increase as $\mu$ is increased. 

%%%%%%%% FIGS %%%%%%%%%%%%%%%%
\begin{figure}[h!]
  \hfill
  \begin{minipage}[t]{.49\textwidth}
    \begin{center}
\includegraphics[width=0.99\textwidth]{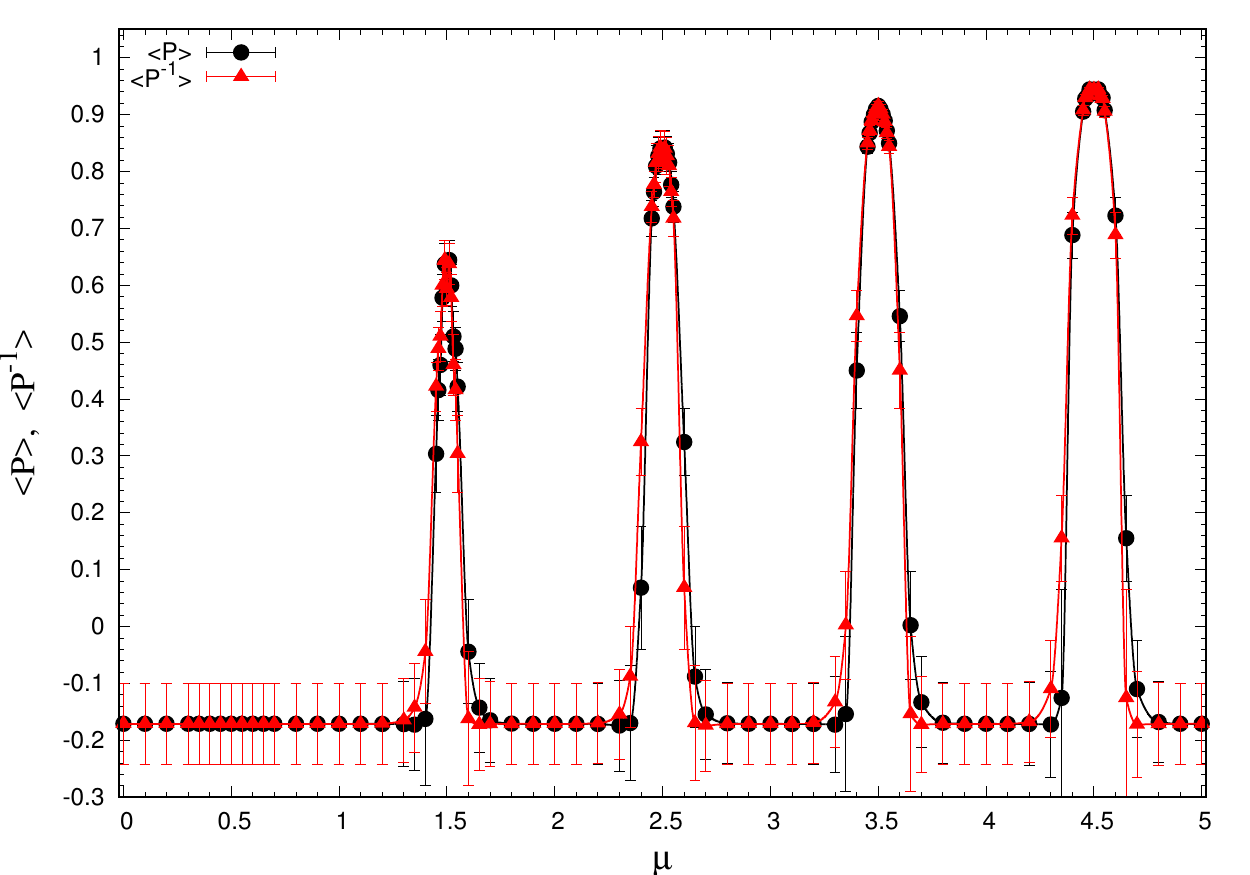}
    \end{center}
  \end{minipage}
  \hfill
  \begin{minipage}[t]{.49\textwidth}
    \begin{center}
\includegraphics[width=0.99\textwidth]{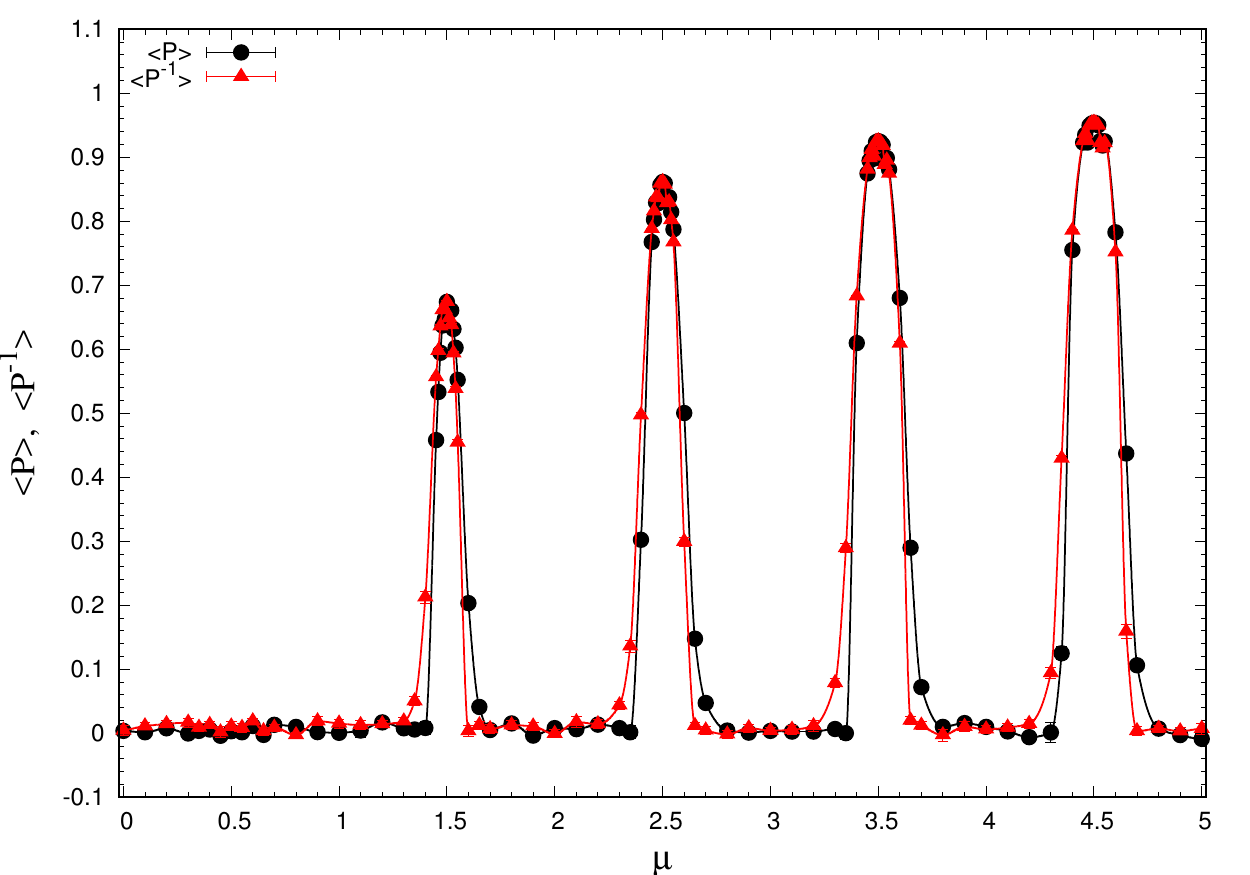}
    \end{center}
  \end{minipage}
  \hfill
\caption{Expectation values of the Polyakov line $\langle P \rangle$ and inverse Polyakov line $\langle P^{-1} \rangle$ as a function of the quark chemical potential $\mu$ for QCD on $S^1 \times S^3$. (See Eq. (\ref{eq:action-multi-level}) for the form of the action.) Here $m = 0$, inverse temperature $\beta = 30$, $N = N_f = 3$ (Left) and $N = N_f = 30$ (Right). The data are obtained through complex Langevin simulations with adaptive Langevin step sizes $\Delta \tau \leq 0.00005$, thermalization steps $N_{\rm therm} = 10000$, generation steps $N_{\rm gen} = 50000$ and with measurements performed with an interval of $100$ steps. The solid lines are to guide the eye.} 
\label{fig:fN-nc3_30-nf3_30-m0-b30}
\end{figure}
%%%%%%%% FIGS %%%%%%%%%%%%%%%%

%%%%%%%%%%%%%%%%%%%%%%%%%%%%%%%%%%%%%%%%%%%%%%%%%%%%%%%%%%%%%%%%%%
\subsection{Pressure $\langle p \rangle$ and Energy $\langle E \rangle$}
%%%%%%%%%%%%%%%%%%%%%%%%%%%%%%%%%%%%%%%%%%%%%%%%%%%%%%%%%%%%%%%%%%

\begin{figure}[h!]
\centering
\includegraphics[width=5.0in]{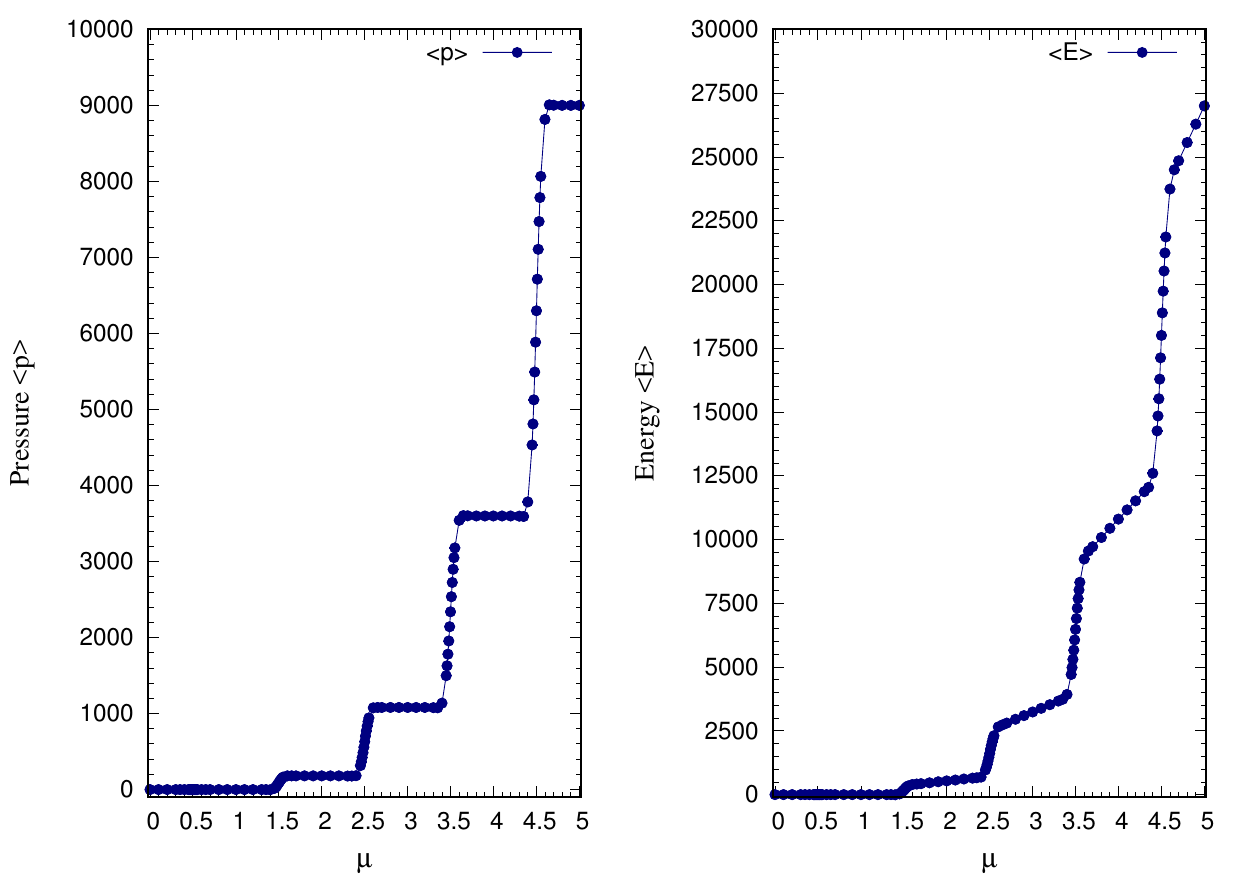}
\caption{(Left) Pressure $\langle p \rangle$ and (Right) energy $\langle E \rangle$ as a function of the quark chemical potential for QCD on $S^1 \times S^3$. (See Eq. (\ref{eq:action-multi-level}) for the form of the action.) Here $N = N_f = 3$, $m = 0$ and inverse temperature $\beta = 30$. The data are obtained through complex Langevin simulations with adaptive Langevin step sizes $\Delta \tau \leq 0.00005$, thermalization steps $N_{\rm therm} = 10000$, generation steps $N_{\rm gen} = 50000$ and with measurements performed every $100$ steps. The solid lines are guide to the eye.}
\label{fig:p-E-nc3-nf3-m0-b30}
\end{figure}

\begin{figure}[h!]
\centering
\includegraphics[width=5.0in]{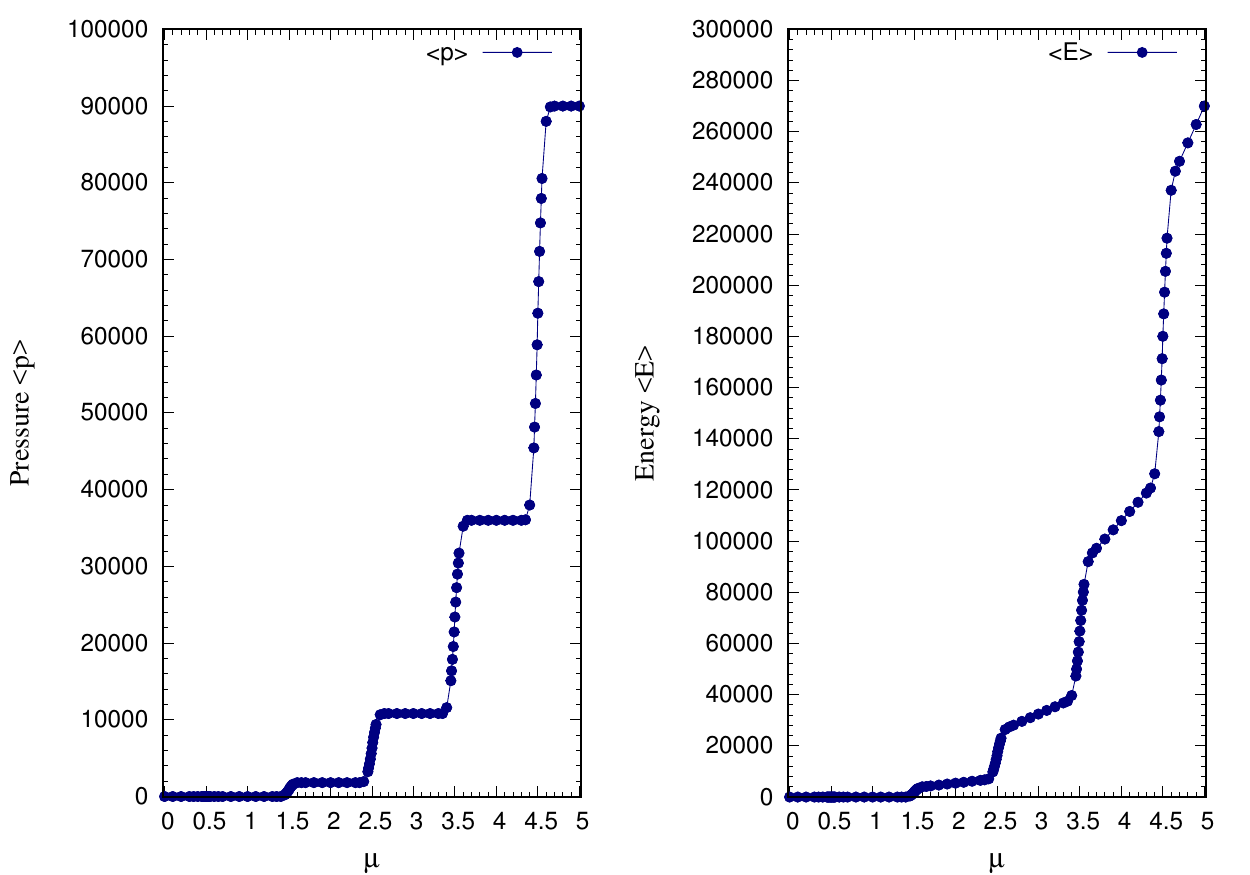}
\caption{(Left) Pressure $\langle p \rangle$ and (Right) energy $\langle E \rangle$ as a function of the quark chemical potential for QCD on $S^1 \times S^3$. (See Eq. (\ref{eq:action-multi-level}) for the form of the action.)  Here $N = N_f = 30$, $m = 0$ and inverse temperature $\beta = 30$. The data are obtained through complex Langevin simulations with adaptive Langevin step sizes $\Delta \tau \leq 0.00005$, thermalization steps $N_{\rm therm} = 10000$, generation steps $N_{\rm gen} = 50000$ and with measurements performed with an interval of $100$ steps. The solid lines are to guide the eye.}
\label{fig:p-E-nc30-nf30-m0-b30}
\end{figure}

In Figs. \ref{fig:p-E-nc3-nf3-m0-b30} and \ref{fig:p-E-nc30-nf30-m0-b30} we provide the pressure multiplied by the 4-volume and energy $\langle E \rangle = - \langle p \rangle + \mu \langle f_N \rangle$. We note that the pressure exhibits a level structure. The energy levels are not horizontal. The factor $\mu$ in front of the fermion number causes the levels to rise linearly with $\mu$.  

%%%%%%%% FIGS %%%%%%%%%%%%%%%%
%%%%%%%% FIGS %%%%%%%%%%%%%%%% 

%%%%%%%% FIGS %%%%%%%%%%%%%%%%
%%%%%%%% FIGS %%%%%%%%%%%%%%%% 

\begin{figure}[h!]
\centering
\includegraphics[width=5.0in]{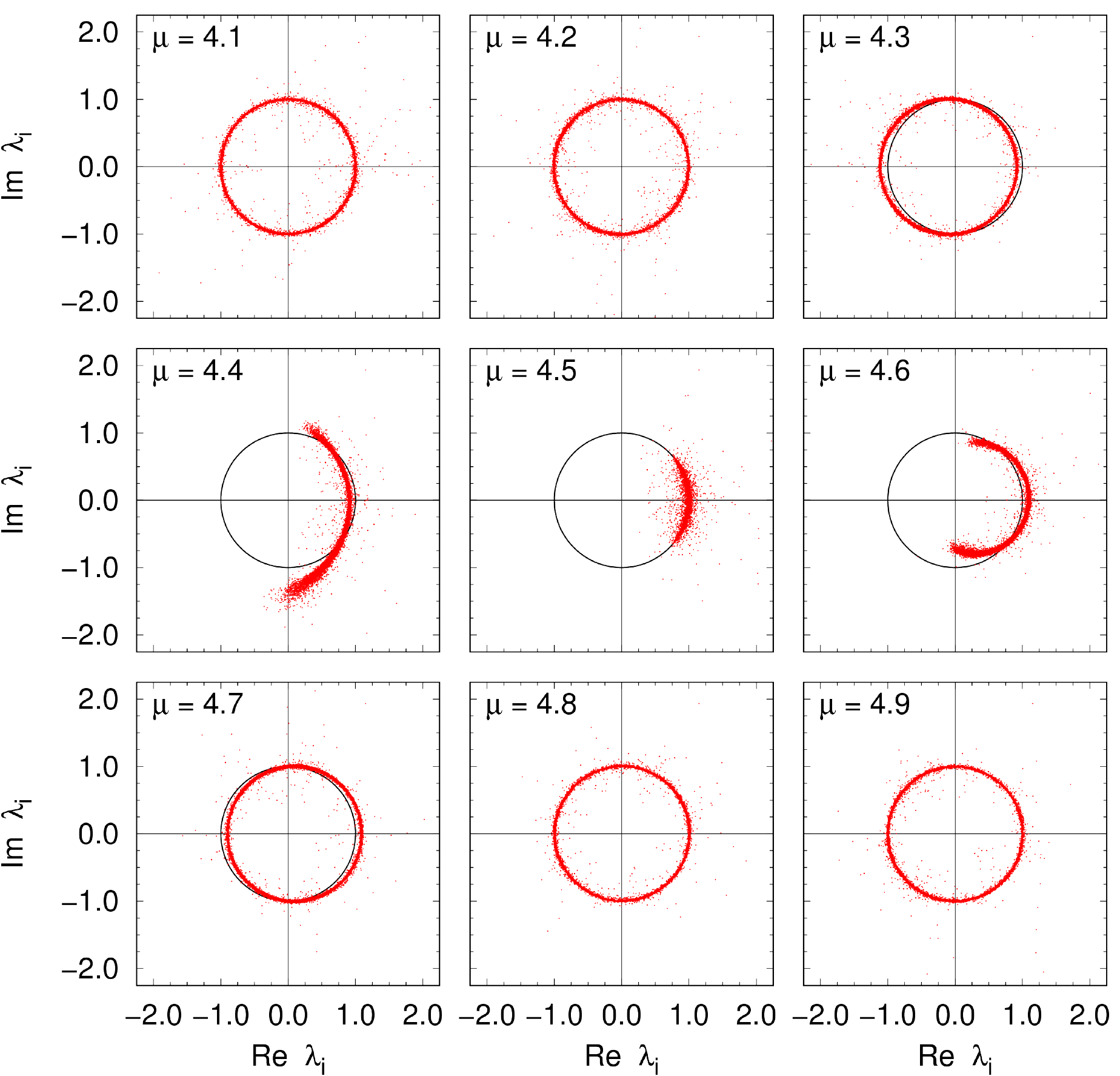}
\caption{The eigenvalue distributions in the confined, deconfined and again confined phases as a function of the quark chemical potential for QCD on $S^1 \times S^3$. (See Eq. (\ref{eq:action-multi-level}) for the form of the action.) Here $N = N_f = 30$, $m = 0$ and inverse temperature $\beta = 30$ (low $T$). The data are obtained through complex Langevin simulations with adaptive Langevin step sizes $\Delta \tau \leq 0.00005$, thermalization steps $N_{\rm therm} = 10000$, generation steps $N_{\rm gen} = 50000$ and with measurements performed with an interval of $100$ steps. The solid unit circles are guide to the eye.}
\label{fig:eigs-nc30-nf30-m0-b30_noise}
\end{figure}

In Fig. \ref{fig:eigs-nc30-nf30-m0-b30_noise} we show the eigenvalue distributions in the confined and deconfined phases as a function of the quark chemical for $N = N_f = 30$ and barious $\mu$ values.

%%%%%%%% FIGS %%%%%%%%%%%%%%%%
%%%%%%%% FIGS %%%%%%%%%%%%%%%% 

%%%%%%%%%%%%%%%%%%%%%%%%%%%%%%%%%%%%%%%%%%%%%%%%%%
\section{Reliability of Complex Langevin Dynamics}
\label{sec:appendix_c}
%%%%%%%%%%%%%%%%%%%%%%%%%%%%%%%%%%%%%%%%%%%%%%%%%%

We would like to justify the use of complex Langevin dynamics for the matrix models we simulated in this work. In Ref. \cite{Nagata:2016vkn,Nagata:2018net} the authors suggested a possible criterion to determine the correct convergence of the complex Langevin method -- the probability distribution of the magnitude of the drift term should fall off exponentially or faster. This criterion can, in general, be violated if the complexified fields develop large imaginary parts (the {\it excursion problem}). In Fig. \ref{fig:prob-dist-histogram} we show the probability distributions $P(u)$ for the magnitude of the drift term
\beq
u = \sqrt{\frac{1}{N^3} \sum_{i=1}^N \left| \frac{\partial S}{\partial \theta_i} \right|^2},
\eeq 
of the single level $SU(N)$ matrix model.

However, in our case the plots hint that the probability distribution falls off like a power law with $u$ even though we have excellent agreements with analytical results. Figs. \ref{fig:fN_N500} and \ref{fig:P_inv_P_N500} show excellent agreement between simulation and analytical data in this model. We also observed a similar fall off behavior in the $ab$-model. We think this needs further investigations and we save it for future work. 

\begin{figure}[h!]
\centering
\includegraphics[width=6.0in]{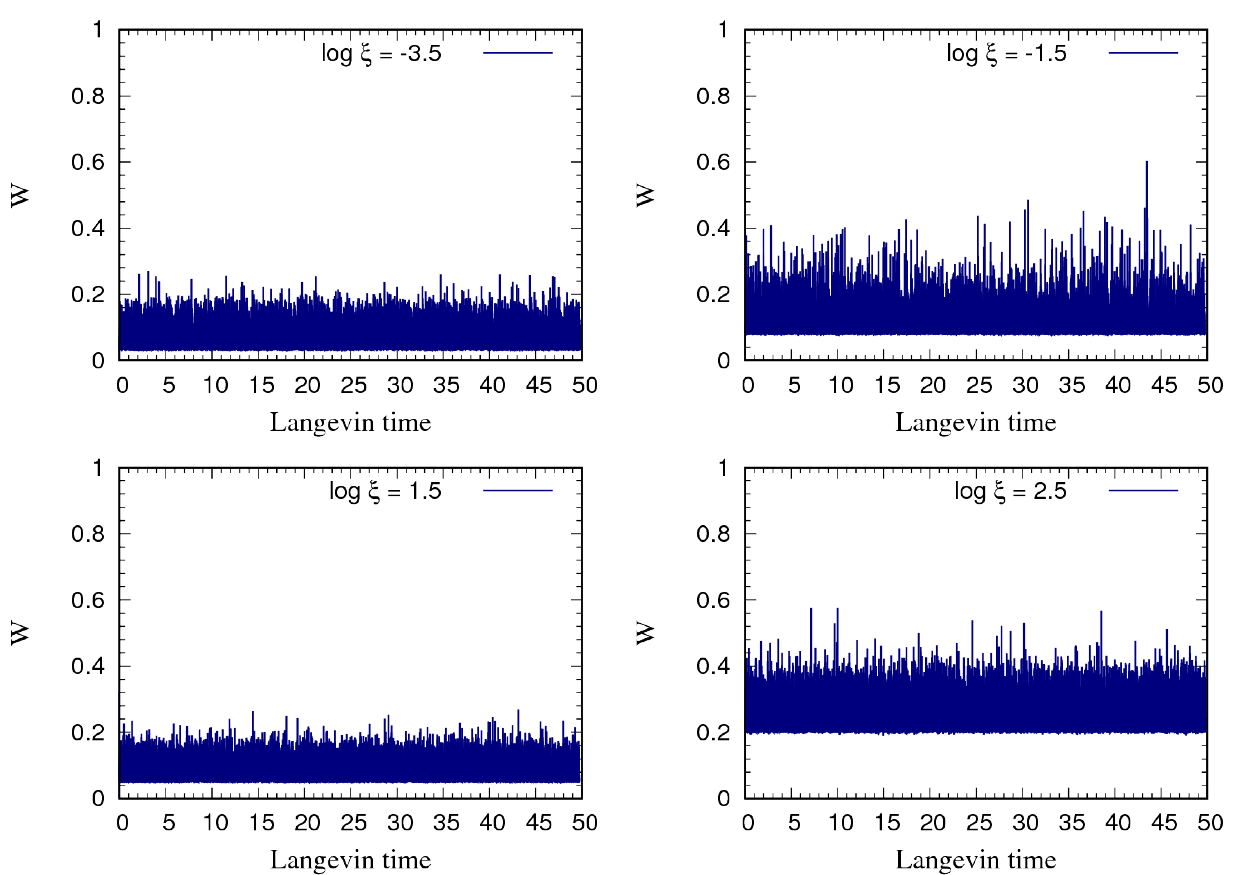}
\caption{The unitarity norm $W$ against the Langevin time for the single level $SU(N)$ matrix model with $N = N_f = 500$ and quark mass $m = 0$. (See Eq. (\ref{eq:act-single-level}) for the form of the action.) The plots are for $\log \xi = -3.5, -1.5, 1.5$ and $2.5$. We used adaptive Langevin step sizes $\Delta \tau \leq 0.00005$ in the simulations.}
\label{fig:unitarity-norm}
\end{figure}

\begin{figure}[htp]

\subfloat[]{\includegraphics[width=3.0in]{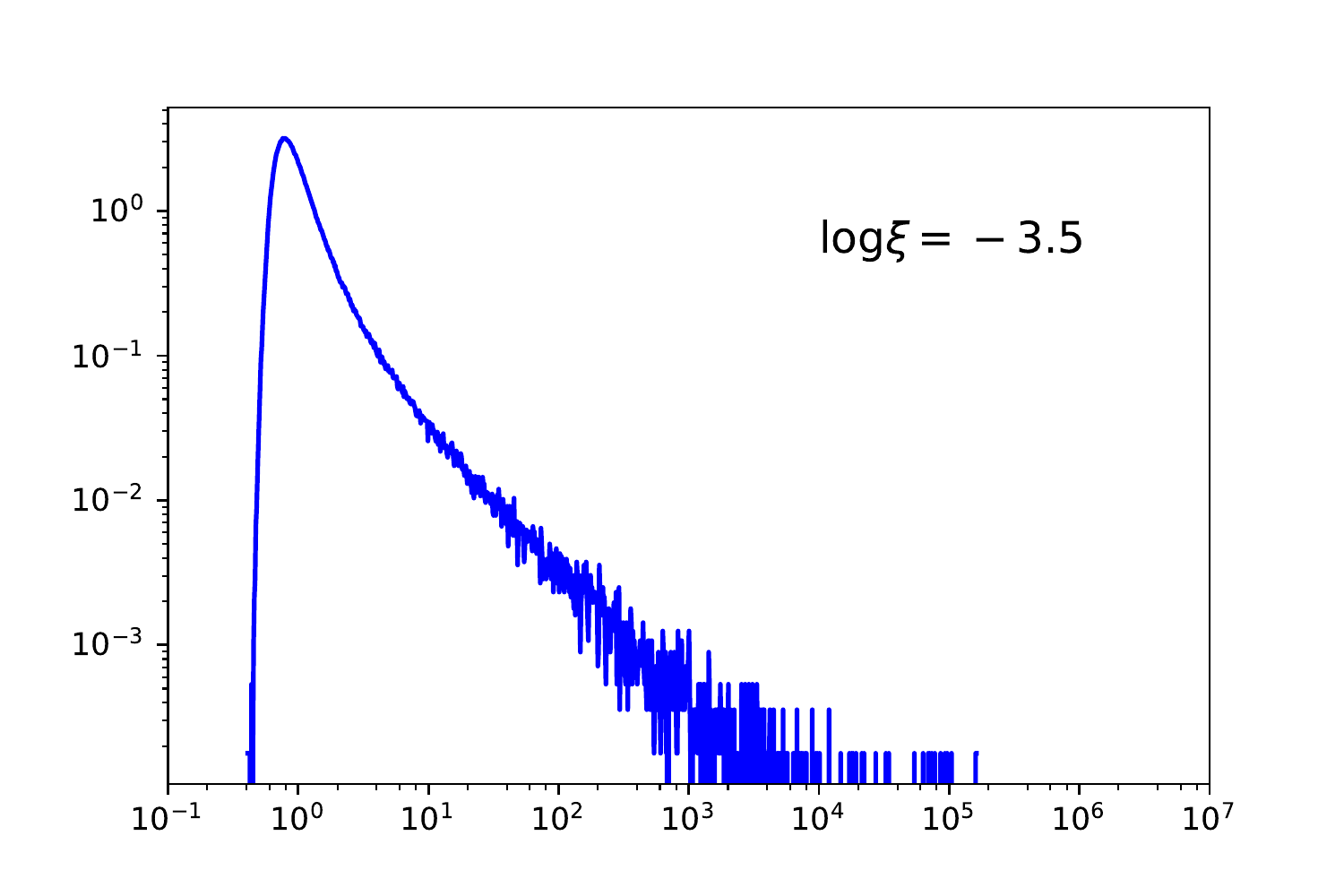}}
\subfloat[]{\includegraphics[width=3.0in]{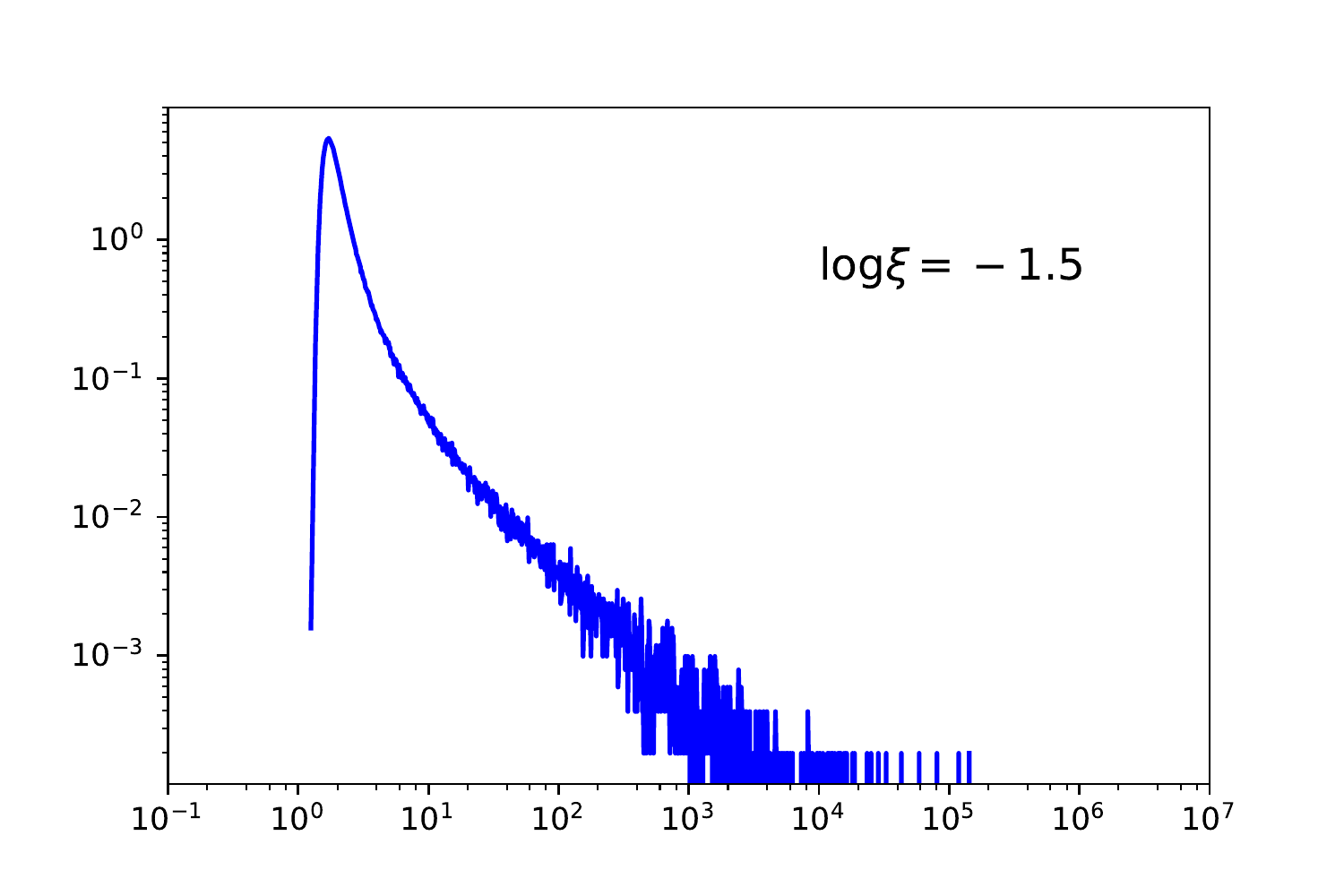}}

\subfloat[]{\includegraphics[width=3.0in]{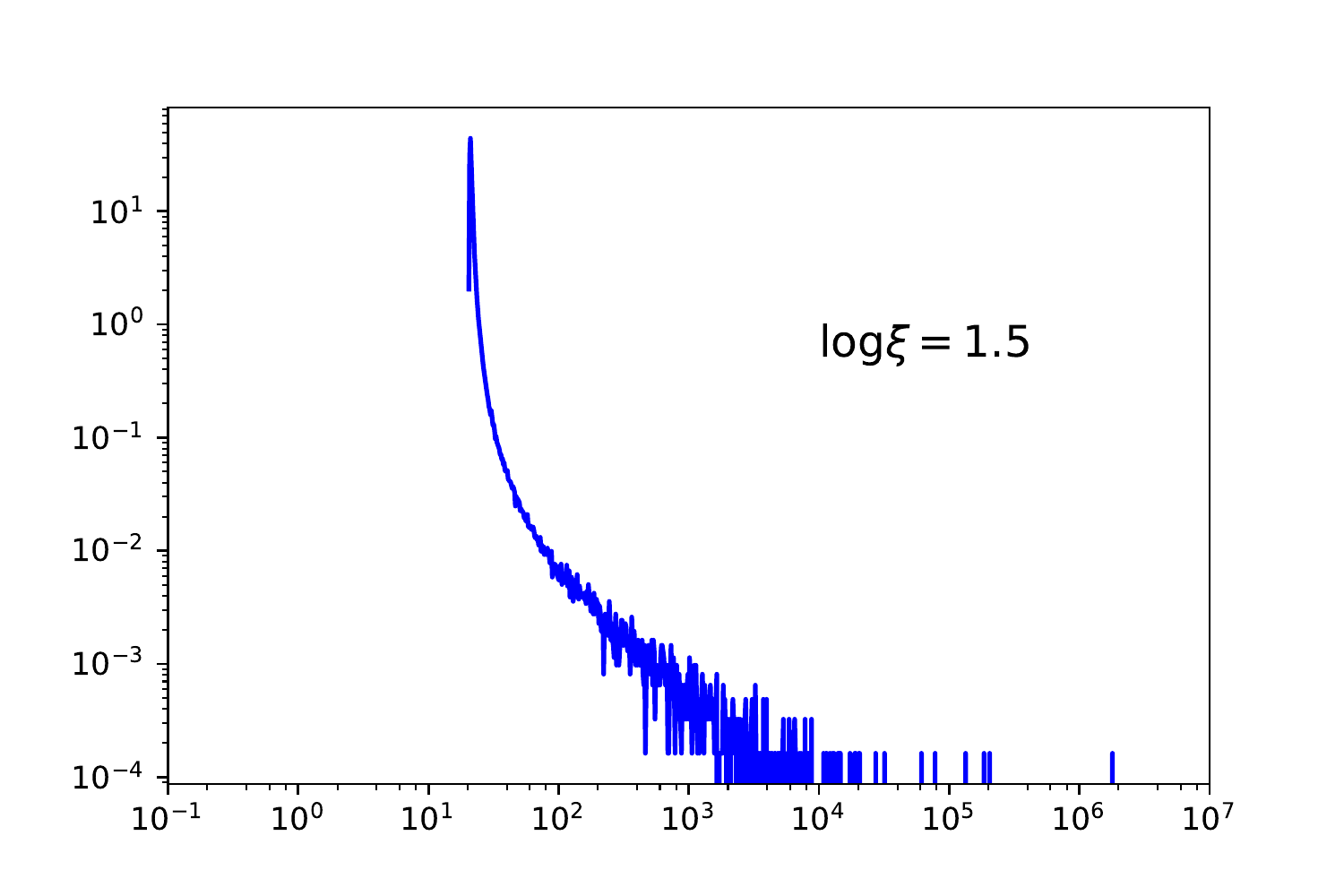}}
\subfloat[]{\includegraphics[width=3.0in]{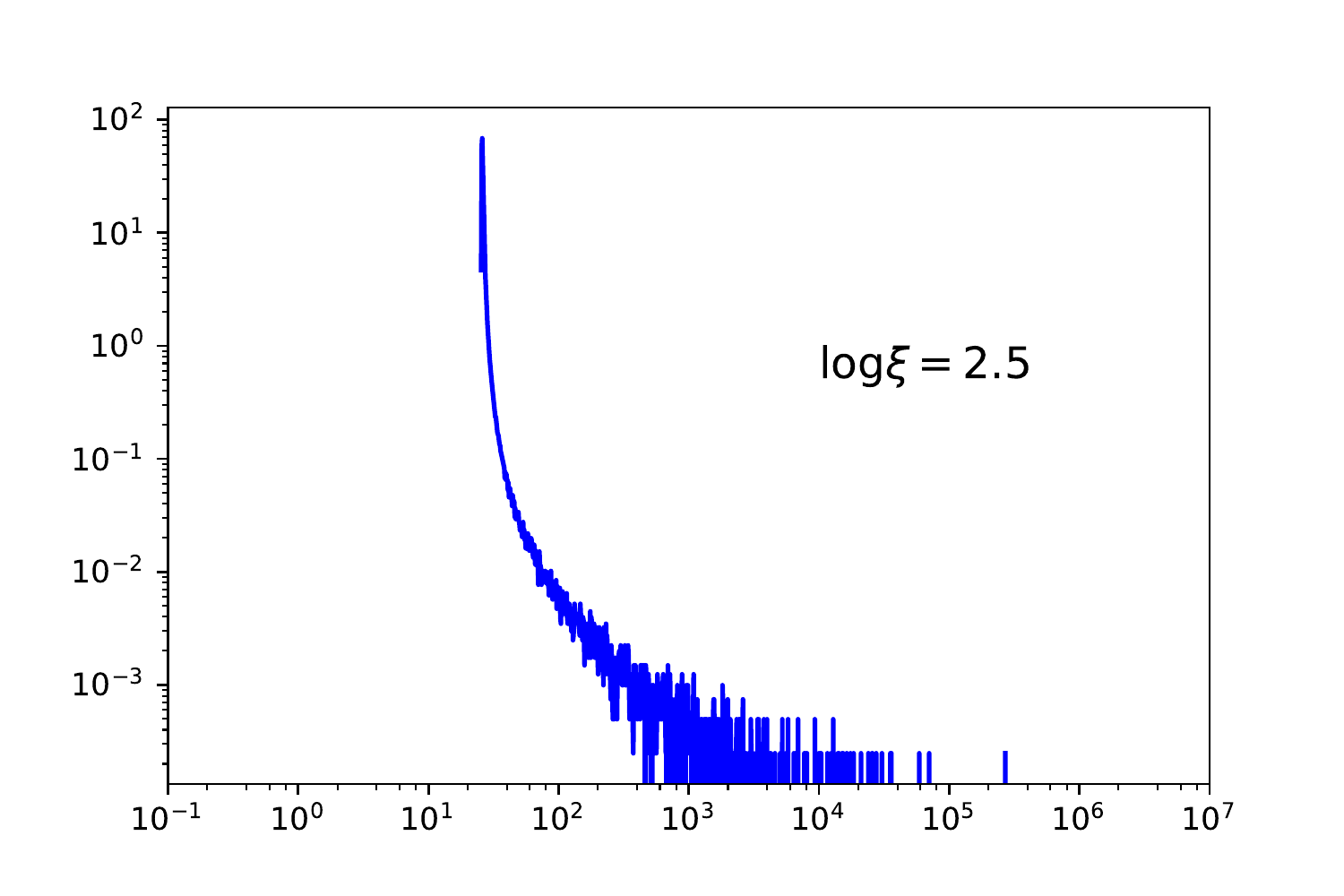}}

\caption{The probability distributions $P(u)$ of the magnitude of the drift term $u$ for the single level matrix model with positive chemical potential plotted in a log-log plot. (See Eq. (\ref{eq:act-single-level}) for the form of the action.) We used adaptive Langevin step sizes $\Delta \tau \leq 0.00005$ in the simulations. The simulation data are for quark mass $m = 0$ and for $N = N_f = 500$. The Langevin evolution is performed for $10^6$ steps.}
\label{fig:prob-dist-histogram}
\end{figure}

\begin{figure}[h!]
\centering
\includegraphics[width=7.0in]{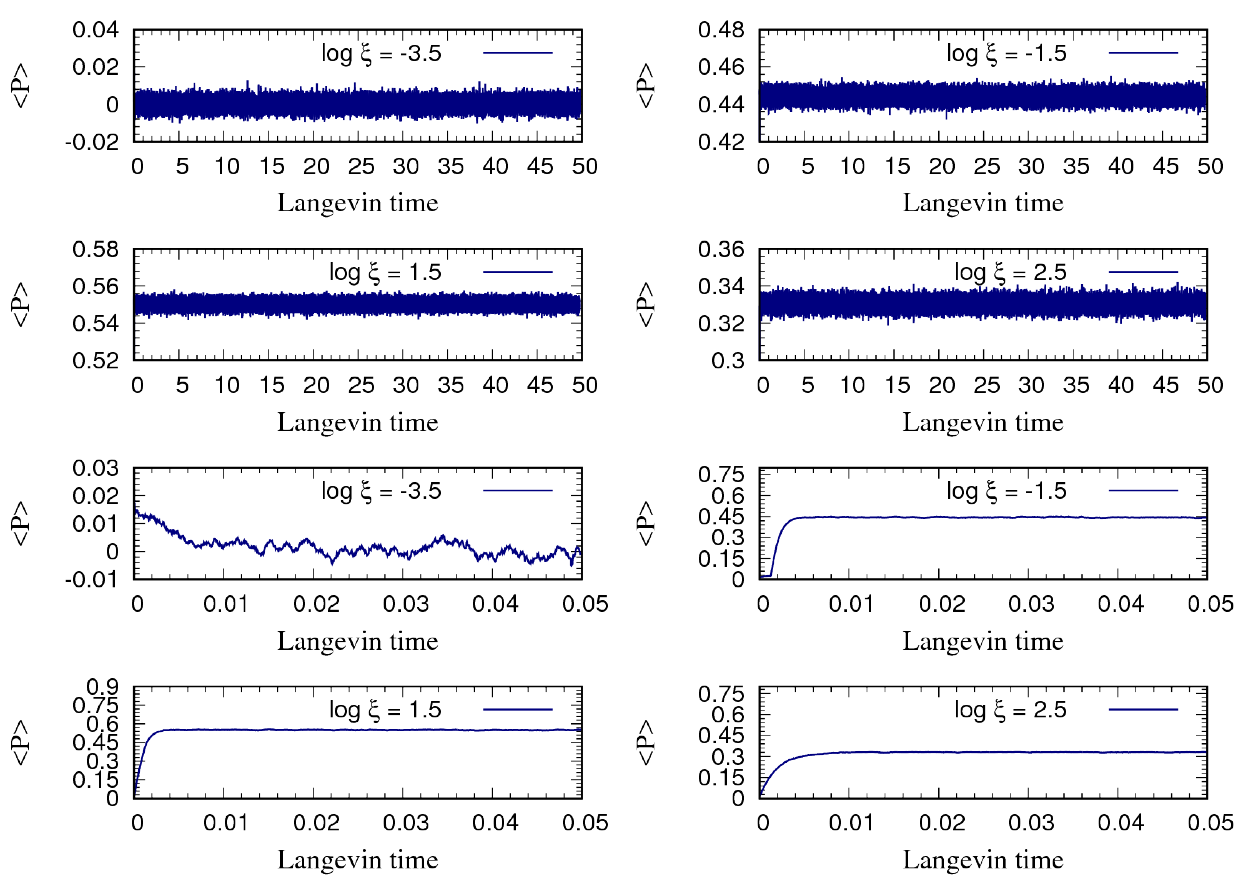}
\caption{The Polyakov loop $\langle P \rangle$ against the Langevin time for the single level $SU(N)$ matrix model with $N = N_f = 500$ and quark mass $m=0$. (See Eq. (\ref{eq:act-single-level}) for the form of the action.) The plots are for $\log \xi = -3.5, -1.5, 1.5$ and $2.5$. We used adaptive Langevin step sizes $\Delta \tau \leq 0.00005$ in the simulations. The bottom four plots show the thermalizations of the observables shown on the top four plots.}
\label{fig:poly-evolution}
\end{figure}

It is desirable to have a well localized distribution of dynamical variables of the theory in the complexified field configuration space. A convenient measure of the size of the distribution in imaginary directions of the field variables is the unitarity norm \cite{Sexty:2013ica} defined as
\beq
W \equiv \frac{1}{N} \Tr \left( \left( U U^\dagger - 1 \right)^2 \right) \geq 0,
\eeq
with the equality reaching when the fields take values in $SU(N)$. In Fig. \ref{fig:unitarity-norm} we show the unitarity norm as a function of Langevin time for the single level $SU(N)$ matrix model with $N = N_f = 500$, quark mass $m = 0$ and inverse temperature $\beta = 30$ (low $T$). We see that the unitarity norm remains bounded in the simulations. In Fig. \ref{fig:poly-evolution} we show the Langevin evolution of the Polyakov line observable for the same set of parameters.

%%%%%%%% FIGS %%%%%%%%%%%%%%%%
%%%%%%%% FIGS %%%%%%%%%%%%%%%% 

\begin{comment}

%%%%%%%% FIGS %%%%%%%%%%%%%%%%
\begin{figure}[h!]
\centering
\includegraphics[width=3.0in]{FIGS/fig_m3p50.pdf}
\includegraphics[width=3.0in]{FIGS/fig_m1p50.pdf}

\includegraphics[width=3.0in]{FIGS/fig_1p50.pdf}
\includegraphics[width=3.0in]{FIGS/fig_2p50.pdf}
\caption{The probability distribution $P(u)$ of the magnitude of the drift term $u$ for the single level matrix model with positive chemical potential plotted in a log-log plot. (See Eq. (\ref{eq:act-single-level}) for the form of the action.) We used adaptive Langevin step sizes $\Delta \tau \leq 0.00005$ in the simulations. The simulation data are for quark mass $m = 0$ and for $N = N_f = 500$. The Langevin evolution is performed for $10^6$ steps.}
\label{fig:prob-dist-histogram}
\end{figure}
%%%%%%%% FIGS %%%%%%%%%%%%%%%% 

\end{comment}

%%%%%%%%%%%%%%%%%%%%%%% FIG %%%%%%%%%%%%%%%%%%%%%%%%%%%%%%%%%%%%%%
%%%%%%%%%%%%%%%%%%%%%%% FIG %%%%%%%%%%%%%%%%%%%%%%%%%%%%%%%%%%%%%%

%%%%%%%% FIGS %%%%%%%%%%%%%%%%
%%%%%%%% FIGS %%%%%%%%%%%%%%%% 

\bibliographystyle{apsrev4-1}
\bibliography{qcds}
\end{document}